\documentstyle[preprint,eqsecnum,aps,psbox]{revtex}
\tightenlines

\newcommand{\ph}{p\,-\,$^3$He\ } 
\begin{document}
\title{p-$^3$He Effective Potentials based on the
		Pauli Corrected Resonating Group
		Method}
\author{S. Gojuki $^1$, S. Oryu$^1$\footnote{Email address: oryu@ph.noda.sut.ac.jp},  and S. A. Sofianos$^2$}
\address{
$^1$Department of Physics, Faculty of Science and Technology,
      Science University of Tokyo, \\
	  2641 Yamazaki, Noda, Chiba 278-8510, Japan}
\address{
$^2$Physics Department, University of  
      South Africa, Pretoria 0003, South Africa}
\date{\today}
\maketitle
\begin{abstract}
Effective interactions that  fit the low energy 
p-$^3$He experimental data  have been constructed.
They are based on
the Resonating Group Method and a modified Orthogonality Condition
Model in which Pauli and partly Pauli forbidden states
are removed from the spectrum.
Partial waves up to $L=3$ have been  considered. The 
LS force acting between the proton and $^3$He has been included 
phenomenologically, while the Coulomb interaction has been 
incorporated  using a renormalization technique for a screened 
Coulomb interaction.  
The potentials are also given in a separable momentum space form, 
obtained using the Ernst-Shakin-Thaler (EST) method. 
In all cases the potentials  generate  phase shifts 
that  fit  well the low energy experimental data.  
\end{abstract}
\vspace{.5cm}
PACS Numbers: 21.45+v, 21.30.-x, 21.60.Gx, 25.10+s

\widetext
\section{Introduction}
$^3$He plays an important role  in a variety of  light-cluster
reactions, especially at low energies.
One such reaction is  $^3$He(d,p)$^4$He which is of
interest in  primordial nucleosynthesis studies 
and in studies related to the abundance of elements in the Universe.
It is also of interest as a fusion reaction, as it generates no neutrons.    
At medium energies, this reaction with polarized spins has been used
to constrain the deuteron D-state probability \cite{Dstate}.
Another interesting aspect of this reaction is that the $^3$He 
nucleus behaves like a neutron,  because the proton pair in $^3$He 
almost forms a closed shell structure and therefore $^3$He 
can serve as a good substitute for a neutron target.

The $^3$He(d,p)$^4$He reaction can be modeled as  a
three-body  p--n--$^3$He problem  which can be handled using the 
Faddeev formalism \cite{interfaddeev1,interfaddeev2,multifaddeev}. This
presupposes a knowledge of the  p--$^3$He and the n--$^3$He 
interactions and of the nuclear potential.
The nucleon-trinucleon interaction has been the subject of several
investigations in the past. Despite the effort made,
the interaction  is far from being understood.  Many
questions concerning the range and shape of the 
potential, as well as its  energy--, parity-, and $\ell$-dependence.
For example,  in the  optical model 
approach the shape is either pre-chosen  (Woods-Saxon\cite{Pot}, Gaussian
\cite{Neu} {\it etc.}) or  is fitted to  specific scattering cross sections
\cite{Oers,Sciav}, or to phase shifts which are used as input
 to construct phase equivalent local potentials  
via inverse scattering methods \cite{Fied}. Some experimental 
data on p--$^3$He elastic scattering up to $\sim 10$ MeV
were analyzed within the framework of a separable potential model, 
phase shifts for the p--$^3$He and n--$^3$H systems were
generated \cite{Pisent}.

The quality  of scattering data used as input  and their reproducibility
is another source of ambiguity for the field practitioner. 
Extensive phase shift analyses for the nucleon-trinucleon
system were performed in the 1960s and 1970s 
\cite{exp1,exp2,exp3,exp4,exp5,exp6,exp7,exp8}.
The most recent was carried out  by Yoshino {\it et al.} 
\cite{yoshino}  in the  $E_p=4.0$-$19.48$\,MeV  energy region.
As was pointed out, in Ref. \cite{Fied}, performing phase shift
analyses and constructing a potential in a specific channel requires 
extreme care to insure that the interaction has  the correct 
energy--, parity-, and $\ell$-dependence.  The mere use of cross sections
is not sufficient to uniquely extract all of these dependencies.
An alternative scheme is to generate the phase shifts from reliable
theoretical models, such as the Resonating Group Model (RGM).
Within the RGM formalism, Reichstein  {\it et al.}\cite{rgm}
extracted the intercluster phase shifts for  this system
while Furutani {\it et al.}\cite{furutani} calculated the phase shifts 
by using the Generator Coordinate Method (GCM).
Their results are in good agreement with the differential 
cross section and the polarization data in the low energy region.
Although the calculated energy spectrum of the $T=1$ resonances fits
equally well the experimental data\cite{exp2},
these experimental data have been replaced 
in the recent article by Tilley {\it et al.}~\cite{level}.
Finally, we mention  that an optical potential approach
has been employed  by  Teshigawara {\it et al.}\cite{cssm} to study
the n--$^3$He system  within the Complex Scaled Siegert Method.

In this paper, we employ  the p--$^3$He nonlocal RGM interaction
\cite{rgm,unified} and its variant version, the  so-called
Orthogonality Condition  Model (OCM)  
in which the Pauli forbidden states (PFS) are removed \cite{ocm,aal}. 
Furthermore,  to reproduce 
the experimental phase shifts 
sufficiently well and to treat the degenerate states
stemming from the absence of  LS forces from the RGM kernel, 
we have introduced  phenomenological LS forces using a technique
similar to that  of Ref. \cite{lsp}.
Tensor forces could also be considered for the 
$^3$S$_1$-$^3$D$_1$, $^1$P$_1$-$^3$P$_1$, $^3$P$_2$-$^3$F$_2$, 
and $^1$D$_2$-$^3$D$_2$ channels. However, the corresponding
experimental mixing parameters were found to be
negligibly small \cite{yoshino} and therefore they have been omitted.

At higher energies, the absorptive part of the phase 
shifts  could be  important, in which case an imaginary part 
should be included in the potential.
However,  in a recent phase  shift analysis of Yoshino {\it et al.}
\cite{yoshino}, neither the reflection parameters for the coupling effects 
in  different particle channels nor the 
imaginary part of the phase shifts  above the break-up threshold 
were found to be significant.  
Therefore, we will not consider the construction 
of complex  p--$^3$He potentials and concentrate instead
on real  potentials that  fit the low energy experimental data well.
 
Finally, we have endeavored to represent the potentials,
that reproduce the phase shifts without Coulomb effects,
in a separable form  with appropriate form factors and parameters
by using the Ernst-Shakin-Thaler (EST) method \cite{EST}. 
Such potentials are useful in solving the three-body  Faddeev equations
in momentum space.
         
In Sec. II we describe our methods to construct the p-$^3$He potential
and in Sec. III we present our results. In Sec. IV  we present
the separable expansions for the  p-$^3$He potentials while
in Sec.  V we present our  discussions and conclusions. 
Finally, some technical details
and formulas are shifted in  Appendices A and B.
%
%
\section{The \ph Potential}
The construction of our  p--$^3$He effective potential  is based on 
the RGM formalism, which is given in configuration space in the paper by
Reichstein {\it et al.} \cite{rgm}.   
In operator form the RGM potential  is  
\begin{equation}
\label{rgmpota}
	V^{\rm RGM} \equiv W+E{\cal N} ,
\end{equation}
where $W$ is the energy independent local and nonlocal potential, 
$E$ is the relative energy between the proton and $^3$He 
while  ${\cal N}$ is the norm (-integral)  kernel (see Appendix A for details).

A common  problem encountered in the RGM formalism is the existence
of the PFS which can be removed  via the OCM technique (see Appendix A). 
The OCM potential is given by
\begin{equation}
	V^{{\rm OCM}}=\frac{1}{\sqrt{1-{\cal N}}}\left (\,h_0+W\,\right)
	\frac{1}{\sqrt{1-{\cal N}}}-h_0\,
\end{equation}
\narrowtext
where $h_0$ is the kinetic energy operator between the proton and $^3$He.
It should be noted that $V^{{\rm OCM}}$ is an energy independent potential. 

The above operator equations can be brought into a more 
concrete form by using  the eigen-values
 $\gamma_i$ and the eigen-functions $\tilde{U}_i$ of the norm kernel,
\widetext
\begin{equation}
\label{ocmp2-1}
	V^{{\rm OCM}}=\sum_{ij=1}^{\infty}| \tilde{U}_i\rangle
	[\frac{1}{\sqrt{1-\gamma_i}}(h_{0ij}+W_{ij})
	\frac{1}{\sqrt{1-\gamma_j}}-h_{0ij}]\langle \tilde{U}_j|
\end{equation}
with
\begin{equation}
\label{eigen1}
         {\cal N}| \tilde{U}_i\rangle=\gamma_i | \tilde{U}_i\rangle
\end{equation}
\narrowtext
and with matrix elements  
$W_{ij}=\langle \tilde{U}_i|\, W\,|\tilde{U}_j\rangle\,$
and 
$ h_{0ij}=\langle \tilde{U}_i|\, h_0 \, |\tilde{U}_j\rangle\,$.
We solved Eq.(\ref{eigen1})
to obtain the eigen-value $\gamma_{i}$ and the eigen-function $\tilde{U}_i$ for each state of the 
p--$^3$He system up to a large number $i=N_s$, beyond which the eigen-functions
$\tilde{U}_i$ become highly oscillatory.
These oscillations make the numerical calculations cumbersome and unstable.

The eigen-functions $\tilde{U}_i$ can be
expressed in terms of harmonic oscillator functions $U_i$ 
(see Eq. (\ref{hofex})). 
Since the potential $V^{OCM}$ becomes infinite for a PFS($\gamma_i=1$)
the corresponding wave function is removed from the
eigen-functions of the Schr\"{o}dinger equation.
For the partly PFS ($0 < \gamma_i <1$) we note that
the eigen-values  $\gamma_i$ converge quickly
to zero as $i$ increases and thus for a  sufficiently large number $N_S$,
\begin{equation}
\label{ocmappa}
       V^{{\rm OCM}} \rightarrow \sum_{ij=N_S}^{\infty}
		| \tilde{U}_i \rangle W_{ij} \langle \tilde{U}_j |.
\end{equation}
Therefore, by using $W=\sum_{ij=1}^{\infty}| \tilde{U}_i 
	\rangle W_{ij} \langle \tilde{U}_j |=V^{RGM}(E=0)$, 
we obtain instead of Eq. (\ref{ocmp2-1})  the modified OCM (MOCM)
potential
\widetext
\begin{eqnarray}\label{ocmpotorg}
	V^{\rm MOCM}& = & W-\sum_{ij=1}^{\infty}| \tilde{U}_i 
	\rangle W_{ij} \langle \tilde{U}_j |
        +\sum_{ij=N_S}^{\infty}|\tilde{U}_i \rangle W_{ij} \langle \tilde{U}_j |\nonumber\\
	&+&\sum_{ij=1}^{N_S}| \tilde{U}_i \rangle
	[\frac{1}{\sqrt{1-\gamma_i}}(h_{0ij}+W_{ij})
	\frac{1}{\sqrt{1-\gamma_j}}-h_{0ij}]\langle \tilde{U}_j |\nonumber\\
	&=&V^{RGM}(E=0)+\sum_{ij=1}^{N_S}| \tilde{U}_i\rangle \biggl[
	\frac{1}{\sqrt{1-\gamma_i}}(h_{0ij}+W_{ij})\frac{1}{\sqrt{1-\gamma_j}}
	-(h_{0ij}+W_{ij})\biggl]\langle \tilde{U}_j|.
\end{eqnarray}
\narrowtext
If the  number of  PFS is  $N_F$ then from  the orthogonality
condition between $\langle \tilde{U}_j|$ and  the physical wave function
$|\psi\rangle$,  given by  Eq. (\ref{renorm}), we obtain
\widetext
\begin{equation}
\biggl(\sum_{ij=1}^{N_F}|\tilde{U}_i\rangle\bigg[\frac{1}{\sqrt{1-\gamma_i}}(h_{0ij}+
W_{ij})\frac{1}{\sqrt{1-\gamma_j}}-(h_{0ij}+W_{ij})\bigg]\langle\tilde{U}_j|\biggl)|\psi\rangle=0.
\end{equation}
\narrowtext
Thus, the part of the potential in Eq.(\ref{ocmpotorg}) has no effect in the 
Schr\"{o}dinger equation.
Consequently, we introduce our MOCM potential as follows,
\widetext
\begin{equation}\label{ocmp3}
V^{\rm MOCM}\equiv V^{RGM}(E=0)+\sum_{ij=N_F+1}^{N_S}|\tilde{U}_i\rangle
\biggl[\frac{1}{\sqrt{1-\gamma_i}}(h_{0ij}+W_{ij})
\frac{1}{\sqrt{1-\gamma_j}}-(h_{0ij}+W_{ij})\biggl]\langle\tilde{U}_j|.
\end{equation}
\narrowtext
For the  $\ell=0$ state of the p-$^3$He system,
the eigenvalues $\gamma_i$  are $\gamma_{1}$= 0.98993, $\gamma_{2}$= 0.12020,
$\gamma_{3}$= 0.01373, and so on,  {\em i.e.}  there are no  PFS
in the system.
For higher partial waves, the absolute values of
$\gamma_{i}$ are much less than one, and therefore $N_F=0$ as well.
Thus,  Eq. (\ref{ocmp3}) is not only free from a diverging
term caused by $\gamma_i=1$, but also from  numerical
errors due to presence of oscillator functions with higher frequencies.

Each term of Eq.(\ref{ocmp3}) can be  easily transformed into its
momentum space using Fourier transforms (see Appendix B).
It is found that the on-shell parts of the OCM scattering amplitude
obtained with the transformed potentials are almost the same as those of the RGM amplitude,
although the off-shell parts differ.
For the spin-triplet P- and D- waves and with $N_S=20$ the resulting phase shift differ slightly.
However, the differences are small enough and do not give rise to any important effects
in the parameter fitting of these triplet channels.

Within the above formalism, a realistic force with an LS component has 
rarely been considered until now \cite{hackenbriich}. As a consequence, 
the intercluster 
p--$^3$He potential of the MOCM, Eq. (\ref{ocmp3}), 
does not include an LS force either. 
Although in  the RGM formalism the NN interaction could 
include an LS and tensor
force, this may necessitate some additional antisymmetrization procedure.
However, if we adopt a proper intrinsic cluster function 
that  includes the effects of these forces, 
we could include these effects in the inter-cluster potentials by means of a folding procedure.
In previous work  McIntyre and Haeberli\cite{lsp} introduced a 
phenomenological LS force to reduce the degeneracy of the L-induced 
potentials.   In the present work we use the same technique but with 
an extended form for the  LS forces. The LS dependence in the 
intercluster potentials for the spin-triplet channels
is introduced via
\widetext
\begin{equation}\label{lspot}
       V^{\rm OCM(LS)}=\{C_0+C_1\mbox{\boldmath $L\cdot S$}
	+C_2(\mbox{\boldmath $L\cdot S$})^2
	+C_3(\mbox{\boldmath $L\cdot S$})^3\}V^{\rm MOCM} ,
\end{equation}
\narrowtext
where $C_0$, $C_1$, $C_2$,and  $C_3$ are parameters adjusted to fit
the experimental phase shifts.
For now on we shall, for simplicity, use from now on the abbreviation OCM instead of MOCM.
The parameters for the  p--$^3$He interactions are shown in 
Table.\ref{hppara}.
%
\begin{table}[h]\centering
\caption{Parameters of the p--$^3$He interactions.}\label{hppara}
\begin{tabular}{c|rrrr}
State & \multicolumn{1}{c}{$C_0$} & \multicolumn{1}{c}{$C_1$} & \multicolumn{1}{c}{$C_2$} & \multicolumn{1}{c}{$C_3$} \\
\hline
$^3$P$_0$ $^3$P$_1$ $^3$P$_2$ & $~1.2080$ & $~0.0788$
 & $-0.0740$ & $-0.0110$ \\
$^3$D$_1$ $^3$D$_2$ $^3$D$_3$ & $~0.9780$ & $~0.0446$ 
& $~0.0616$ & $-0.0090$ \\
$^3$F$_2$ $^3$F$_3$ $^3$F$_4$ & $~1.2000$ & $~0.0360$ 
& $~0.0084$ & $~0.0000$\\
\end{tabular}
\end{table}

Tensor forces should be considered, 
in principle, for the   $^3$S$_1$-$^3$D$_1$, $^1$P$_1$-$^3$P$_1$, 
$^3$P$_2$-$^3$F$_2$, and $^1$D$_2$-$^3$D$_2$ coupled channels.
However, as mentioned in the introduction,
in the phase-shift analysis by Yoshino {\it et. al.}
\cite{yoshino} it was  found that 
the mixing parameters for the  p--$^3$He interaction are very small;
{\it i.e.}, 
$\varepsilon_1^{+} = -0.2719\pm0.0097$\,deg,\
$\varepsilon_1^{-} = -0.4001\pm0.3490$\,deg,\
$\varepsilon_2^{+} = -0.0269\pm0.2510$\,deg,\
and $\varepsilon_2^{-} = 0.0827\pm0.0129$\,deg 
at $E_{p}=19.48$\,MeV for the $^3$S$_1$-$^3$D$_1$,
$^1$P$_1$-$^3$P$_1$, 
$^1$D$_2$-$^3$D$_2$,
and $^3$P$_2$-$^3$F$_2$ partial waves, respectively.  Therefore, 
tensor forces were  omitted.
%
%
\section{Phase Shift Calculations}
The  p-$^3$He phase shifts were calculated using the program ``GSE"
developed in Refs. \cite{gse1,gse2}.
The GSE method is one generalization of Bateman's method in which
the on-shell and half-off-shell properties of the t-matrix are exact 
when the on-shell momentum is chosen to be the Bateman parameter.
 In order to take into account the Coulomb interaction for the
p-$^3$He system, Reichstein {\it et al.} introduced a kind of Coulomb force
in the RGM formalism which takes the form of an Error function \cite{rgm}.
In momentum representation, the Error function can be expanded in the
short range region in terms of Gaussian functions 
while  in the long range region it can be expressed  in terms  of
a Coulomb potential between the $^3$He cluster and the proton.
In order to obtain the phase-shift modification by the Coulomb interaction,
we adopt a screened Coulomb potential of the Yukawa type.
The difference between the pure Coulomb and the screened Coulomb potential in
the long range region is corrected by means of a renormalization 
technique\cite{renorm}.

The screened Coulomb potential is defined by
\begin{equation}
	V^{R}(r)=\frac{ZZ'e^2}{r}\exp (-r/R)
\end{equation}
where $R$ is called the screened Coulomb range parameter which,
in our case, has the value of  
$R=810.753$\,fm for S-wave,
$R=1128.677$\,fm for P-wave,
$R=1439.062$\,fm for D-wave,
and $R=10158.163$\,fm for F-wave. 
The partial wave expansion of $V^{R}$, in momentum space, is given by
\begin{equation}\label{screenedcoulomb}
	V_\ell^R(k,k')=2\pi\frac{ZZ'e^2}{kk'}
	Q_\ell\biggl{(}\frac{k^2+k'^2+1/R^2}{2kk'}\biggl{)}
\end{equation}
where $Q_\ell(x)$ is the Legendre function of the second kind.
The calculated screened Coulomb phase shift $\delta_\ell^{R}(k)$
using Eq. (\ref{screenedcoulomb}) is given by
\begin{equation}
	\delta_\ell^{R}(k)=\sigma_\ell(k)-\phi(k,R)
\end{equation}
where $\sigma_\ell(k)$ is the Coulomb phase, $\phi(k,R)$ is
the renormalization phase  given by $\phi(k,R)=\eta(\ln 2kR-\gamma)+\cdots$,
$\eta$ is the Sommerfeld parameter $\eta=\mu ZZ'e^2/k$, $\mu$ is a reduced 
mass, and $\gamma$ is the  Euler constant $\gamma=0.5772156\cdots$.

\begin{figure}\centering
\psbox[size=0.5#1]{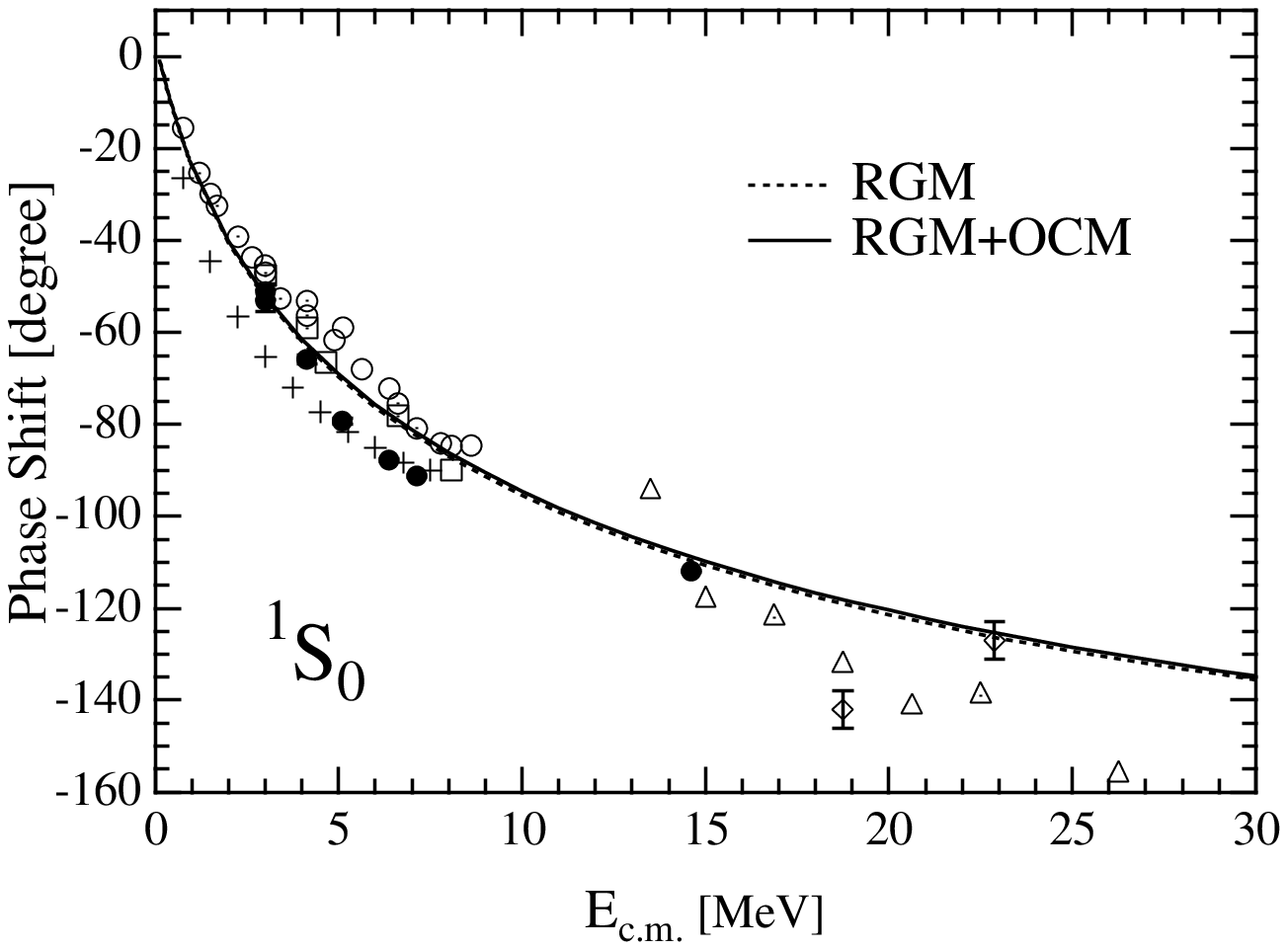}
\psbox[size=0.5#1]{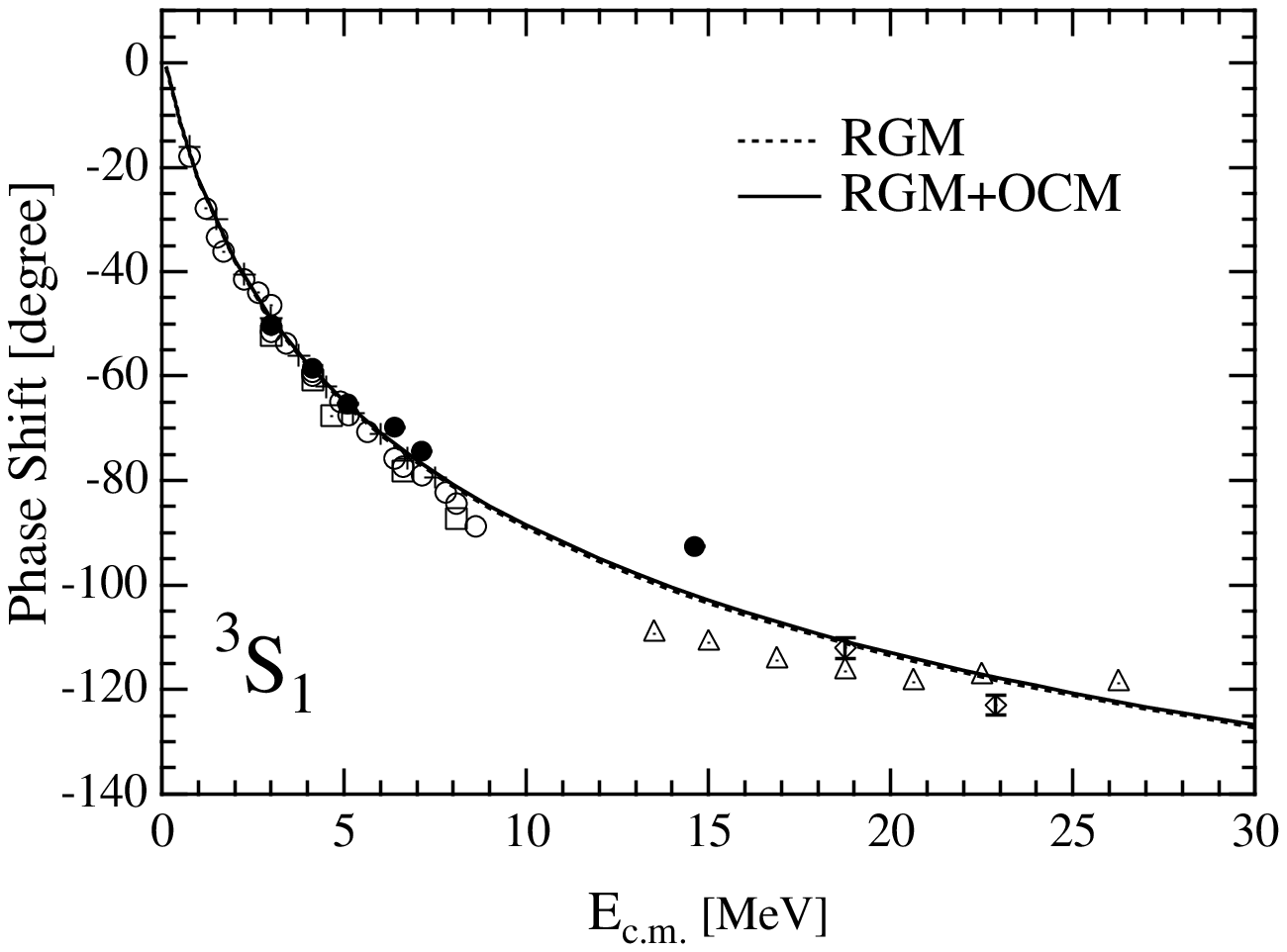}
\caption{p-$^3$He phase shifts for the $^1$S$_0$  and  $^3$S$_1$ 
partial waves.
The experimental data are: 
$\circ$ Tombrello [12] \                     
$\Box$ Mc Sherry and Baker [14],\            
$\triangle$  Morales {\it et al.} [16], \    
$\diamond$   M\"{u}ller {\it et al.} [17], \ 
 $+$ Beltramin {\it et al.} [10], and        
$\bullet$ Yoshino {\it et al.} [19].        
The dashed line denotes the RGM and          
the solid line the  RGM+OCM results, respectively.         
}
\label{hpswave}
\end{figure}

\begin{figure}\centering
\psbox[size=0.5#1]{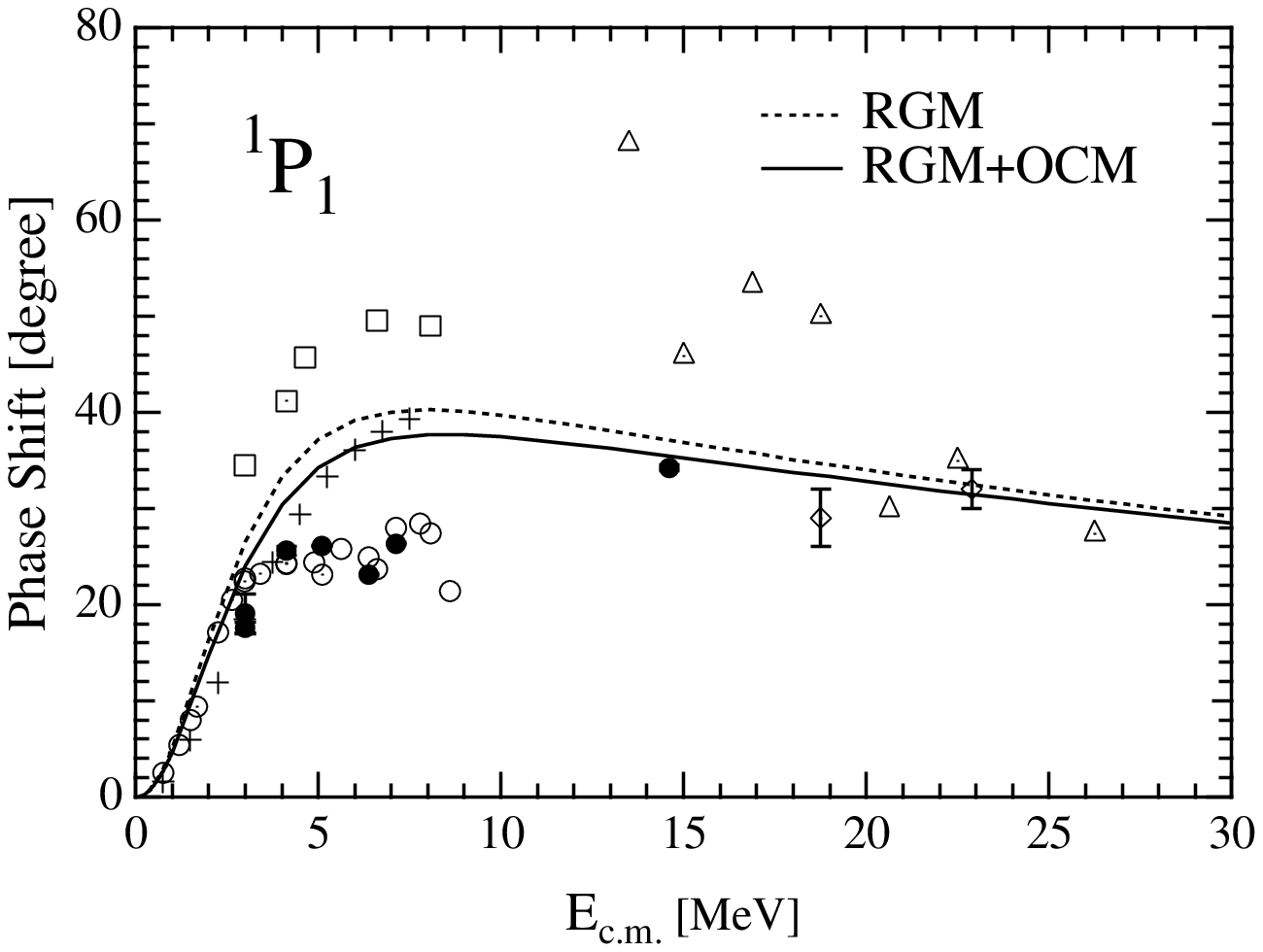}
\psbox[size=0.5#1]{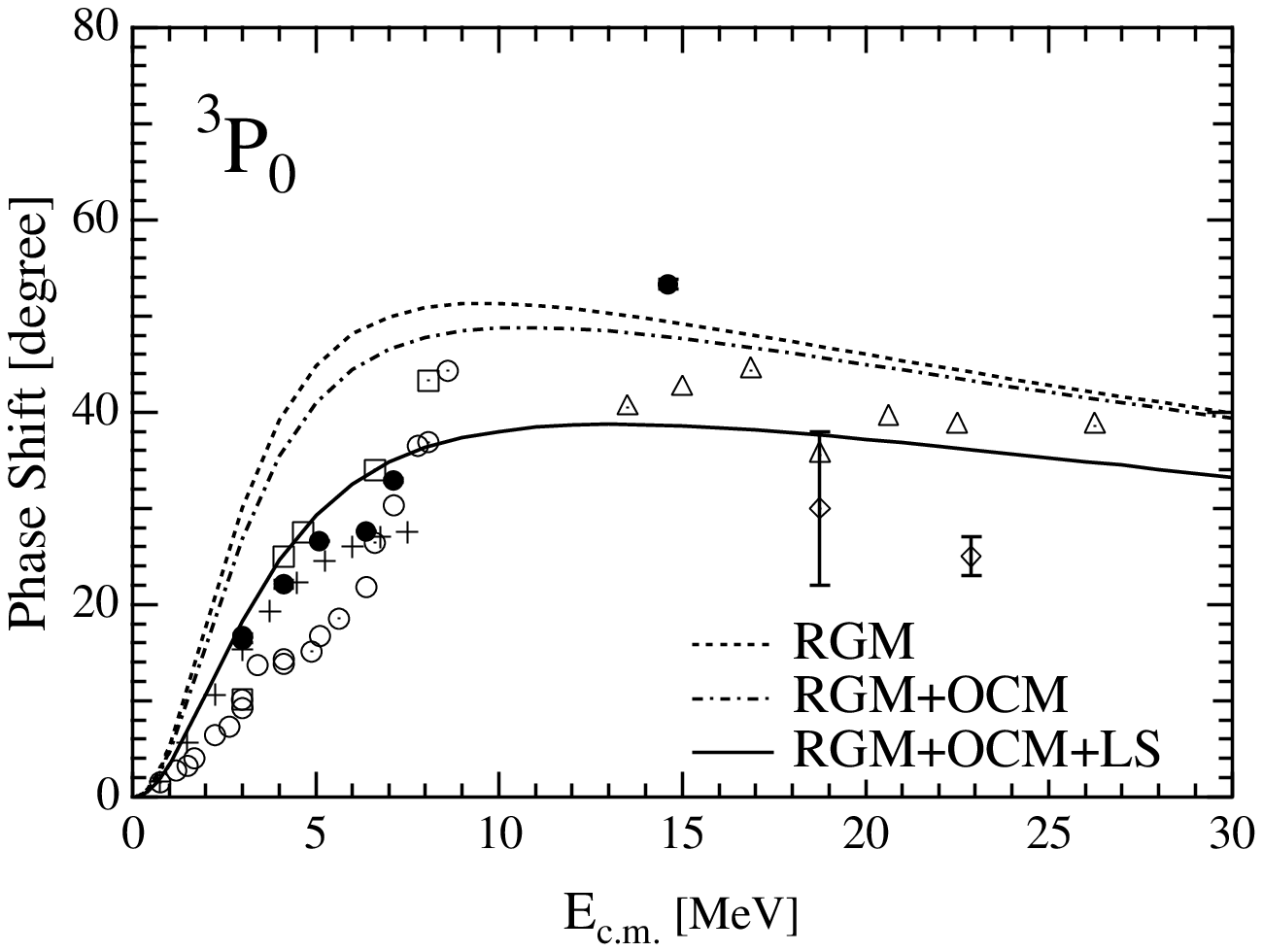}
\psbox[size=0.5#1]{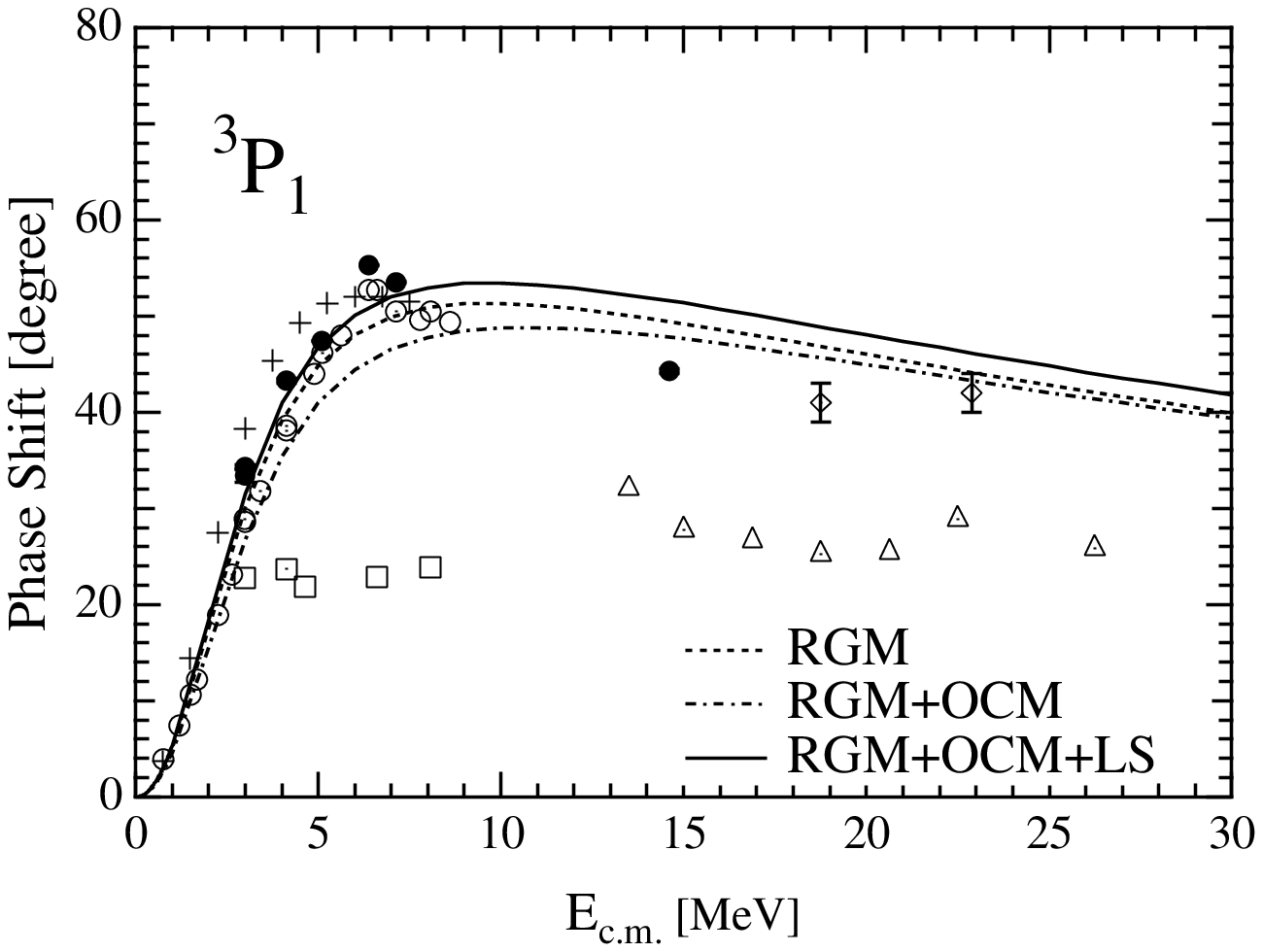}
\psbox[size=0.5#1]{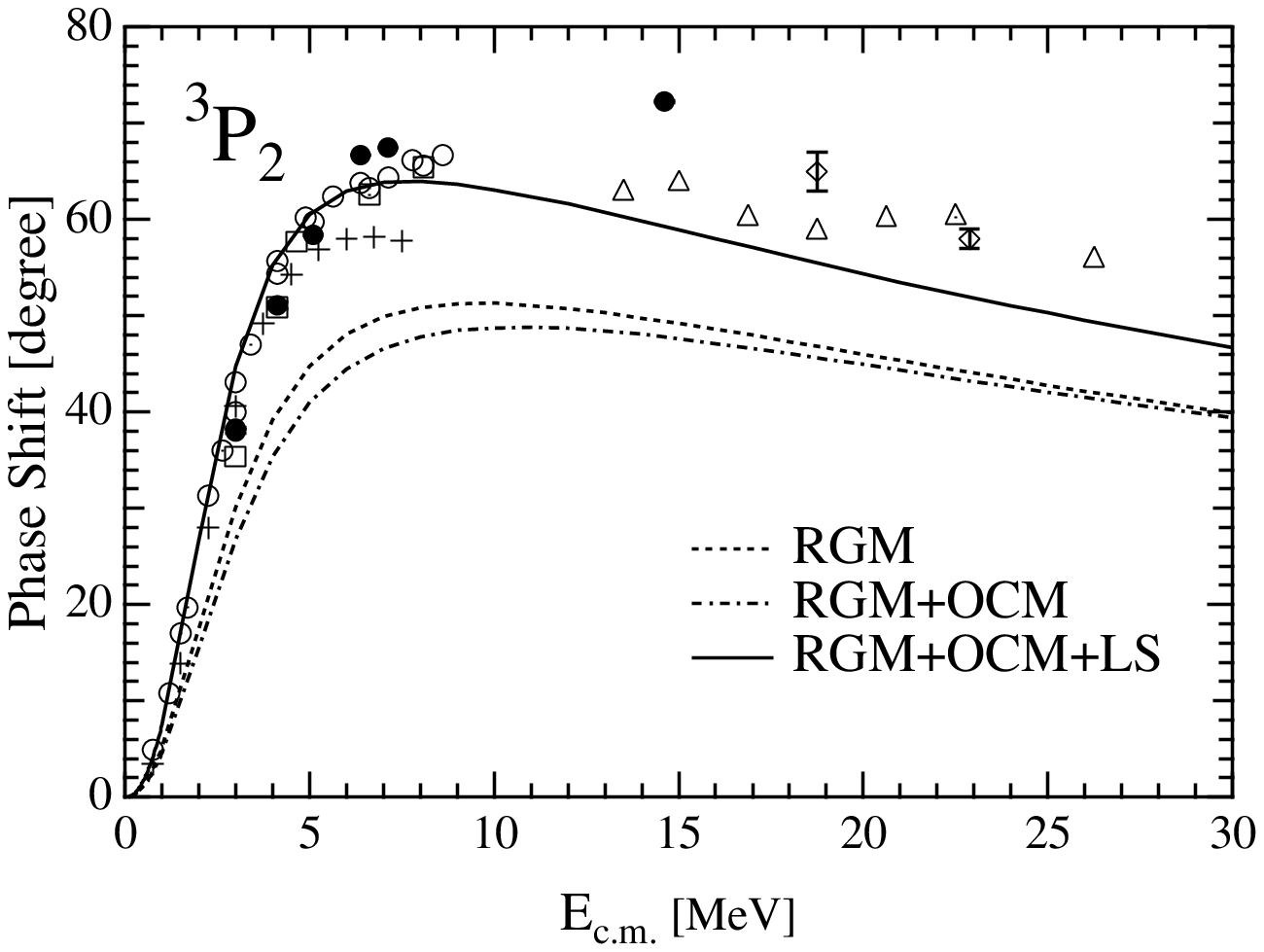}
\caption{p-$^3$He phase shifts for the $^1$P$_1$, $^3$P$_0$, $^3$P$_1$,
and $^3$P$_2$ partial waves.
 Symbols for the experimental data are the same as 
in Fig. 1.  For the singlet channel the dashed line denotes the RGM
and the solid line the RGM+OCM results.  For the triplet 
channel the dashed line denotes the RGM, the dashed-dotted line the 
RGM+OCM, and the solid lines the RGM+OCM+LS results, respectively.
}
\label{hppwave}
\end{figure}

\begin{figure}\centering
\psbox[size=0.5#1]{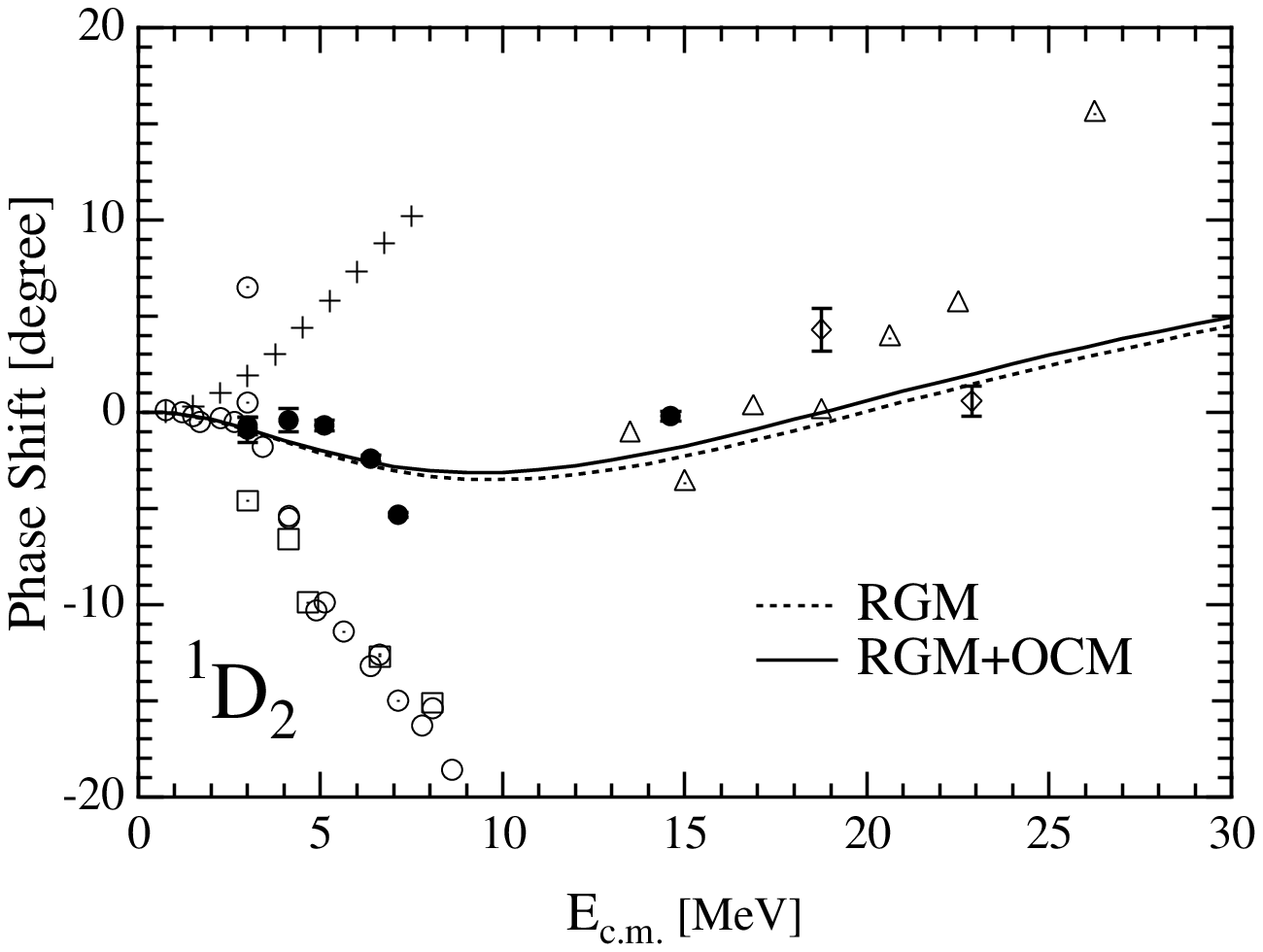}
\psbox[size=0.5#1]{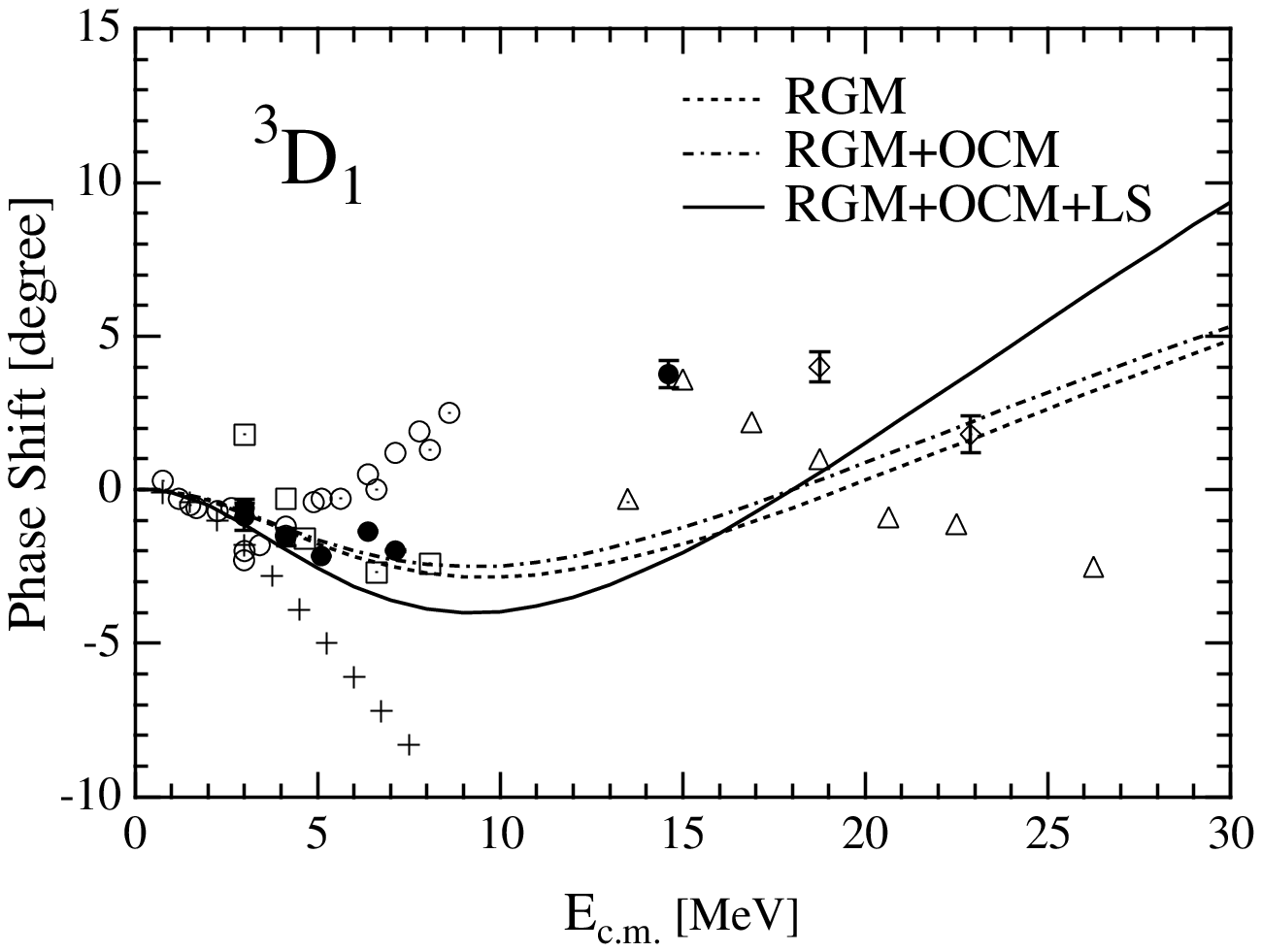}
\psbox[size=0.5#1]{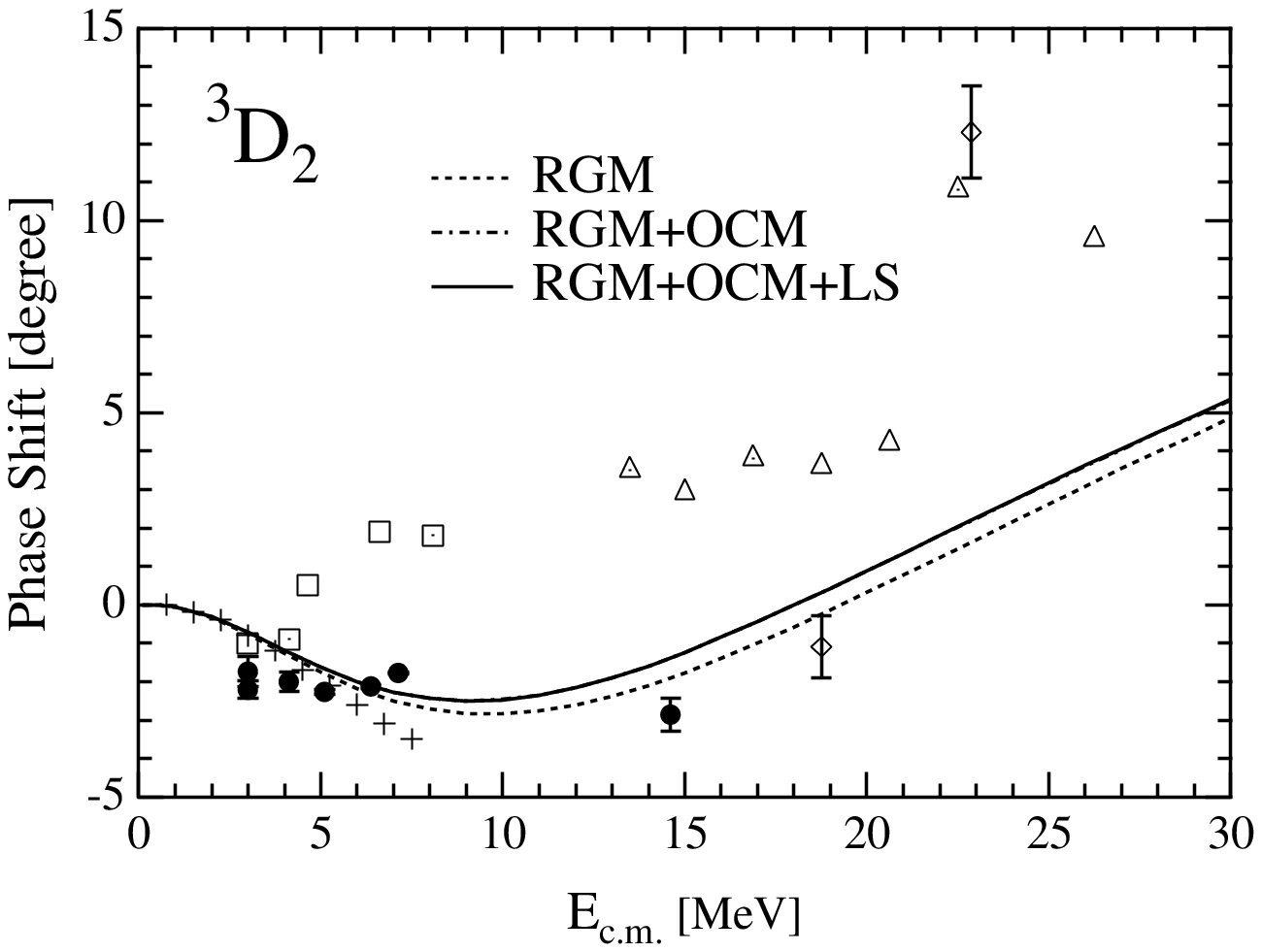}
\psbox[size=0.5#1]{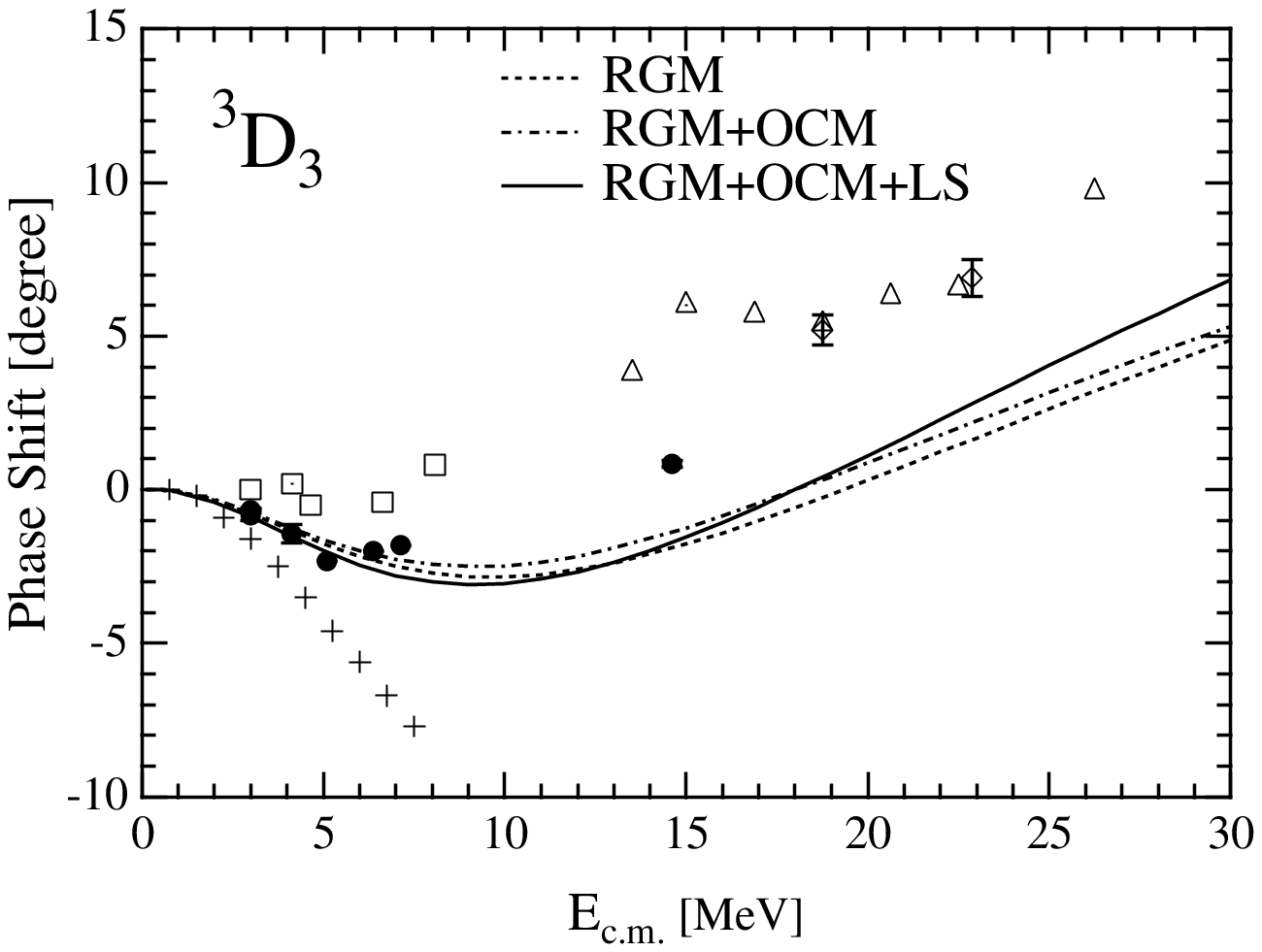}
\caption{p-$^3$He phase shifts for the $^1$D$_2$, $^3$D$_1$, $^3$D$_2$, 
and $^3$D$_3$ partial waves.  The notation is the same as in Fig. 1 and 2.}
\label{hpdwave}
\end{figure}

\begin{figure}\centering
\psbox[size=0.5#1]{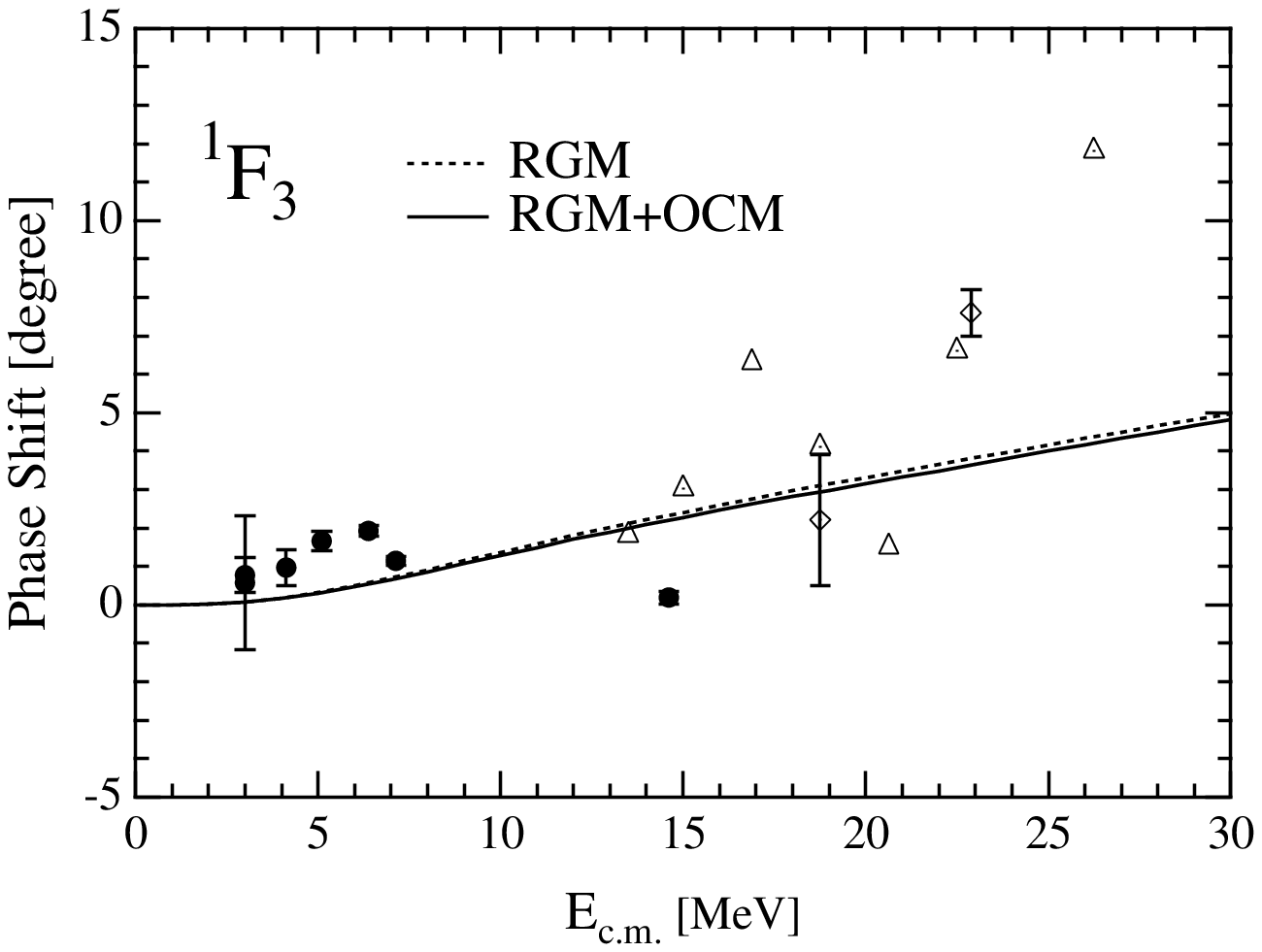}
\psbox[size=0.5#1]{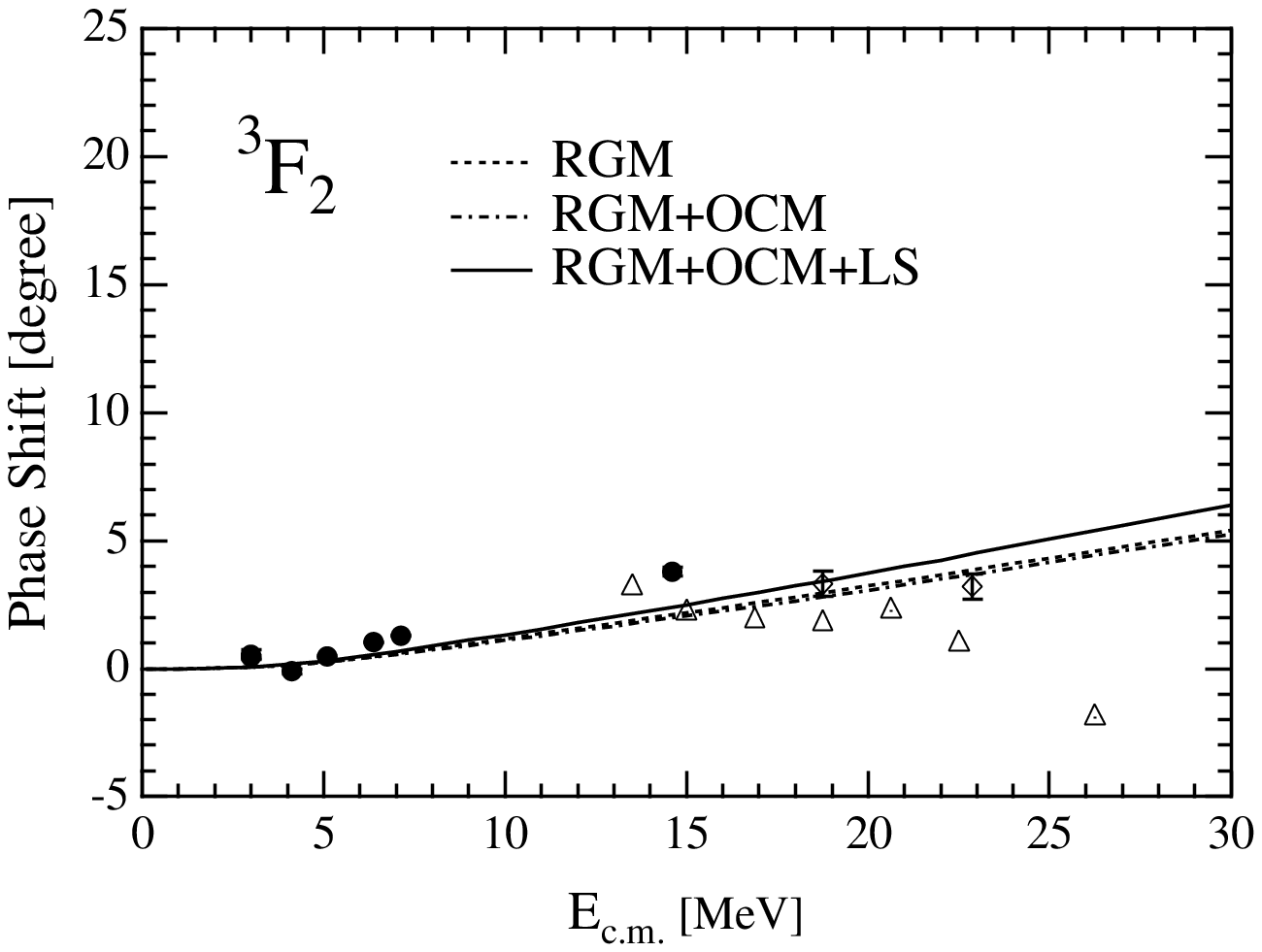}
\psbox[size=0.5#1]{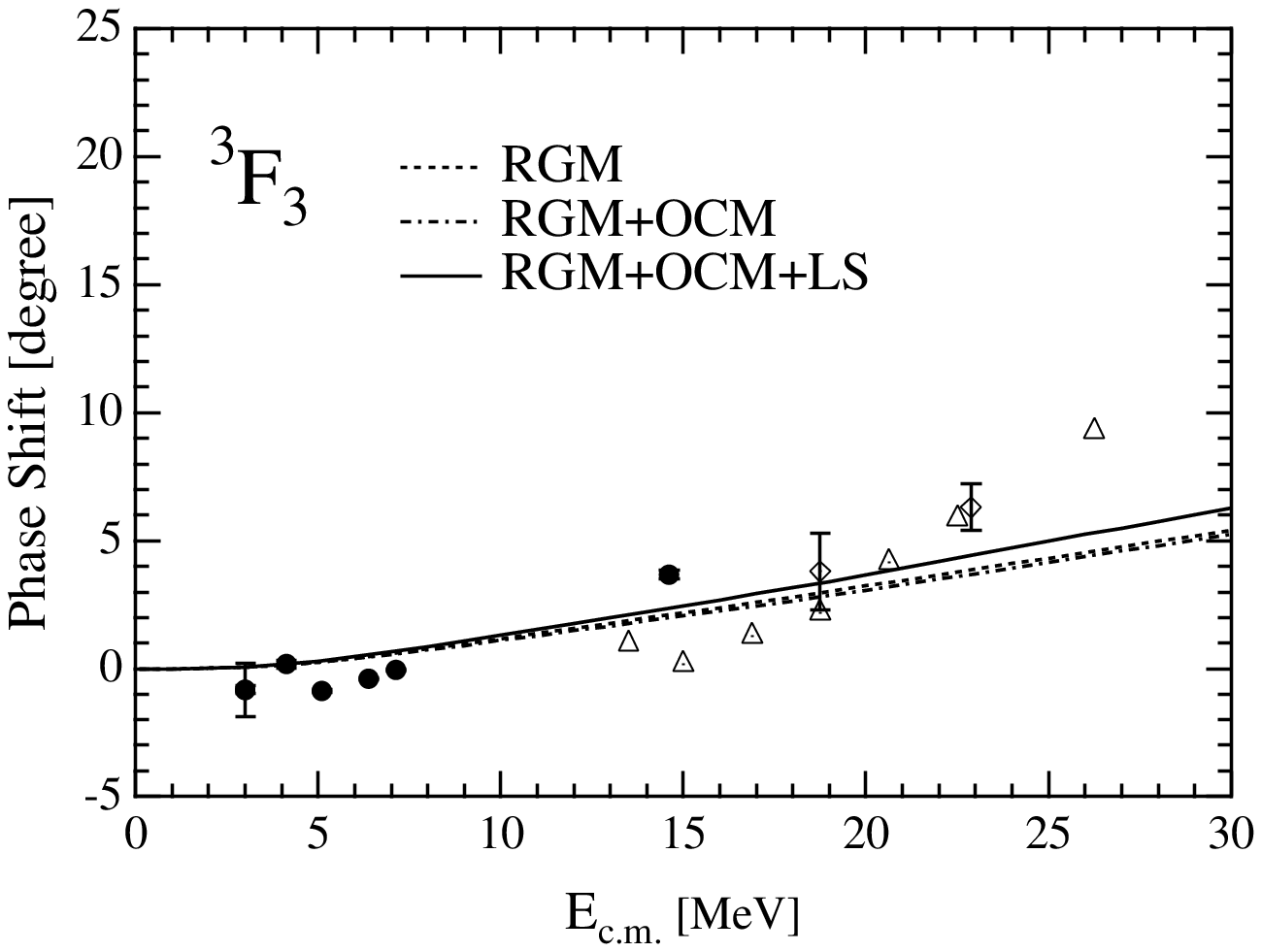}
\psbox[size=0.5#1]{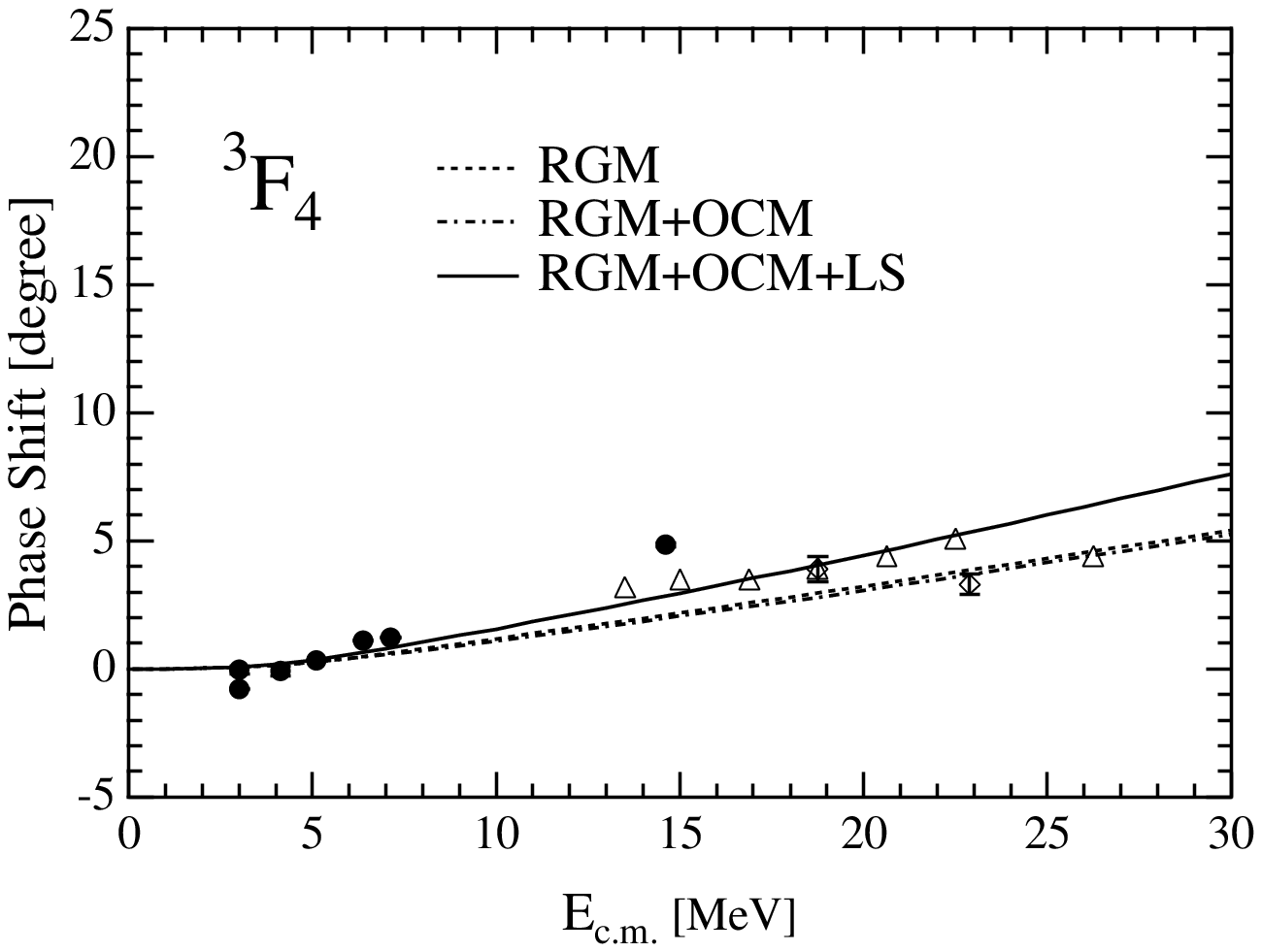}
\caption{p-$^3$He phase shifts for the $^1$F$_3$, $^3$F$_2$, $^3$F$_3$, 
$^3$F$_4$ partial waves.  The notation is the same as in Fig. 1 and 2.}
 \label{hpfwave}
\end{figure}

In order to calculate the Coulomb modified nuclear phase shifts 
$\delta_\ell^{SC}$, 
we first  put the total potential with the short range  nuclear part 
$V^{\rm OCM(LS)}$
and the Coulomb part into partial wave form,
\begin{equation}
	     V_\ell(k,k')=V_\ell^{{\rm OCM(LS)}}(k,k') + V_\ell^R(k,k').
\end{equation}
Then, the calculated total phase shift is given by
\begin{equation}
	\delta_\ell^{(R)} \equiv \delta_\ell^{SR}+\delta_\ell^{R}(k)
	                 =\delta_\ell^{SR}(k)+\sigma_\ell(k)-\phi(k,R).
\end{equation}
Consequently the Coulomb modified nuclear phase shift $\delta_\ell^{SR}(k)$ is 
\begin{eqnarray}
\label{coulombmodif}
	\delta_\ell^{SR}(k) & = & \delta_\ell^{(R)}-\delta_\ell^{R}(k)\nonumber\\
                            & = & \delta_\ell^{(R)}(k)+\phi(k,R)-\sigma_\ell(k).
\end{eqnarray}
The genuine Coulomb modified nuclear phase shifts are obtained
via Eq. (\ref{coulombmodif}) in the long range limit,
\begin{eqnarray}
           \delta_\ell^{SC}(k)&=&\lim_{R\rightarrow 
	\infty}[\delta_\ell^{(R)}+\phi(k,R)]-\sigma_\ell(k)\nonumber\\
		&=&\delta_\ell(k)-\sigma_\ell(k),
\end{eqnarray}
where $\delta_\ell(k)$ is the total phase shift generated by the nuclear 
plus Coulomb potentials.
%
The results, for the $^1$S$_0$, $^3$S$_1$ partial waves, are shown in Fig. 
\ref{hpswave} and compared with the results of the phase shift analyses
\cite{Pisent,yoshino,exp1,exp2,exp4,exp5,exp6,exp7}. It is seen that
the phase shifts for these channels are well reproduced.
This  confirms the  on-shell equivalence of the OCM and the RGM potentials.
%
%
In Fig.  \ref{hppwave}, the phase shifts for the  $^1$P$_0$, $^3$P$_0$, 
$^3$P$_1$, $^3$P$_2$ partial waves are shown. The spin singlet $^1$P$_0$ 
phases for the RGM and the RGM+OCM potentials are obtained without an 
LS force.  In contrast, the $^3$P$_0$, $^3$P$_1$, $^3$P$_2$ partial waves 
are spin triplet,
and therefore the LS force is taken into account. It is seen that our results 
are in very good agreement with the experimental data, especially 
for energies below 10\,MeV. This demonstrates
the importance of the LS force in the potential.
%
%
In Fig. \ref{hpdwave} the phase shifts for $^1$D$_2$, $^3$D$_1$, 
$^3$D$_2$, and $^3$D$_3$ partial waves are shown.   The spin-singlet
$^1$D$_2$ partial wave is calculated using the RGM+OCM potential 
without inclusion of an LS force.  Once again, the results are
in good agreement with the experimental data
of Yoshino {\it et al.} \cite{yoshino} in the lower energy region.
It should be noted that, unlike for the P wave results,
the  LS force is less important for the  $^3$D$_1$, 
$^3$D$_2$, and  $^3$D$_3$ partial waves.
%
In Fig. \ref{hpfwave}, the  $^1$F$_3$, $^3$F$_2$, $^3$F$_3$, and  $^3$F$_4$ 
partial wave phase shifts are shown. Again, the $^1$F$_3$  is a spin singlet. 
The results were obtained using the RGM and  RGM+OCM potentials only.  
In order to obtain the relevant constants for the LS force in 
the $^3$F$_2$, $^3$F$_3$, and  $^3$F$_4$ channels, we employed 
the latest  experimental data by  Yoshino {\it et al.}\cite{yoshino}.
%

\section{ Separable Expansion of the \ph Potentials }

In order to be useful the potential $V^{OCM(LS)}$ must be converted
into a separable expansion form. To achieve this we
employed  the EST separable expansion method\cite{EST}
which is briefly described below.

The  EST rank $N$ separable potential $V^{\rm EST}(p,p')$ is defined 
in terms of  the form factor $g_i(p)$ and  the coupling 
constant $\lambda_{ij}$
\begin{equation}\label{sepV}
	V^{\rm EST}(p,p')=\sum_{i,j=1}^{N}g_i(p)\,\lambda_{ij}\,g_j(p')\,.
\end{equation}
The form factor is given by the R-matrix,
\begin{equation}
		g_i(p)=\langle p | V | \Psi (E_i) \rangle \equiv R(p,k_i;E_i),
\end{equation}
where  $E_i$ is a fixed energy point and
$k_i$ is the on-shell momentum at this energy.
The expansion energies used are $E_1$=1MeV for the rank-1 case,
and $E_1$=1MeV, $E_2$=10MeV, and $E_3$=15MeV for the rank-3 case.

The R-matrix satisfies the  equation
\widetext
\begin{eqnarray}
	R(p,k_i;E_i) &=& V(p,k_i)
             +{\cal P}\int_0^{\infty} V(p,p')G_0(p';E_i)
		R(p',k_i;E_i)\frac{p'^2}{2\pi^2}\,{\rm d}p'
\nonumber\\
	&=& V(p,k_i)+{\cal P}\int_0^{\infty}\frac{V(p,p')
		R(p',k_i;E_i)}{E_i-p'^2/2\mu}\frac{p'^2}{2\pi^2}\,{\rm d}p'
\end{eqnarray}
\narrowtext
where ${\cal P}$ denotes the principal value integral.
For practical reasons  the form factor $g_i(p)$ is expanded
in terms of  polynomials,
\begin{equation}
\label{formfactor}
	g_n(p)=e^{-\beta p^2}p^l \sum_{m=1}^{20}a_{n,m}p^{2(m-1)}\,,
\end{equation}
where $\beta=1/0.825$ and $a_{n,m}$ are fitting  parameters.

The coupling constant $\lambda_{ij}$  is defined by 
\widetext
\begin{equation}
\label{laminv}
	(\lambda^{-1})_{ij}= \langle \Psi(E_i) | V | \Psi(E_j) \rangle 
	= R(k_i,k_j;E_i) + {\cal P}\int_0^{\infty}R(k_i,p;E_i)
	G_0(p;E_i)R(p,k_j;E_i)\frac{p^2}{2\pi^2}\,{\rm d}p.
\end{equation}
\narrowtext
\subsection{Singlet channel}
The singlet p--$^3$He potential does not contain an LS force.
We endeavored to construct  two different  p--$^3$He potentials,
one of rank-1 and the other of rank-3. The rank-1 parameters 
for the form factor are the same as those of the first term  of the rank-3
but the coupling constant is different.

\begin{figure}[h]\centering
\psbox[size=0.5#1]{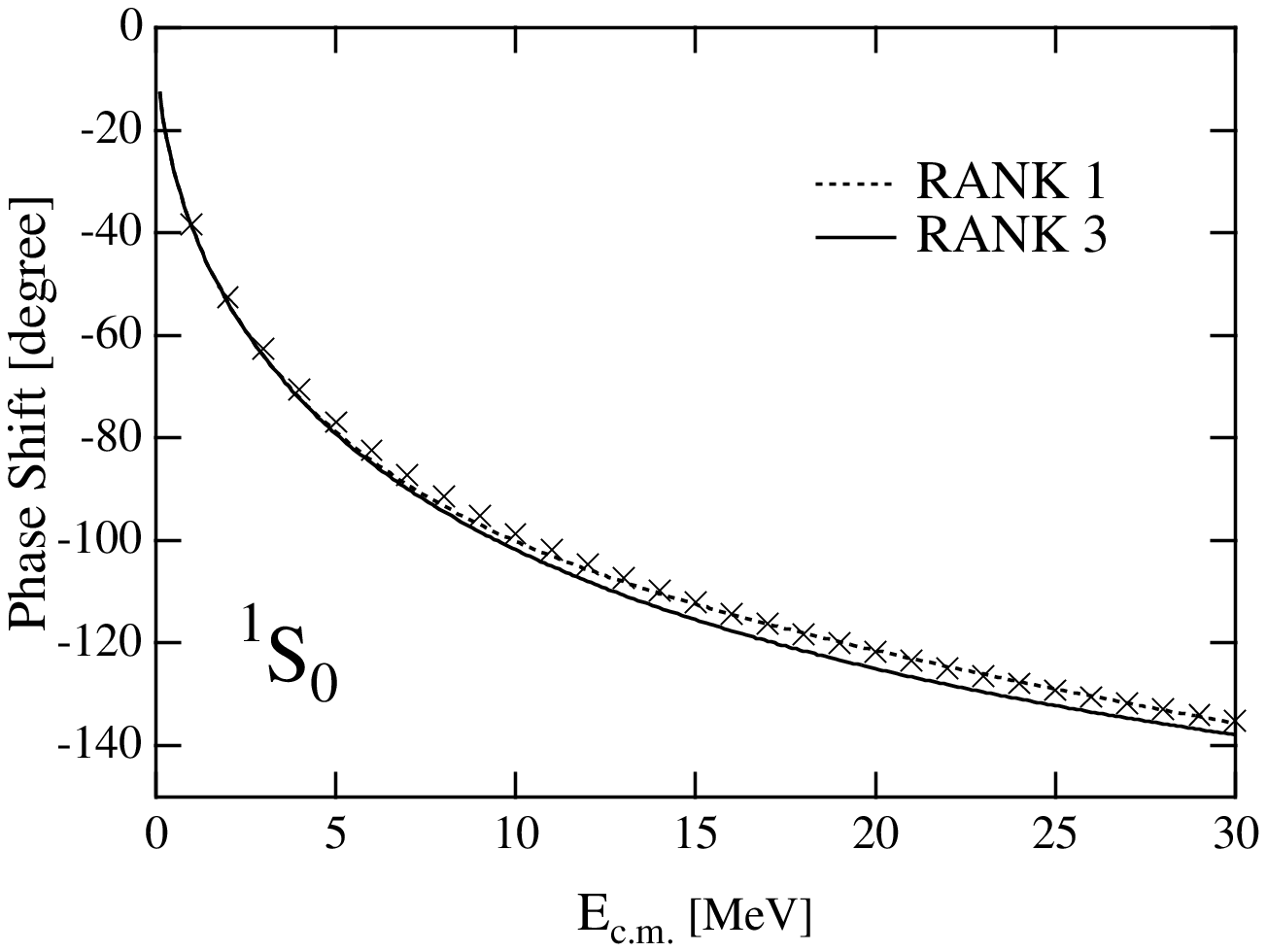}
\caption{Phase shifts for the $^1$S$_0$ partial wave without the 
Coulomb effects. 
The crosses denote the RGM+OCM results, the dashed line the rank-1
result, and the solid line the rank-3 result, respectively.\label{1s0mocmps}}
\end{figure}

The parameters for these potentials for the $^1$S$_0$ state are given in Tables II and III while the
phase shift results are presented in Fig. \ref{1s0mocmps}. It is  seen that 
the difference between the  phase shifts, especially in the lower energy 
region, is very small.
%
\begin{table}
\caption{Parameters $a_{n,m}$  of  the form factor of Eq. (4.4) 
for the $^1$S$_0$ potential. The first column lists the rank-1 parameters
with $\lambda_{11}=2.976586$. }
\begin{tabular}{lll}
\multicolumn{1}{c}{$a_{1,m}$}  & \multicolumn{1}{c}{$a_{2,m}$} 
		& \multicolumn{1}{c}{$a_{3,m}$}\\
\hline
~0.81290761$\times10^1$   & -0.37821814$\times10^2$    &-0.13596682$\times10^2$\\
-0.13818233$\times10^2$   & ~0.45121340$\times10^2$    &~0.14012550$\times10^2$\\
~0.26993988$\times10^2$   & -0.86975431$\times10^2$    &-0.27226075$\times10^2$\\
-0.29584805$\times10^2$   & ~0.78262662$\times10^2$    &~0.21711608$\times10^2$\\
~0.22158774$\times10^2$   & -0.47493102$\times10^2$    &-0.11624436$\times10^2$\\
-0.12179712$\times10^2$   & ~0.20523082$\times10^2$    &~0.42622121$\times10^1$\\
~0.52046713$\times10^1$   & -0.68468420$\times10^1$    &-0.11659454$\times10^1$\\
-0.17809791$\times10^1$   & ~0.18744208$\times10^1$    &~0.26053262            \\
~0.49289681               & -0.44089876                &-0.54300629$\times10^{-1}$\\
-0.11013105               & ~0.90888217$\times10^{-1}$ &~0.11621725$\times10^{-1}$\\
~0.19710935$\times10^{-1}$& -0.16177816$\times10^{-1}$ &-0.23952017$\times10^{-2}$\\
-0.27963987$\times10^{-2}$& ~0.23984064$\times10^{-2}$ &~0.41701960$\times10^{-3}$\\
~0.31072739$\times10^{-3}$& -0.28510226$\times10^{-3}$ &-0.56522712$\times10^{-4}$\\
-0.26665392$\times10^{-4}$& ~0.26325182$\times10^{-4}$ &~0.57510182$\times10^{-5}$\\
~0.17360057$\times10^{-5}$& -0.18369762$\times10^{-5}$ &-0.43035052$\times10^{-6}$\\
-0.83650320$\times10^{-7}$& ~0.94090317$\times10^{-7}$ &~0.23179820$\times10^{-7}$\\
~0.28751338$\times10^{-8}$& -0.34035371$\times10^{-8}$ &-0.86968112$\times10^{-9}$\\
-0.66353945$\times10^{-10}$&~0.81822792$\times10^{-10}$&~0.21475347$\times10^{-10}$\\
~0.91778681$\times10^{-12}$&-0.11672988$\times10^{-11}$&-0.31250844$\times10^{-12}$\\
-0.57298542$\times10^{-14}$&~0.74478363$\times10^{-14}$&~0.20237810$\times10^{-14}$\\
\end{tabular}
\end{table}

\begin{table}
\caption{Coupling constants $\lambda_{ij}=\lambda_{ji}$ of the rank--3 
$^1$S$_0$ potential.}
\begin{tabular}{c}
$\lambda_{11}=$-0.12888040$\times10^2$\\
$\lambda_{12}=$-0.12924909$\times10^2$\\
$\lambda_{13}=$~0.25638703$\times10^2$\\ 
$\lambda_{22}=$-0.11179525$\times10^2$\\
$\lambda_{23}=$~0.21826395$\times10^2$\\
$\lambda_{33}=$-0.42790529$\times10^2$\\
\end{tabular}
\end{table}
%
The results for the $^1$P$_1$ channel are given in Tables IV and V
while the phase shifts are given in Fig. 6. As can be seen from this 
figure the rank-1 potential fails to reproduce the results beyond 
$\sim 4$\,MeV.
Similar results were obtained for the $^1$D$_2$ 
and $^1$F$_3$ channels. These results are presented in Tables VI-IX 
and plotted in Figs. 7-8. 
%
%
\begin{table}
\caption{Parameters $a_{n,m}$  of  the form factor of Eq. (4.4) 
for the $^1$P$_1$ potential. The first column exhibits the 
rank-1 parameters with $\lambda_{11}=-0.29317504$.}
\begin{tabular}{lll}
\multicolumn{1}{c}{$a_{1,m}$}  & \multicolumn{1}{c}{$a_{2,m}$} 
		& \multicolumn{1}{c}{$a_{3,m}$}\\
\hline
-0.94249747$\times10^1$   &-0.80751728$\times10^1$    &-0.54943593$\times10^1$    \\
~0.16850309$\times10^2$    &~0.72033423$\times10^1$    &~0.28748548$\times10^1$    \\
-0.25259235$\times10^2$    &-0.90782035$\times10^1$    &-0.33647532$\times10^1$    \\
~0.22361127$\times10^2$    &~0.46110606$\times10^1$    &~0.41169528                \\
-0.14005952$\times10^2$    &-0.13188315$\times10^1$    &~0.60064787                \\
~0.67212849$\times10^1$    &~0.16478322                &-0.41395211                \\
-0.26035140$\times10^1$    &-0.27097734$\times10^{-1}$ &~0.86574730$\times10^{-1}$ \\
~0.83231138                &~0.41703584$\times10^{-1}$ &~0.28862915$\times10^{-1}$ \\
-0.22037742                &-0.29197374$\times10^{-1}$ &-0.28632910$\times10^{-1}$ \\
~0.47917437$\times10^{-1}$ &~0.11484498$\times10^{-1}$ &~0.11370739$\times10^{-1}$ \\
-0.84391784$\times10^{-2}$ &-0.29953237$\times10^{-2}$ &-0.29062500$\times10^{-2}$ \\
~0.11860729$\times10^{-2}$ &~0.55358368$\times10^{-3}$ &~0.52610797$\times10^{-3}$ \\
-0.13104955$\times10^{-3}$ &-0.74563678$\times10^{-4}$ &-0.69658876$\times10^{-4}$ \\
~0.11203792$\times10^{-4}$ &~0.73865049$\times10^{-5}$ &~0.68075554$\times10^{-5}$ \\
-0.72718918$\times10^{-6}$ &-0.53624860$\times10^{-6}$ &-0.48896135$\times10^{-6}$ \\
~0.34933248$\times10^{-7}$ &~0.28090702$\times10^{-7}$ &~0.25397873$\times10^{-7}$ \\
-0.11963972$\times10^{-8}$ &-0.10294665$\times10^{-8}$ &-0.92452666$\times10^{-9}$ \\
~0.27487008$\times10^{-10}$&~0.24947694$\times10^{-10}$&~0.22283655$\times10^{-10}$\\
-0.37800739$\times10^{-12}$&-0.35784010$\times10^{-12}$&-0.31823024$\times10^{-12}$\\
~0.23427534$\times10^{-14}$&~0.22924616$\times10^{-14}$&~0.20314053$\times10^{-14}$\\

\end{tabular}
\end{table}
%
%
\begin{table}
\caption{Coupling constants $\lambda_{ij}=\lambda_{ji}$ of the rank-3 
$^1$P$_1$ potential.}
\begin{tabular}{c}
$\lambda_{11}=$-0.86234010$\times10^1$\\
$\lambda_{12}=$~0.26099258$\times10^2$\\
$\lambda_{13}=$-0.24411307$\times10^2$\\
$\lambda_{22}=$-0.88300863$\times10^2$\\
$\lambda_{23}=$~0.85424554$\times10^2$\\
$\lambda_{33}=$-0.83781354$\times10^2$\\
\end{tabular}
\end{table}

%
\begin{figure}[h]\centering
\psbox[size=0.5#1]{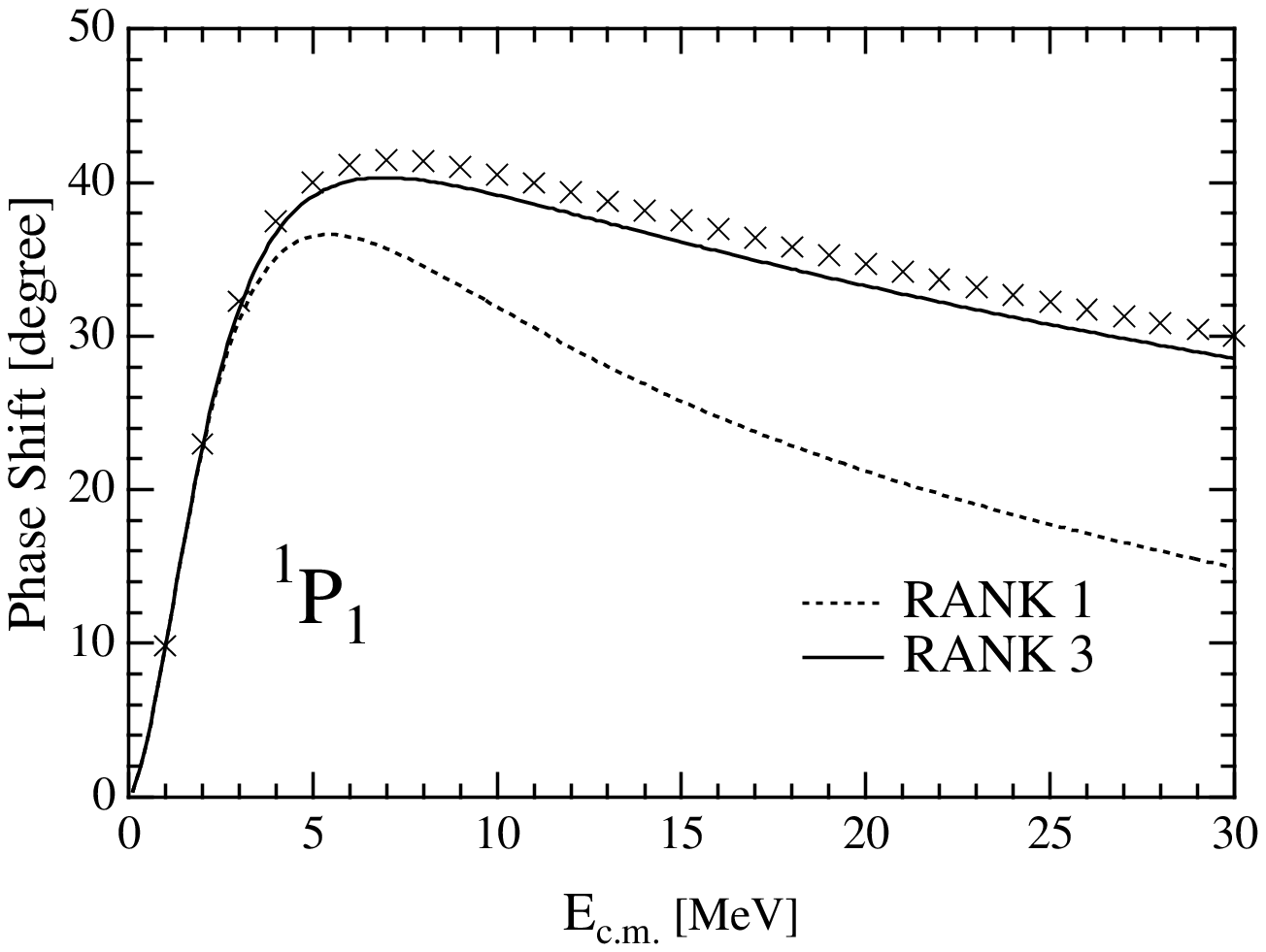}
\caption{Phase shifts for the $^1$P$_1$ partial wave without 
Coulomb effects. 
The crosses denote the RGM+OCM result, the dashed line  the rank-1 result, 
and the solid line the rank-3 result, respectively.}
\end{figure}
%
%
\begin{table}
\caption{Parameters $a_{n,m}$ of the form factor of Eq. (4.4) 
for the $^1$D$_2$ potential.  The first column gives the rank-1 parameters
with $\lambda_{11}=$~0.49171217$\times10^2$}
\begin{tabular}{lll}
\multicolumn{1}{c}{$a_{1,m}$} & \multicolumn{1}{c}{$a_{2,m}$} &\multicolumn{1}{c}{$a_{3,m}$}\\
\hline
~0.68105470                &~0.19763270$\times10^1$   &~0.16194259$\times10^1$    \\
-0.19186565$\times10^1$    &-0.54138849$\times10^1$   &-0.45399709$\times10^1$    \\
~0.27889420$\times10^1$    &~0.67678908$\times10^1$   &~0.50617192$\times10^1$    \\
-0.25046661$\times10^1$    &-0.50530126$\times10^1$   &-0.33071540$\times10^1$    \\
~0.15881034$\times10^1$    &~0.26186515$\times10^1$   &~0.14542318$\times10^1$    \\
-0.76207628                &-0.10174428$\times10^1$   &-0.45838231                \\
~0.28729893                &~0.31100450               &~0.10636485                \\
-0.86461811$\times10^{-1}$ &-0.76634068$\times10^{-1}$&-0.17716414$\times10^{-1}$ \\
~0.20858825$\times10^{-1}$ &~0.15393708$\times10^{-1}$&~0.18712562$\times10^{-2}$ \\
-0.40260171$\times10^{-2}$ &-0.25251882$\times10^{-2}$&-0.43762336$\times10^{-4}$ \\
~0.61860736$\times10^{-3}$ &~0.33681168$\times10^{-3}$&-0.26015437$\times10^{-4}$ \\
-0.75152221$\times10^{-4}$ &-0.36220702$\times10^{-4}$&~0.57462271$\times10^{-5}$ \\
~0.71556851$\times10^{-5}$ &~0.31045680$\times10^{-5}$&-0.69663479$\times10^{-6}$ \\
-0.52783394$\times10^{-6}$ &-0.20899733$\times10^{-6}$&~0.57050021$\times10^{-7}$ \\
~0.29679582$\times10^{-7}$ &~0.10841787$\times10^{-7}$&-0.32922530$\times10^{-8}$ \\
-0.12425007$\times10^{-8}$ &-0.42224769$\times10^{-9}$&~0.13358706$\times10^{-9}$ \\
~0.37349976$\times10^{-10}$&~0.11883187$\times10^{-10}$&-0.37020130$\times10^{-11}$\\
-0.75924107$\times10^{-12}$&-0.22720832$\times10^{-12}$&~0.65926451$\times10^{-13}$\\
~0.93178858$\times10^{-14}$&~0.26314644$\times10^{-14}$&-0.66816378$\times10^{-15}$\\
-0.51997857$\times10^{-16}$&-0.13885275$\times10^{-16}$&~0.28516854$\times10^{-17}$\\
\end{tabular}
\end{table}
%
\begin{table}
\caption{Coupling constants $\lambda_{ij}=\lambda_{ji}$ for the rank-3 
$^1$D$_2$ potential.}
\begin{tabular}{c}
$\lambda_{11}=$~0.23985939$\times10^4$\\
$\lambda_{12}=$-0.13537297$\times10^4$\\   
$\lambda_{13}=$~0.66626420$\times10^3$\\
$\lambda_{22}=$~0.74471421$\times10^3$\\
$\lambda_{23}=$-0.34111641$\times10^3$\\
$\lambda_{33}=$~0.13709213$\times10^3$\\
\end{tabular}
\end{table}
%
%
\begin{figure}[h]\centering
\psbox[size=0.5#1]{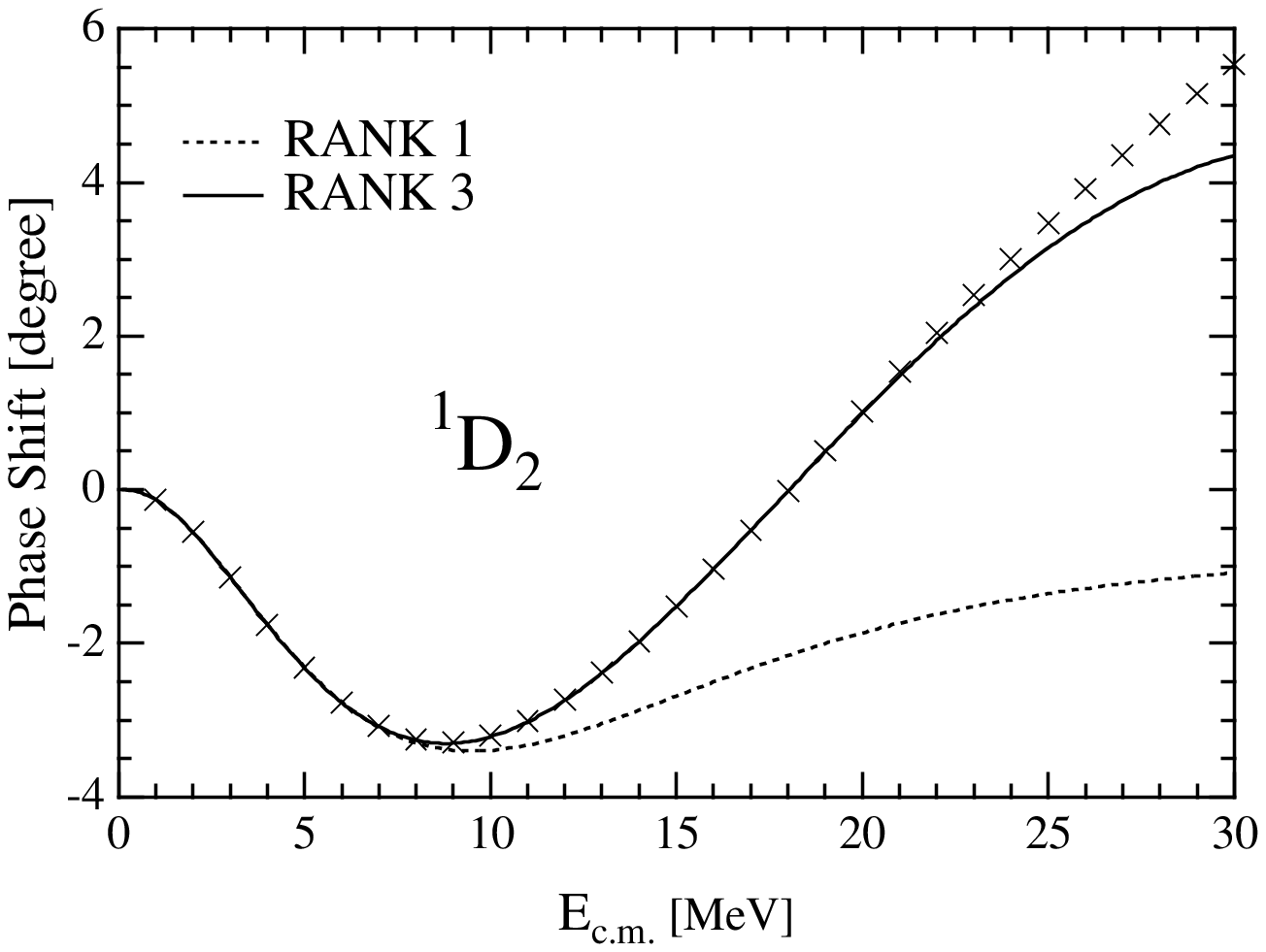}
\caption{Phase shifts for the $^1$D$_2$ partial waves without 
Coulomb effects. 
The crosses denote the RGM+OCM result, the dashed line the rank-1 result, 
and the solid line  the rank-3 result, respectively.}
\end{figure}
%
\begin{table}
\caption{
Parameters $a_{n,m}$ of the form factor of Eq. (4.4) 
for the $^1$F$_3$ potential.  The first column lists the rank-1 parameters
with $\lambda_{11}=$-0.157512965$\times10^4$.}
\begin{tabular}{lll}
\multicolumn{1}{c}{$a_{1,m}$}&\multicolumn{1}{c}{$a_{2,m}$}
&\multicolumn{1}{c}{$a_{3,m}$}\\
\hline
-0.10309579                &-0.10491693$\times10^1$    &-0.10971956$\times10^1$    \\
~0.25614578                &~0.21551217$\times10^1$    &~0.19720076$\times10^1$    \\
-0.35470074                &-0.26919503$\times10^1$    &-0.23113579$\times10^1$    \\
~0.30827988                &~0.20398472$\times10^1$    &~0.15842458$\times10^1$    \\
-0.18968846                &-0.10780556$\times10^1$    &-0.73781489                \\
~0.88041294$\times10^{-1}$ &~0.42456486                &~0.24470228                \\
-0.31866032$\times10^{-1}$ &-0.12903376                &-0.57371456$\times10^{-1}$ \\
~0.91323183$\times10^{-2}$ &~0.30698165$\times10^{-1}$ &~0.84797207$\times10^{-2}$ \\
-0.20840227$\times10^{-2}$ &-0.57388079$\times10^{-2}$ &-0.30810984$\times10^{-3}$ \\
~0.37892090$\times10^{-3}$ &~0.84146200$\times10^{-3}$ &-0.20824797$\times10^{-3}$ \\
-0.54767188$\times10^{-4}$ &-0.96337919$\times10^{-4}$ &~0.63773382$\times10^{-4}$ \\
~0.62640005$\times10^{-5}$ &~0.85660575$\times10^{-5}$ &-0.10605018$\times10^{-4}$ \\
-0.56298706$\times10^{-6}$ &-0.58902484$\times10^{-6}$ &~0.11988500$\times10^{-5}$ \\
~0.39349866$\times10^{-7}$ &~0.31325432$\times10^{-7}$ &-0.97107918$\times10^{-7}$ \\
-0.21061797$\times10^{-8}$ &-0.13004858$\times10^{-8}$ &~0.56943638$\times10^{-8}$ \\
~0.84350422$\times10^{-10}$&~0.42992976$\times10^{-10}$&-0.23915836$\times10^{-9}$ \\
-0.24379678$\times10^{-11}$&-0.11462659$\times10^{-11}$&~0.69843623$\times10^{-11}$\\
~0.47882660$\times10^{-13}$&~0.23655180$\times10^{-13}$&-0.13400069$\times10^{-12}$\\
-0.57032237$\times10^{-15}$&-0.32743658$\times10^{-15}$&~0.15094646$\times10^{-14}$\\
~0.31009923$\times10^{-17}$&~0.21752827$\times10^{-17}$&-0.75136464$\times10^{-17}$\\
\end{tabular}
\end{table}
%
\begin{table}
\caption{Coupling constants $\lambda_{ij}=\lambda_{ji}$ for the rank-3 
$^1$F$_3$ potential.}
\begin{tabular}{c}
$\lambda_{11}=$-0.11253575$\times10^6$\\
$\lambda_{12}=$~0.26062364$\times10^5$\\
$\lambda_{13}=$-0.14514498$\times10^5$\\
$\lambda_{22}=$-0.63200447$\times10^4$\\
$\lambda_{23}=$~0.35959447$\times10^4$\\
$\lambda_{33}=$-0.20743460$\times10^4$\\
\end{tabular}
\end{table}
%
%
\begin{figure}[h]\centering
\psbox[size=0.5#1]{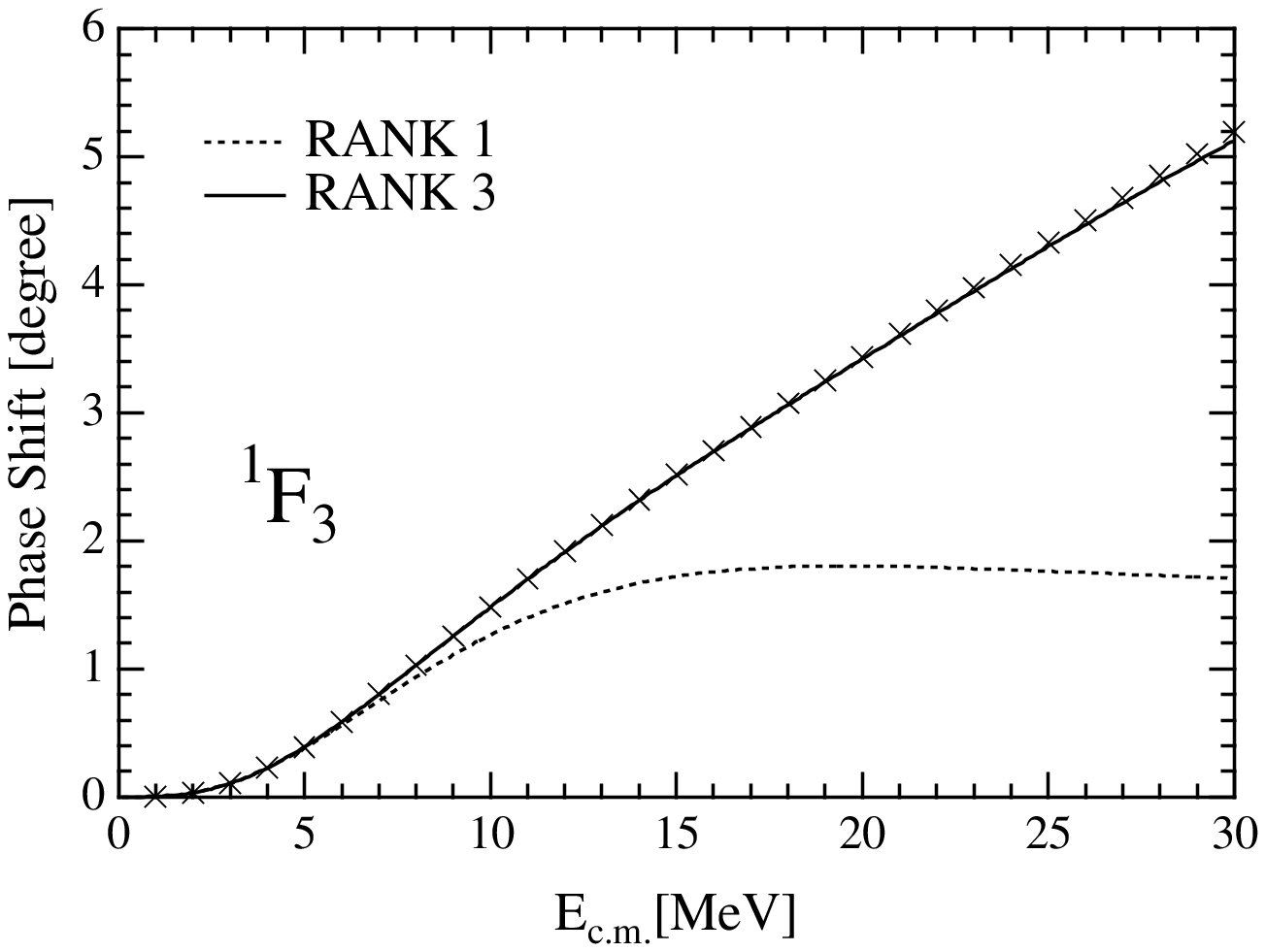}
\caption{Phase shifts for $^1$F$_3$ partial wave without Coulomb effects. 
The crosses denote the RGM+OCM result, the dashed line the rank-1 result, 
and the solid line the rank-3 result, respectively.}
\end{figure}
\subsection{Triplet Channel}
%
The parameters for the spin--triplet partial waves
$^3$S$_1$
($^3$P$_0$,$^3$P$_1$, $^3$P$_2$),
($^3$D$_1$, $^3$D$_2$,$^3$D$_3$),
and
($^3$F$_2$, $^3$F$_3$,$^3$F$_4$), 
 are given in Tables X-XXIX
while the phase shifts are given in Figs. 9-18.
In the case of the $^3$S$_1$ partial wave the rank-3 separable potential 
reproduces quite well the RGM+OCM phase shift.
In the other cases the rank-3 separable potential reproduces equally well 
the RGM+OCM+LS phase shifts.

%
\begin{table}
\caption{
Parameters $a_{n,m}$ of the form factor of Eq. (4.4) 
for the $^3$S$_1$ potential.  The first column gives the rank-1 
parameters with $\lambda_{11}=$-0.25598940$\times10^1$}
\begin{tabular}{lll}
\multicolumn{1}{c}{$a_{1,m}$} & \multicolumn{1}{c}{$a_{2,m}$} & \multicolumn{1}{c}{$a_{3,m}$}\\
\hline
~0.74418704$\times10^1$    &-0.13197287$\times10^3$    &-0.16817810$\times10^2$    \\
-0.99634394$\times10^1$    &~0.10665018$\times10^3$    &~0.11212182$\times10^2$    \\
~0.23859818$\times10^2$    &-0.30740737$\times10^3$    &-0.35992162$\times10^2$    \\
-0.27331253$\times10^2$    &~0.28417959$\times10^3$    &~0.28919328$\times10^2$    \\
~0.20995060$\times10^2$    &-0.17596997$\times10^3$    &-0.15737689$\times10^2$    \\
-0.11699246$\times10^2$    &~0.75921030$\times10^2$    &~0.56602713$\times10^1$    \\
~0.50516284$\times10^1$    &-0.25132501$\times10^2$    &-0.14815321$\times10^1$    \\
-0.17466475$\times10^1$    &~0.68293763$\times10^1$    &~0.30459200                \\
~0.48893164                &-0.16061242$\times10^1$    &-0.56883879$\times10^{-1}$ \\
-0.11058936                &~0.33443137                &~0.11406159$\times10^{-1}$ \\
~0.20040057$\times10^{-1}$ &-0.60527031$\times10^{-1}$ &-0.23718437$\times10^{-2}$ \\
-0.28772181$\times10^{-2}$ &~0.91323541$\times10^{-2}$ &~0.42859938$\times10^{-3}$ \\
~0.32324420$\times10^{-3}$ &-0.11020710$\times10^{-2}$ &-0.60155849$\times10^{-4}$ \\
-0.28013348$\times10^{-4}$ &~0.10296990$\times10^{-3}$ &~0.62866235$\times10^{-5}$ \\
~0.18394221$\times10^{-5}$ &-0.72504967$\times10^{-5}$ &-0.47999987$\times10^{-6}$ \\
-0.89281320$\times10^{-7}$ &~0.37402881$\times10^{-6}$ &~0.26272042$\times10^{-7}$ \\
~0.30874015$\times10^{-8}$ &-0.13612727$\times10^{-7}$ &-0.99973788$\times10^{-9}$ \\
-0.71610131$\times10^{-10}$&~0.32922428$\times10^{-9}$ &~0.25040321$\times10^{-10}$\\
~0.99454633$\times10^{-12}$&-0.47290138$\times10^{-11}$&-0.37035448$\times10^{-12}$\\
-0.62302608$\times10^{-14}$&~0.30443574$\times10^{-13}$&~0.24485301$\times10^{-14}$\\
\end{tabular}
\end{table}
%
%
\begin{table}
\caption{Coupling constants $\lambda_{ij}=\lambda_{ji}$ for the rank-3 
 $^3$S$_1$  potential.}
\begin{tabular}{l}
$\lambda_{11}=$~0.25332051$\times10^1$\\
$\lambda_{12}=$-0.44177961\\ 
$\lambda_{13}=$~0.33554529$\times10^1$\\
$\lambda_{22}=$-0.19258003\\ 
$\lambda_{23}=$~0.11500883$\times10^1$\\
$\lambda_{33}=$-0.67841091$\times10^1$\\
\end{tabular}
\end{table}
%
%
\begin{figure}[h]\centering
\psbox[size=0.5#1]{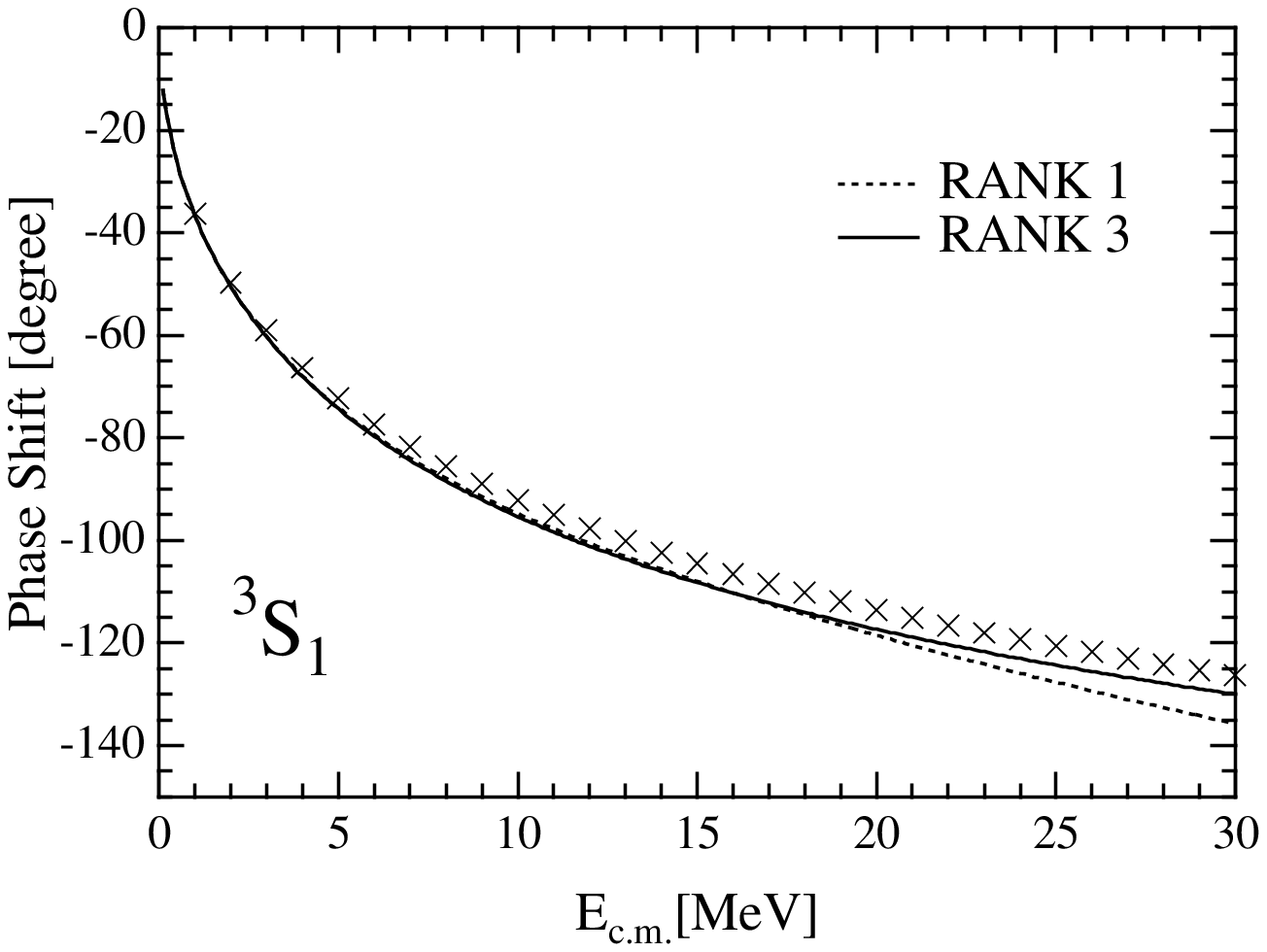}
\caption{Phase shifts for the $^3$S$_1$ partial wave without 
Coulomb effects.  The crosses denote the RGM+OCM result,
the dashed line the rank-1 result, and the solid line the rank-3 result, respectively.}
\end{figure}
%
%
\begin{table}
\caption{
Parameters $a_{n,m}$ of the form factor of Eq. (4.4) 
for the $^3$P$_0$ potential.  The first column lists the 
rank-1 parameters with  $\lambda_{11}=$-0.42316727.}
\begin{tabular}{lll}
\multicolumn{1}{c}{$a_{1,m}$} & \multicolumn{1}{c}{$a_{2,m}$} & \multicolumn{1}{c}{$a_{3,m}$}\\
\hline
-0.66322212$\times10^1$    &-0.79077911$\times10^1$    &-0.59125005$\times10^1$\\
~0.10444487$\times10^2$    &~0.55415998$\times10^1$    &~0.24492255$\times10^1$\\
-0.17647766$\times10^2$    &-0.86712354$\times10^1$    &-0.38446872$\times10^1$\\
~0.16035080$\times10^2$    &~0.44944025$\times10^1$    &~0.77187900            \\
-0.10440381$\times10^2$    &-0.17971337$\times10^1$    &~0.35090483$\times10^{-1}$\\
~0.52745557$\times10^1$    &~0.78617775                &~0.12686539\\
-0.22025734$\times10^1$    &-0.51578766                &-0.30787262\\
~0.77506575                &~0.31027724                &~0.24158041\\
-0.22810778                &-0.13698661                &-0.11373402\\
~0.54968268$\times10^{-1}$ &~0.43590322$\times10^{-1}$ &~0.36741567$\times10^{-1}$\\
-0.10611771$\times10^{-1}$ &-0.10160062$\times10^{-1}$ &-0.85776771$\times10^{-2}$\\
~0.16122722$\times10^{-2}$ &~0.17572875$\times10^{-2}$ &~0.14804394$\times10^{-2}$\\
-0.18998460$\times10^{-3}$ &-0.22687698$\times10^{-3}$ &-0.19056704$\times10^{-3}$\\
~0.17117673$\times10^{-4}$ &~0.21835886$\times10^{-4}$ &~0.18287608$\times10^{-4}$\\
-0.11593237$\times10^{-5}$ &-0.15527041$\times10^{-5}$ &-0.12969524$\times10^{-5}$\\
~0.57638824$\times10^{-7}$ &~0.80073563$\times10^{-7}$ &~0.66726978$\times10^{-7}$\\
-0.20292626$\times10^{-8}$ &-0.28985088$\times10^{-8}$ &-0.24102829$\times10^{-8}$\\
~0.47658438$\times10^{-10}$&~0.69528124$\times10^{-10}$&~0.57703914$\times10^{-10}$\\
-0.66681851$\times10^{-12}$&-0.98848850$\times10^{-12}$&-0.81884000$\times10^{-12}$\\
~0.41875253$\times10^{-14}$&~0.62812081$\times10^{-14}$&~0.51931988$\times10^{-14}$\\
\end{tabular}
\end{table}
%
%
\begin{table}
\caption{Coupling constants $\lambda_{ij}=\lambda_{ji}$ for the rank-3
$^3$P$_0$ potential.}
\begin{tabular}{c}
$\lambda_{11}=$-0.32152588$\times10^2$\\
$\lambda_{12}=$~0.93930066$\times10^2$\\
$\lambda_{13}=$-0.90374767$\times10^2$\\
$\lambda_{22}=$-0.29056296$\times10^3$\\
$\lambda_{23}=$~0.28341791$\times10^3$\\
$\lambda_{33}=$-0.27758530$\times10^3$\\
\end{tabular}
\end{table}
%
%
\begin{figure}[h]\centering
\psbox[size=0.5#1]{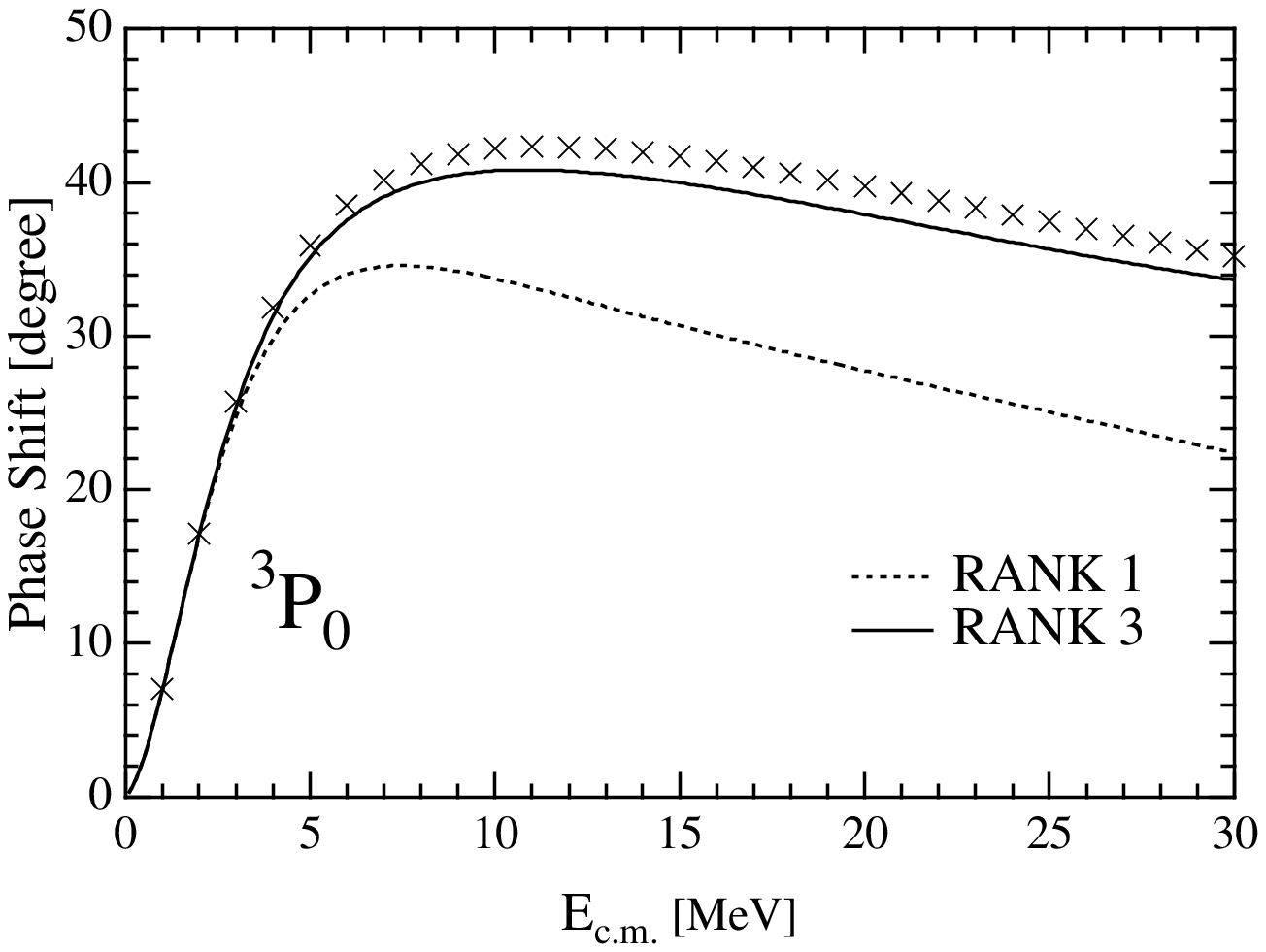}
\caption{Phase shifts for the $^3$P$_0$ state without Coulomb effects. 
The crosses denote the RGM+OCM+LS result,
the dashed line the rank-1 result, and the solid line the rank-3 result, respectively.}
\end{figure}
%
%
\begin{table}
\caption{
Parameters $a_{n,m}$ of the form factor of Eq. (4.4) 
for the $^3$P$_1$ potential.  The first column exhibits the rank-1 parameters
with  $\lambda_{11}=$-0.15329984}
\begin{tabular}{lll}
\multicolumn{1}{c}{$a_{1,m}$}&\multicolumn{1}{c}{$a_{2,m}$}&\multicolumn{1}{c}{$a_{3,m}$}\\
\hline
-0.11983147$\times10^2$    &-0.13550093$\times10^2$    &-0.90367514$\times10^1$\\
~0.17014168$\times10^2$    &~0.73213312$\times10^1$    &~0.24098258$\times10^1$\\
-0.28697894$\times10^2$    &-0.11569940$\times10^2$    &-0.40100766$\times10^1$\\
~0.25268582$\times10^2$    &~0.43479794$\times10^1$    &-0.58369026\\
-0.16200758$\times10^2$    &-0.11052072$\times10^1$    &~0.96722200\\
~0.81469516$\times10^1$    &~0.52139426                &-0.13854134\\
-0.34392479$\times10^1$    &-0.63361772                &-0.38566501\\
~0.12378490$\times10^1$    &~0.47703657                &~0.35442445\\
-0.37446386                &-0.22713933                &-0.17258809\\
~0.92646826$\times10^{-1}$ &~0.74443476$\times10^{-1}$ &~0.56334788$\times10^{-1}$\\
-0.18290727$\times10^{-1}$ &-0.17583617$\times10^{-1}$ &-0.13201035$\times10^{-1}$\\
~0.28292359$\times10^{-2}$ &~0.30615614$\times10^{-2}$ &~0.22815526$\times10^{-2}$\\
-0.33808968$\times10^{-3}$ &-0.39667201$\times10^{-3}$ &-0.29380436$\times10^{-3}$\\
~0.30793834$\times10^{-4}$ &~0.38250521$\times10^{-4}$ &~0.28190681$\times10^{-4}$\\
-0.21030263$\times10^{-5}$ &-0.27223565$\times10^{-5}$ &-0.19982189$\times10^{-5}$\\
~0.10522448$\times10^{-6}$ &~0.14041899$\times10^{-6}$ &~0.10271462$\times10^{-6}$\\
-0.37222405$\times10^{-8}$ &-0.50808201$\times10^{-8}$ &-0.37053918$\times10^{-8}$\\
~0.87716868$\times10^{-10}$&~0.12175644$\times10^{-9}$ &~0.88550576$\times10^{-10}$\\
-0.12300073$\times10^{-11}$&-0.17282223$\times10^{-11}$&-0.12535112$\times10^{-11}$\\
~0.77324366$\times10^{-14}$&~0.10955268$\times10^{-13}$&~0.79236297$\times10^{-14}$\\
\end{tabular}
\end{table}
%
%
\begin{table}
\caption{Coupling constants $\lambda_{ij}=\lambda_{ji}$ for the rank-3 
$^3$P$_1$ potential.}
\begin{tabular}{c}
$\lambda_{11}=$-0.10202649$\times10^2$\\
$\lambda_{12}=$~0.31235498$\times10^2$\\
$\lambda_{13}=$-0.33737459$\times10^2$\\
$\lambda_{22}=$-0.10318492$\times10^3$\\
$\lambda_{23}=$~0.11347250$\times10^3$\\
$\lambda_{33}=$-0.12541685$\times10^3$\\
\end{tabular}
\end{table}
%
%
\begin{figure}[h]\centering
\psbox[size=0.5#1]{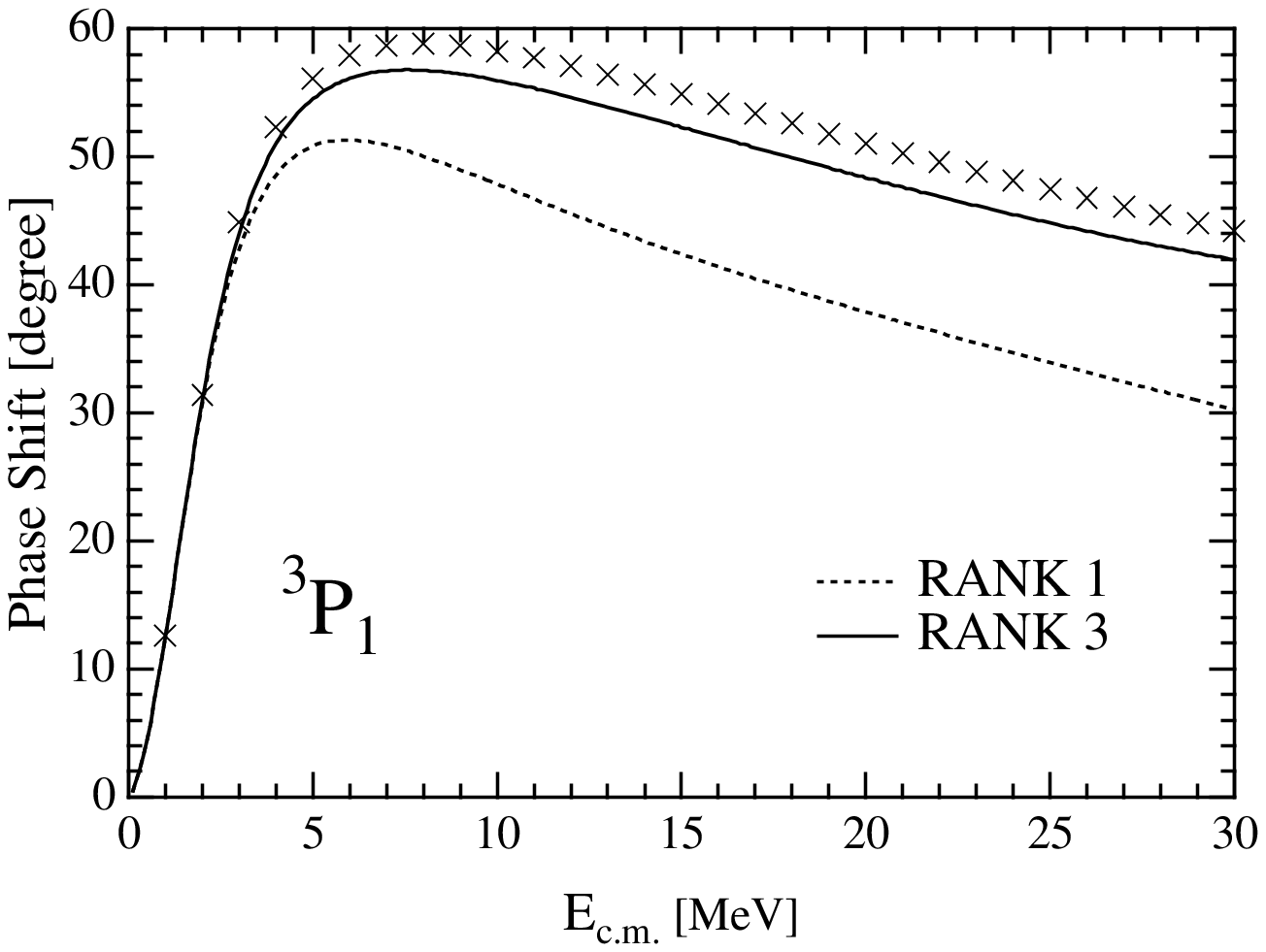}
\caption{Phase shifts for the $^3$P$_1$ partial wave without Coulomb effects. 
The crosses denote the RGM+OCM+LS result,
the dashed line the rank-1 result, and the solid line the rank-3  result, respectively.}
\end{figure}
%
%
%
\begin{table}
\caption{
Parameters $a_{n,m}$ of the form factor of Eq. (4.4) 
for the $^3$P$_2$ potential.  The first column presents the rank-1 parameters
with $\lambda_{11}=$-0.66395138$\times10^{-1}$}
\begin{tabular}{lll}
\multicolumn{1}{c}{$a_{1,m}$}&\multicolumn{1}{c}{$a_{2,m}$}
&\multicolumn{1}{c}{$a_{3,m}$}\\
\hline
-0.18802681$\times10^2$    &-0.20111944$\times10^2$    &-0.11866155$\times10^2$\\
~0.24695039$\times10^2$    &~0.86908371$\times10^1$    &~0.19942254$\times10^1$\\
-0.41611877$\times10^2$    &-0.13903725$\times10^2$    &-0.36405995$\times10^1$\\
~0.35688295$\times10^2$    &~0.31275599$\times10^1$    &-0.22992102$\times10^1$\\
-0.22579637$\times10^2$    &~0.31524135                &~0.20651004$\times10^1$\\
~0.11307515$\times10^2$    &-0.49795244$\times10^{-1}$ &-0.47700417\\
-0.48182932$\times10^1$    &-0.68502022                &-0.42575885\\
~0.17672518$\times10^1$    &~0.64872070                &~0.44814211\\
-0.54652963                &-0.32730654                &-0.22367571\\
~0.13797319                &~0.10952640                &~0.73558687$\times10^{-1}$\\
-0.27690659$\times10^{-1}$ &-0.26106999$\times10^{-1}$ &-0.17285137$\times10^{-1}$\\
~0.43380605$\times10^{-2}$ &~0.45661980$\times10^{-2}$ &~0.29907200$\times10^{-2}$\\
-0.52342569$\times10^{-3}$ &-0.59304311$\times10^{-3}$ &-0.38526921$\times10^{-3}$\\
~0.48023975$\times10^{-4}$ &~0.57258765$\times10^{-4}$ &~0.36964876$\times10^{-4}$\\
-0.32978152$\times10^{-5}$ &-0.40774761$\times10^{-5}$ &-0.26191716$\times10^{-5}$\\
~0.16568211$\times10^{-6}$ &~0.21032277$\times10^{-6}$ &~0.13454055$\times10^{-6}$\\
-0.58782654$\times10^{-8}$ &-0.76069363$\times10^{-8}$ &-0.48484250$\times10^{-8}$\\
~0.13880285$\times10^{-9}$ &~0.18213056$\times10^{-9}$ &~0.11569505$\times10^{-9}$\\
-0.19485870$\times10^{-11}$&-0.25815323$\times10^{-11}$&-0.16344262$\times10^{-11}$\\
~0.12253427$\times10^{-13}$&~0.16330410$\times10^{-13}$&~0.10302405$\times10^{-13}$\\
\end{tabular}
\end{table}
%
%
\begin{table}
\caption{Coupling constants $\lambda_{ij}=\lambda_{ji}$ for the rank-3 
$^3$P$_2$ potential.}
\begin{tabular}{c}
$\lambda_{11}=$-0.41423805$\times10^1$\\
$\lambda_{12}=$~0.13294740$\times10^2$\\
$\lambda_{13}=$-0.16229175$\times10^2$\\
$\lambda_{22}=$-0.46727996$\times10^2$\\
$\lambda_{23}=$~0.58280329$\times10^2$\\
$\lambda_{33}=$-0.73111775$\times10^2$\\
\end{tabular}
\end{table}
%
%
\begin{figure}[h]\centering
\psbox[size=0.5#1]{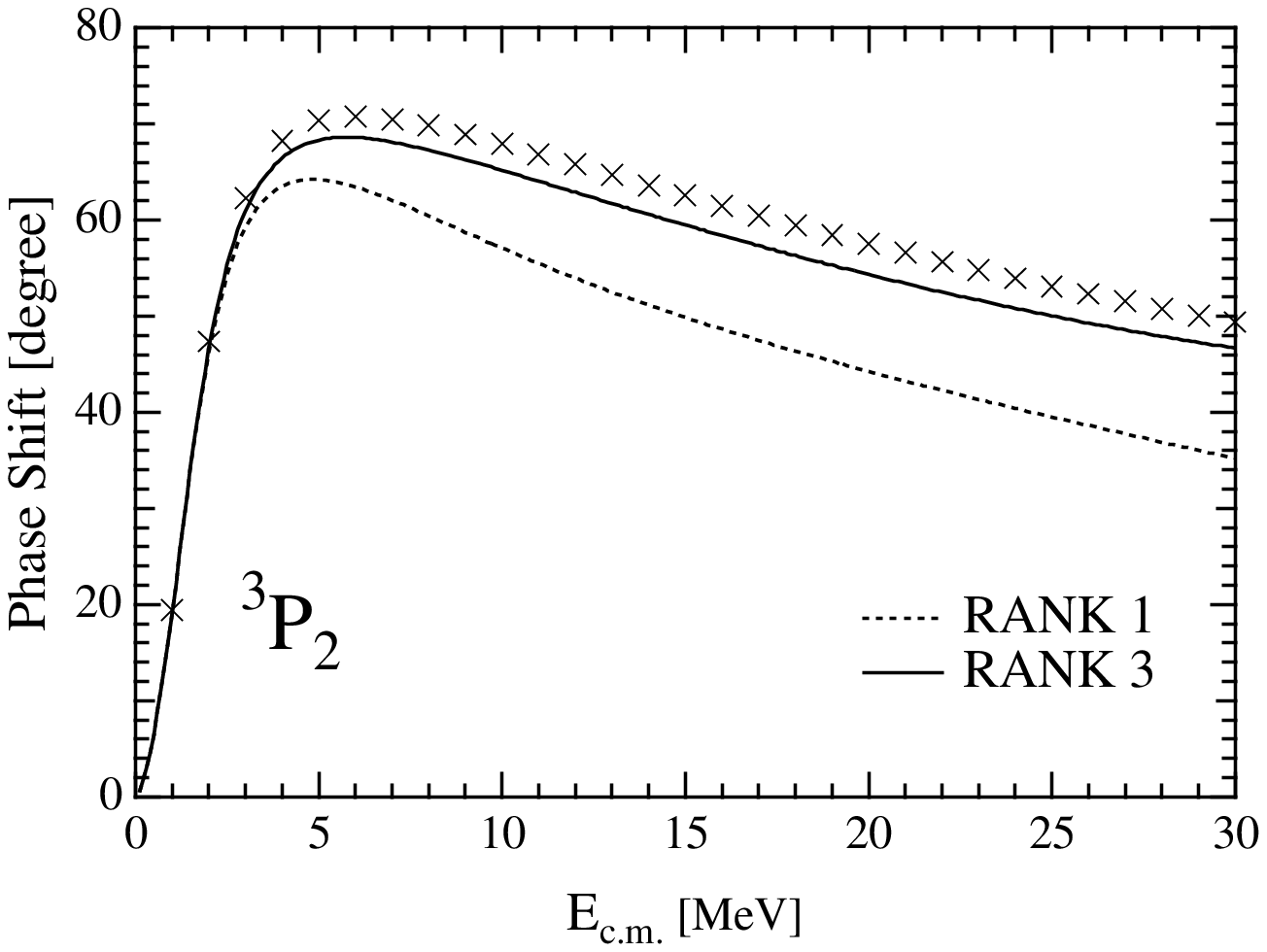}
\caption{Phase shifts for $^3$P$_2$ state without Coulomb effects. 
The crosses denote the RGM+OCM+LS result,
the dashed line the rank-1 result, and the solid line the rank-3 result, respectively.}
\end{figure}
%
%
%
\begin{table}
\caption{
Parameters $a_{n,m}$ of the form factor of Eq. (4.4) 
for the $^3$D$_1$ potential.  The first column lists the rank-1 parameters
with $\lambda_{11}=$~0.38702146$\times10^2$.}
\begin{tabular}{lll}
\multicolumn{1}{c}{$a_{1,m}$}&\multicolumn{1}{c}
		{$a_{2,m}$}&\multicolumn{1}{c}{$a_{3,m}$}\\
\hline
~0.89936516                &~0.26418769$\times10^1$    &~0.23149927$\times10^1$\\
-0.26186344$\times10^1$    &-0.76699755$\times10^1$    &-0.69187833$\times10^1$\\
~0.40420417$\times10^1$    &~0.10171559$\times10^2$    &~0.80438855$\times10^1$\\
-0.37254532$\times10^1$    &-0.77481279$\times10^1$    &-0.53201948$\times10^1$\\
~0.24045068$\times10^1$    &~0.40778272$\times10^1$    &~0.23638415$\times10^1$\\
-0.11674331$\times10^1$    &-0.15957158$\times10^1$    &-0.74115241\\
~0.44371340                &~0.48775235                &~0.16543078\\
-0.13423928                &-0.11887084                &-0.23546429$\times10^{-1}$\\
~0.32471944$\times10^{-1}$ &~0.23244826$\times10^{-1}$ &~0.82565419$\times10^{-3}$\\
-0.62682883$\times10^{-2}$ &-0.36286167$\times10^{-2}$ &~0.57391815$\times10^{-3}$\\
~0.96078361$\times10^{-3}$ &~0.44617253$\times10^{-3}$ &-0.17882613$\times10^{-3}$\\
-0.11613293$\times10^{-3}$ &-0.42259918$\times10^{-4}$ &~0.30875386$\times10^{-4}$\\
~0.10972859$\times10^{-4}$ &~0.29721861$\times10^{-5}$ &-0.36884776$\times10^{-5}$\\
-0.80105503$\times10^{-6}$ &-0.14441536$\times10^{-6}$ &~0.32121255$\times10^{-6}$\\
~0.44458731$\times10^{-7}$ &~0.39214969$\times10^{-8}$ &-0.20613138$\times10^{-7}$\\
-0.18322050$\times10^{-8}$ &~0.15747854$\times10^{-10}$&~0.96582033$\times10^{-9}$\\
~0.54076554$\times10^{-10}$&-0.61093792$\times10^{-11}$&-0.32154572$\times10^{-10}$\\
-0.10765494$\times10^{-11}$&~0.24518321$\times10^{-12}$&~0.72109404$\times10^{-12}$\\
~0.12907316$\times10^{-13}$&-0.45959801$\times10^{-14}$&-0.97789252$\times10^{-14}$\\
-0.70194793$\times10^{-16}$&~0.35260306$\times10^{-16}$&~0.60733413$\times10^{-16}$\\
\end{tabular}
\end{table}
%
%
\begin{table}
\caption{Coupling constants $\lambda_{ij}=\lambda_{ji}$ for the rank-3 
$^3$D$_1$ potential.}
\begin{tabular}{c}
$\lambda_{11}=$-0.12403227$\times10^5$\\
$\lambda_{12}=$~0.72374384$\times10^4$\\
$\lambda_{13}=$-0.33925562$\times10^4$\\
$\lambda_{22}=$-0.42322025$\times10^4$\\
$\lambda_{23}=$~0.19982358$\times10^4$\\
$\lambda_{33}=$-0.95274067$\times10^3$\\
\end{tabular}
\end{table}
%
%
\begin{figure}[h]
\centering
\psbox[size=0.5#1]{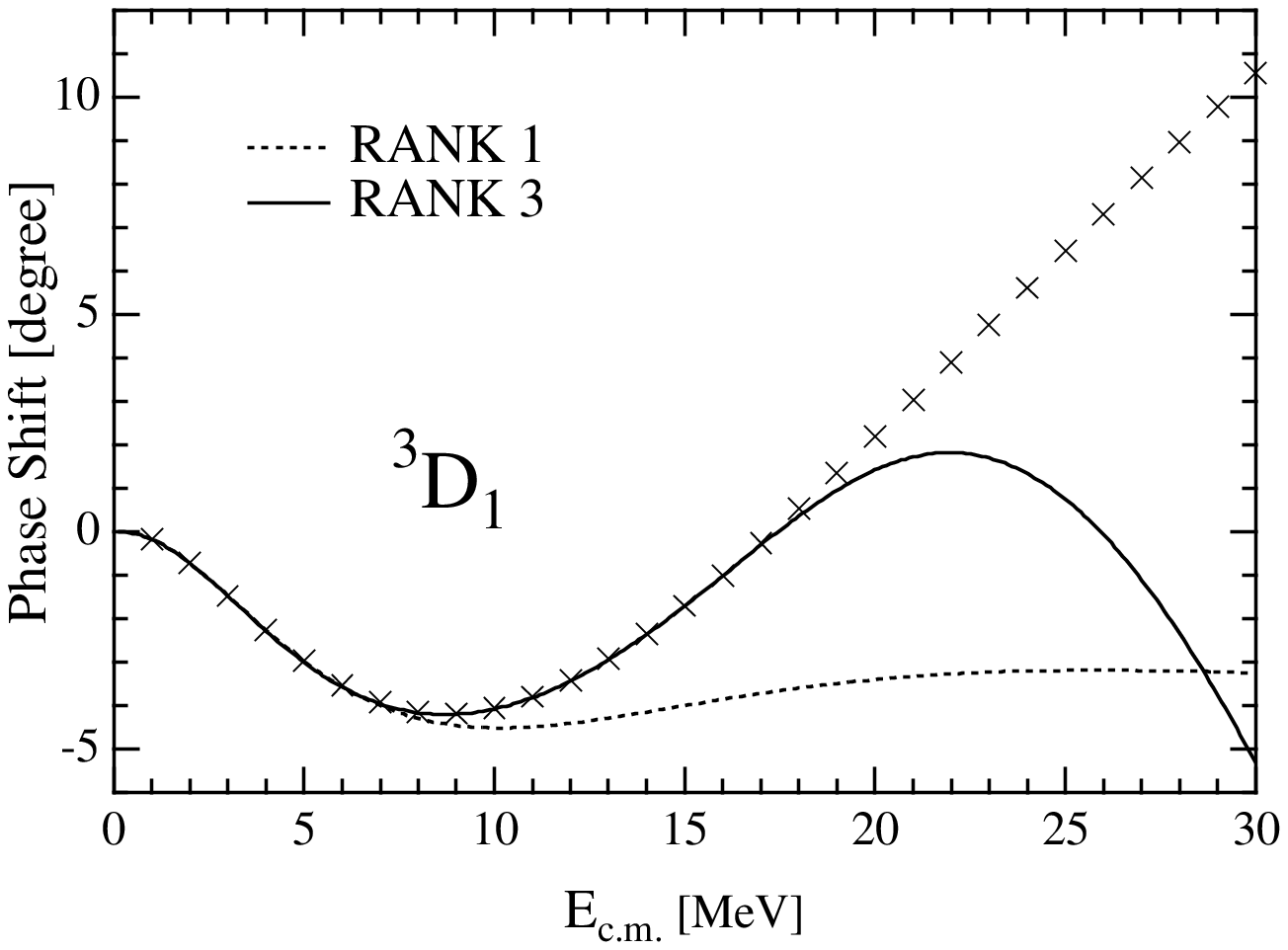}
\caption{Phase shifts for $^3$D$_1$ state without Coulomb effects. 
The crosses denote the RGM+OCM+LS result,
the dashed line the rank-1 result, and the solid line the rank-3 result, respectively.}
\end{figure}
%
%
\begin{table}
\caption{
Parameters $a_{n,m}$ of the form factor of Eq. (4.4) 
for the $^3$D$_2$ potential.  The first column exhibits the rank-1 parameters
with $\lambda_{11}=$~0.59024361$\times10^2$}
\begin{tabular}{lll}
\multicolumn{1}{c}{$a_{1,m}$}&\multicolumn{1}{c}{$a_{2,m}$}&\multicolumn{1}{c}{$a_{3,m}$}\\
\hline
~0.57066711                &~0.16484822$\times10^1$    &~0.14099209$\times10^1$\\
-0.16583375$\times10^1$    &-0.47749159$\times10^1$    &-0.42072799$\times10^1$\\
~0.25409617$\times10^1$    &~0.62850269$\times10^1$    &~0.48779676$\times10^1$\\
-0.23332252$\times10^1$    &-0.47697116$\times10^1$    &-0.32222764$\times10^1$\\
~0.15018139$\times10^1$    &~0.25024490$\times10^1$    &~0.14316862$\times10^1$\\
-0.72790438                &-0.97809048                &-0.45159892\\
~0.27639429                &~0.29955475                &~0.10332838\\
-0.83595442$\times10^{-1}$ &-0.73538742$\times10^{-1}$ &-0.161848780$\times10^{-1}$\\
~0.20227441$\times10^{-1}$ &~0.14601065$\times10^{-1}$ &~0.127725469$\times10^{-2}$\\
-0.39078222$\times10^{-2}$ &-0.23406505$\times10^{-2}$ &~0.124271741$\times10^{-3}$\\
~0.59974333$\times10^{-3}$ &~0.30037269$\times10^{-3}$ &-0.610339620$\times10^{-4}$\\
-0.72616864$\times10^{-4}$ &-0.30435345$\times10^{-4}$ &~0.111994241$\times10^{-4}$\\
~0.68758985$\times10^{-5}$ &~0.23895692$\times10^{-5}$ &-0.133580683$\times10^{-5}$\\
-0.50325849$\times10^{-6}$ &-0.14163274$\times10^{-6}$ &~0.113449284$\times10^{-6}$\\
~0.28016277$\times10^{-7}$ &~0.60927632$\times10^{-8}$ &-0.701058725$\times10^{-8}$\\
-0.11587213$\times10^{-8}$ &-0.17749233$\times10^{-9}$ &~0.313672923$\times10^{-9}$\\
~0.34341671$\times10^{-10}$&~0.29770498$\times10^{-11}$&-0.991409110$\times10^{-11}$\\
-0.68698690$\times10^{-12}$&-0.11337993$\times10^{-13}$&~0.210314951$\times10^{-12}$\\
~0.82832450$\times10^{-14}$&-0.48655006$\times10^{-15}$&-0.269557305$\times10^{-14}$\\
-0.45347221$\times10^{-16}$&~0.63794448$\times10^{-17}$&~0.158651456$\times10^{-16}$\\
\end{tabular}
\end{table}
%
%
\begin{table}
\caption{Coupling constants $\lambda_{ij}=\lambda_{ji}$ for the rank-3 
$^3$D$_2$ potential.}
\begin{tabular}{c}
$\lambda_{11}=$~0.28519285$\times10^5$\\
$\lambda_{12}=$-0.17108157$\times10^5$\\
$\lambda_{13}=$~0.84546586$\times10^4$\\
$\lambda_{22}=$~0.10247694$\times10^5$\\
$\lambda_{23}=$-0.50401399$\times10^4$\\
$\lambda_{33}=$~0.24628297$\times10^4$\\
\end{tabular}
\end{table}
%
%
\begin{figure}[h]\centering
\psbox[size=0.5#1]{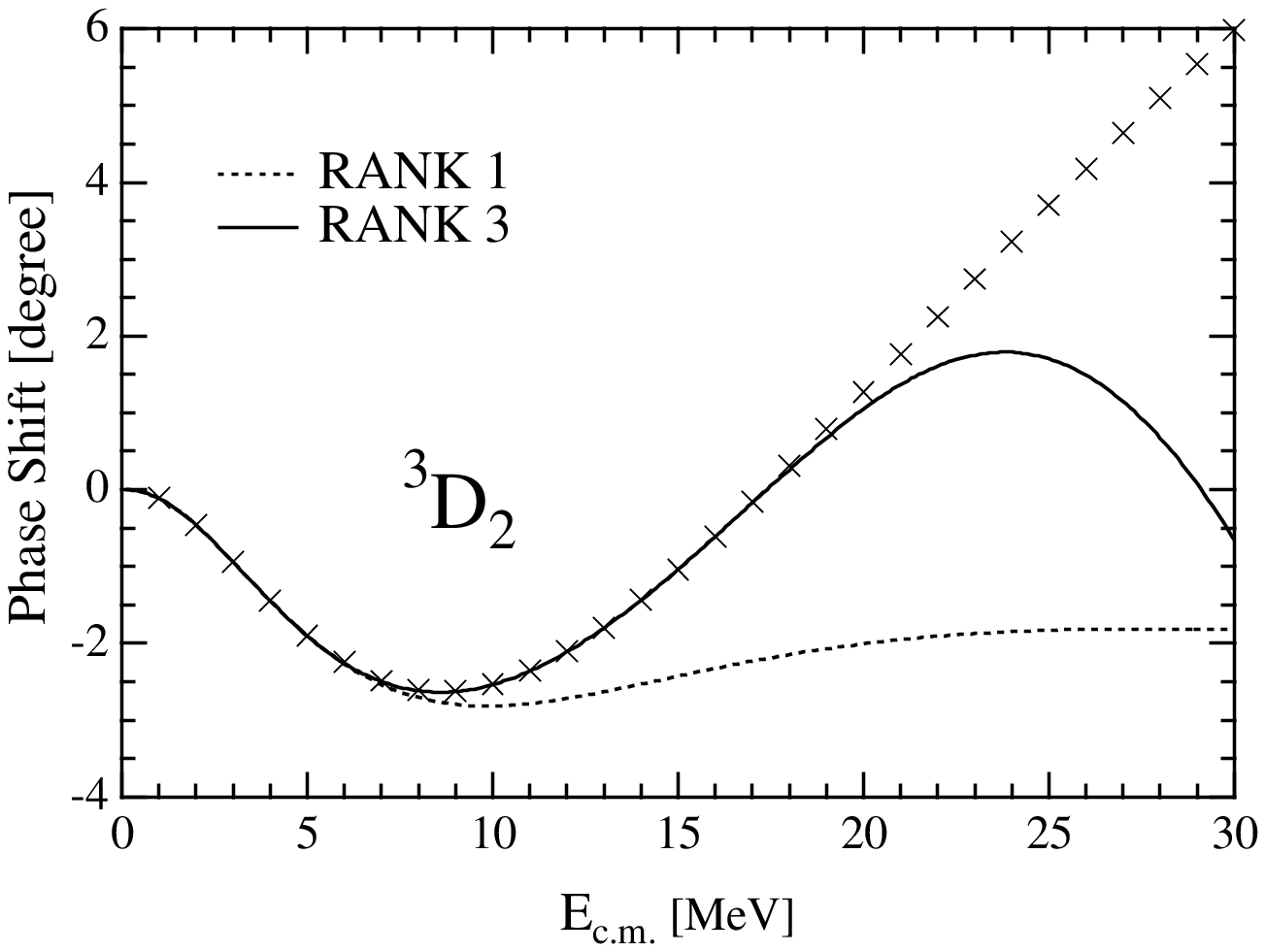}
\caption {Phase shifts for $^3$D$_2$ state without Coulomb effects. 
The crosses denote the RGM+OCM+LS result,
the dashed line the rank-1 result, and the solid line the rank-3 result, respectively.}
\end{figure}
%
%
%
\begin{table}
\caption{
Parameters $a_{n,m}$ of the form factor of Eq. (4.4) 
for the $^3$D$_3$ potential. The first column gives the rank-1 parameters
with  $\lambda_{11}=$~0.48683315$\times10^2$}
\begin{tabular}{lll}
\multicolumn{1}{c}{$a_{1,m}$}&\multicolumn{1}{c}{$a_{2,m}$}&\multicolumn{1}{c}{$a_{3,m}$}\\
\hline
~0.70053005                &~0.20372307$\times10^1$    &~0.17586107$\times10^1$\\
-0.20371819$\times10^1$    &-0.59058643$\times10^1$    &-0.52507453$\times10^1$\\
~0.31301669$\times10^1$    &~0.77955339$\times10^1$    &~0.60941222$\times10^1$\\
-0.28784057$\times10^1$    &-0.59247131$\times10^1$    &-0.40278443$\times10^1$\\
~0.18547114$\times10^1$    &~0.31124541$\times10^1$    &~0.17900376$\times10^1$\\
-0.89957466                &-0.12173452$\times10^1$    &-0.563742149\\
~0.34172242                &~0.37272087                &~0.128054107\\
-0.10337205                &-0.91321241$\times10^{-1}$ &-0.195064865$\times10^{-1}$\\
~0.25011910$\times10^{-1}$ &~0.18051377$\times10^{-1}$ &~0.129543369$\times10^{-2}$\\
-0.48311038$\times10^{-2}$ &-0.28709573$\times10^{-2}$ &~0.243205612$\times10^{-3}$\\
~0.74115882$\times10^{-3}$ &~0.36379018$\times10^{-3}$ &-0.950679435$\times10^{-4}$\\
-0.89691472$\times10^{-4}$ &-0.36150992$\times10^{-4}$ &~0.169973636$\times10^{-4}$\\
~0.84868416$\times10^{-5}$ &~0.27543762$\times10^{-5}$ &-0.203045225$\times10^{-5}$\\
-0.62064686$\times10^{-6}$ &-0.15550511$\times10^{-6}$ &~0.174498404$\times10^{-6}$\\
~0.34516816$\times10^{-7}$ &~0.61282950$\times10^{-8}$ &-0.109761505$\times10^{-7}$\\
-0.14259032$\times10^{-8}$ &-0.14661306$\times10^{-9}$ &~0.502037641$\times10^{-9}$\\
~0.42202574$\times10^{-10}$&~0.10222286$\times10^{-11}$&-0.162758205$\times10^{-10}$\\
-0.84290188$\times10^{-12}$&~0.50688694$\times10^{-13}$&~0.355038528$\times10^{-12}$\\
~0.10144471$\times10^{-13}$&-0.15433806$\times10^{-14}$&-0.468513806$\times10^{-14}$\\
-0.55417256$\times10^{-16}$&~0.14088408$\times10^{-16}$&~0.283746588$\times10^{-16}$\\
\end{tabular}
\end{table}
%
%
\begin{table}
\caption{Coupling constants $\lambda_{ij}=\lambda_{ji}$ for the rank-3 
$^3$D$_3$ potential.}
\begin{tabular}{c}
$\lambda_{11}=$~0.17437597$\times10^6$\\
$\lambda_{12}=$-0.10347752$\times10^6$\\
$\lambda_{13}=$~0.50147622$\times10^5$\\
$\lambda_{22}=$~0.61393167$\times10^5$\\
$\lambda_{23}=$-0.29733363$\times10^5$\\
$\lambda_{33}=$~0.14387562$\times10^5$\\
\end{tabular}
\end{table}
%
%
\begin{figure}[h]\centering
\psbox[size=0.5#1]{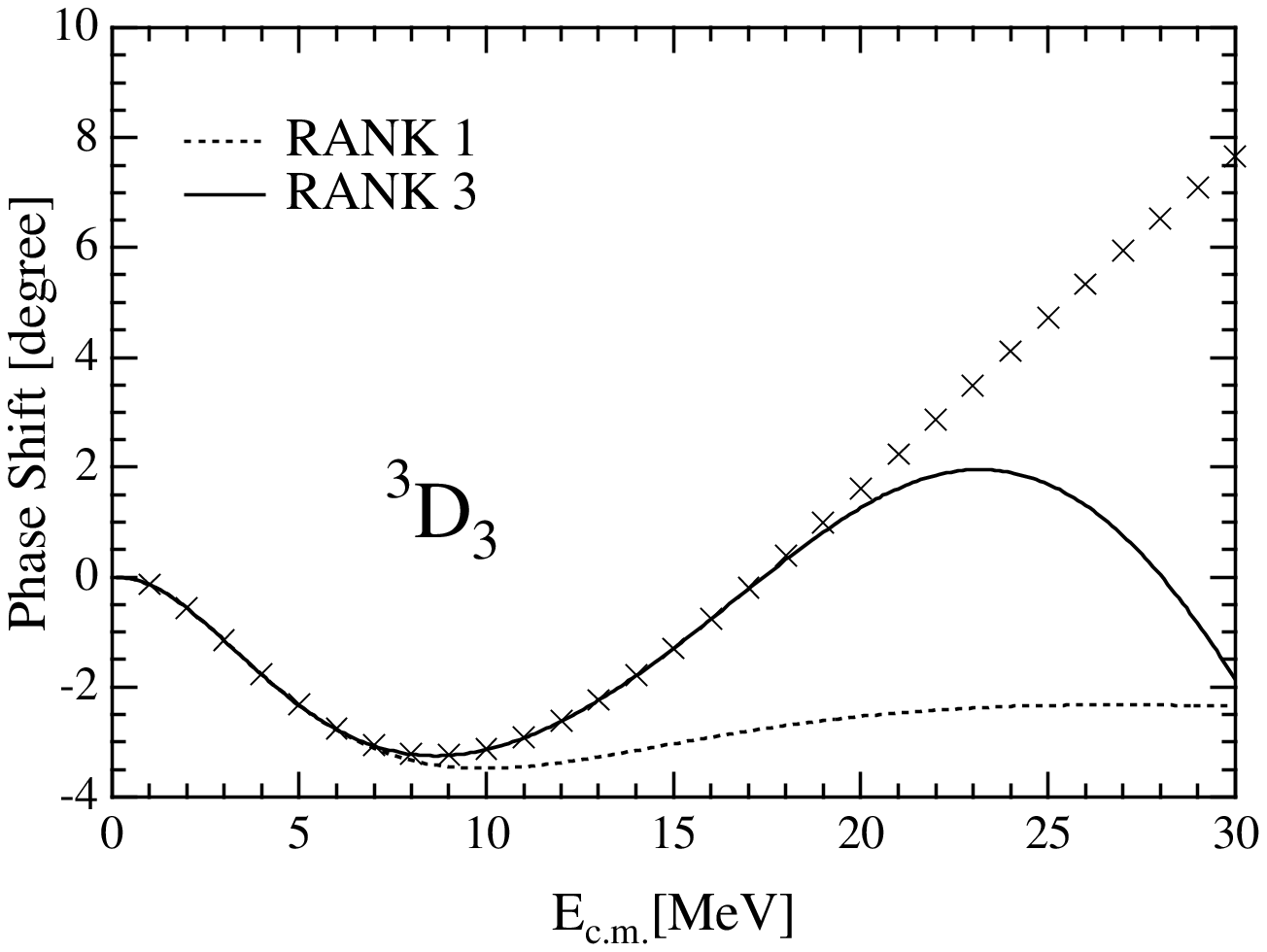}
\caption{Phase shifts for $^3$D$_3$ state without Coulomb effects. 
The crosses denote the RGM+OCM+LS result,
the dashed line the rank-1 result, and the solid line the rank-3 result, respectively.}
\end{figure}
%
%
\begin{table}
\caption{Parameters $a_{n,m}$ of the form factor of Eq. (4.4) 
for the $^3$F$_2$ potential. The first column gives the rank-1 parameters
with  $\lambda_{11}=$-0.15280632$\times10^4$}
\begin{tabular}{lll}
\multicolumn{1}{c}{$a_{1,m}$}&\multicolumn{1}{c}{$a_{2,m}$}&\multicolumn{1}{c}{$a_{3,m}$}\\
\hline
-0.10519807                &-0.10795104$\times10^1$    &-0.11657860$\times10^1$\\
~0.26681674                &~0.22362235$\times10^1$    &~0.20871658$\times10^1$\\
-0.38458973                &-0.29306786$\times10^1$    &-0.25641087$\times10^1$\\
~0.33899788                &~0.22258120$\times10^1$    &~0.17311373$\times10^1$\\
-0.20956819                &-0.11544199$\times10^1$    &-0.759879540\\
~0.96873287$\times10^{-1}$ &~0.42971181                &~0.208280611\\
-0.34663036$\times10^{-1}$ &-0.11609305                &-0.224855812$\times10^{-1}$\\
~0.97602881$\times10^{-2}$ &~0.21859658$\times10^{-1}$ &-0.890945285$\times10^{-2}$\\
-0.21786339$\times10^{-2}$ &-0.24236811$\times10^{-2}$ &~0.551995844$\times10^{-2}$\\
~0.38647216$\times10^{-3}$ &-0.54141248$\times10^{-5}$ &-0.161738180$\times10^{-2}$\\
-0.54465017$\times10^{-4}$ &~0.60393519$\times10^{-4}$ &~0.316664790$\times10^{-3}$\\
~0.60817595$\times10^{-5}$ &-0.12963497$\times10^{-4}$ &-0.447135268$\times10^{-4}$\\
-0.53523779$\times10^{-6}$ &~0.16249029$\times10^{-5}$ &~0.466785193$\times10^{-5}$\\
~0.36794342$\times10^{-7}$ &-0.13891113$\times10^{-6}$ &-0.362189145$\times10^{-6}$\\
-0.19476964$\times10^{-8}$ &~0.83947022$\times10^{-8}$ &~0.207493943$\times10^{-7}$\\
~0.77622991$\times10^{-10}$&-0.35813659$\times10^{-9}$ &-0.862103751$\times10^{-9}$\\
-0.22468227$\times10^{-11}$&~0.10514171$\times10^{-10}$&~0.251419959$\times10^{-10}$\\
~0.44459234$\times10^{-13}$&-0.20104072$\times10^{-12}$&-0.485787313$\times10^{-12}$\\
-0.53628444$\times10^{-15}$&~0.22387162$\times10^{-14}$&~0.556075874$\times10^{-14}$\\
~0.29648478$\times10^{-17}$&-0.10924776$\times10^{-16}$&-0.284419168$\times10^{-16}$\\
\end{tabular}
\end{table}
%
%
\begin{table}
\caption{Coupling constants $\lambda_{ij}=\lambda_{ji}$ for the rank-3 
$^3$F$_2$ potential.}
\begin{tabular}{c}
$\lambda_{11}=$-0.22056858$\times10^6$\\
$\lambda_{12}=$~0.53156082$\times10^5$\\
$\lambda_{13}=$-0.29476815$\times10^5$\\
$\lambda_{22}=$-0.13088707$\times10^5$\\
$\lambda_{23}=$~0.73245374$\times10^4$\\
$\lambda_{33}=$-0.41218725$\times10^4$\\
\end{tabular}
\end{table}
%
%
\begin{figure}[h]\centering
\psbox[size=0.5#1]{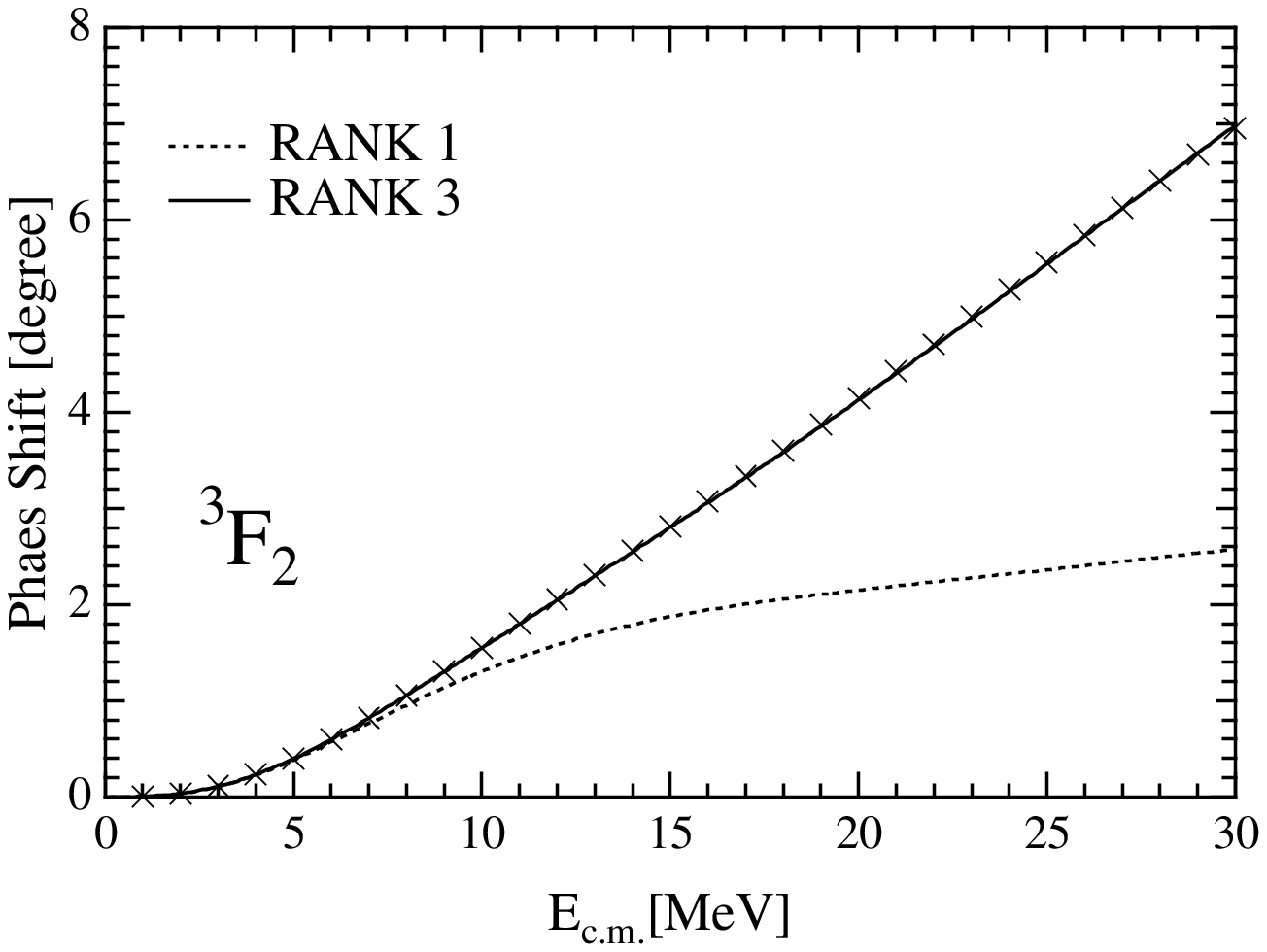}
\caption{Phase shifts for $^3$F$_2$ state without Coulomb effects. 
The crosses denote the RGM+OCM+LS result,
the dashed line the rank-1 result, and the solid line the rank-3 result, respectively.}
\end{figure}
%
%
\begin{table}
\caption{Parameters $a_{n,m}$ of the form factor of Eq. (4.4) 
for the $^3$F$_3$ potential. The first column lists the rank-1 parameters
with $\lambda_{11}=$-0.15533123$\times10^4$}  
\begin{tabular}{lll}
\multicolumn{1}{c}{$a_{1,m}$}&\multicolumn{1}{c}{$a_{2,m}$}&\multicolumn{1}{c}{$a_{3,m}$}\\
\hline
-0.10355351                &-0.10619702$\times10^1$    &-0.11465411$\times10^1$\\
~0.26271214                &~0.22009693$\times10^1$    &~0.20540811$\times10^1$\\
-0.37869541                &-0.28848496$\times10^1$    &-0.25239699$\times10^1$\\
~0.33382672                &~0.21914494$\times10^1$    &~0.17046664$\times10^1$\\
-0.20638383                &-0.11368554$\times10^1$    &-0.748665750\\
~0.95407758$\times10^{-1}$ &~0.42333402                &~0.205500714\\
-0.34141350$\times10^{-1}$ &-0.11445199                &-0.223823711$\times10^{-1}$\\
~0.96143050$\times10^{-2}$ &~0.21585398$\times10^{-1}$ &-0.867190389$\times10^{-2}$\\
-0.21462844$\times10^{-2}$ &-0.24058411$\times10^{-2}$ &~0.540397503$\times10^{-2}$\\
~0.38078040$\times10^{-3}$ &-0.75437723$\times10^{-6}$ &-0.158522915$\times10^{-2}$\\
-0.53669961$\times10^{-4}$ &~0.58646658$\times10^{-4}$ &~0.310495776$\times10^{-3}$\\
~0.59937965$\times10^{-5}$ &-0.12652563$\times10^{-4}$ &-0.438486050$\times10^{-4}$\\
-0.52756756$\times10^{-6}$ &~0.15883035$\times10^{-5}$ &~0.457758155$\times10^{-5}$\\
~0.36271728$\times10^{-7}$ &-0.13585744$\times10^{-6}$ &-0.355154343$\times10^{-6}$\\
-0.19202569$\times10^{-8}$ &~0.82111556$\times10^{-8}$ &~0.203430516$\times10^{-7}$\\
~0.76537251$\times10^{-10}$&-0.35024827$\times10^{-9}$ &-0.845015123$\times10^{-9}$\\
-0.22155859$\times10^{-11}$&~0.10278242$\times10^{-10}$&~0.246353323$\times10^{-10}$\\
~0.43844231$\times10^{-13}$&-0.19638883$\times10^{-12}$&-0.475781212$\times10^{-12}$\\
-0.52889751$\times10^{-15}$&~0.21844895$\times10^{-14}$&~0.544288742$\times10^{-14}$\\
~0.29241713$\times10^{-17}$&-0.10642051$\times10^{-16}$&-0.278159322$\times10^{-16}$\\
\end{tabular}
\end{table}
%
%
\begin{table}
\caption{The coupling constants $\lambda_{ij}=\lambda_{ji}$ for the rank-3 
$^3$F$_3$ potential.}
\begin{tabular}{c}
$\lambda_{11}=$-0.22415203$\times10^6$\\
$\lambda_{12}=$~0.54047498$\times10^5$\\
$\lambda_{13}=$-0.29976913$\times10^5$\\
$\lambda_{22}=$-0.13315252$\times10^5$\\
$\lambda_{23}=$~0.74528575$\times10^4$\\
$\lambda_{33}=$-0.41950064$\times10^4$\\
\end{tabular}
\end{table}
%
%
\begin{figure}[h]\centering
\psbox[size=0.5#1]{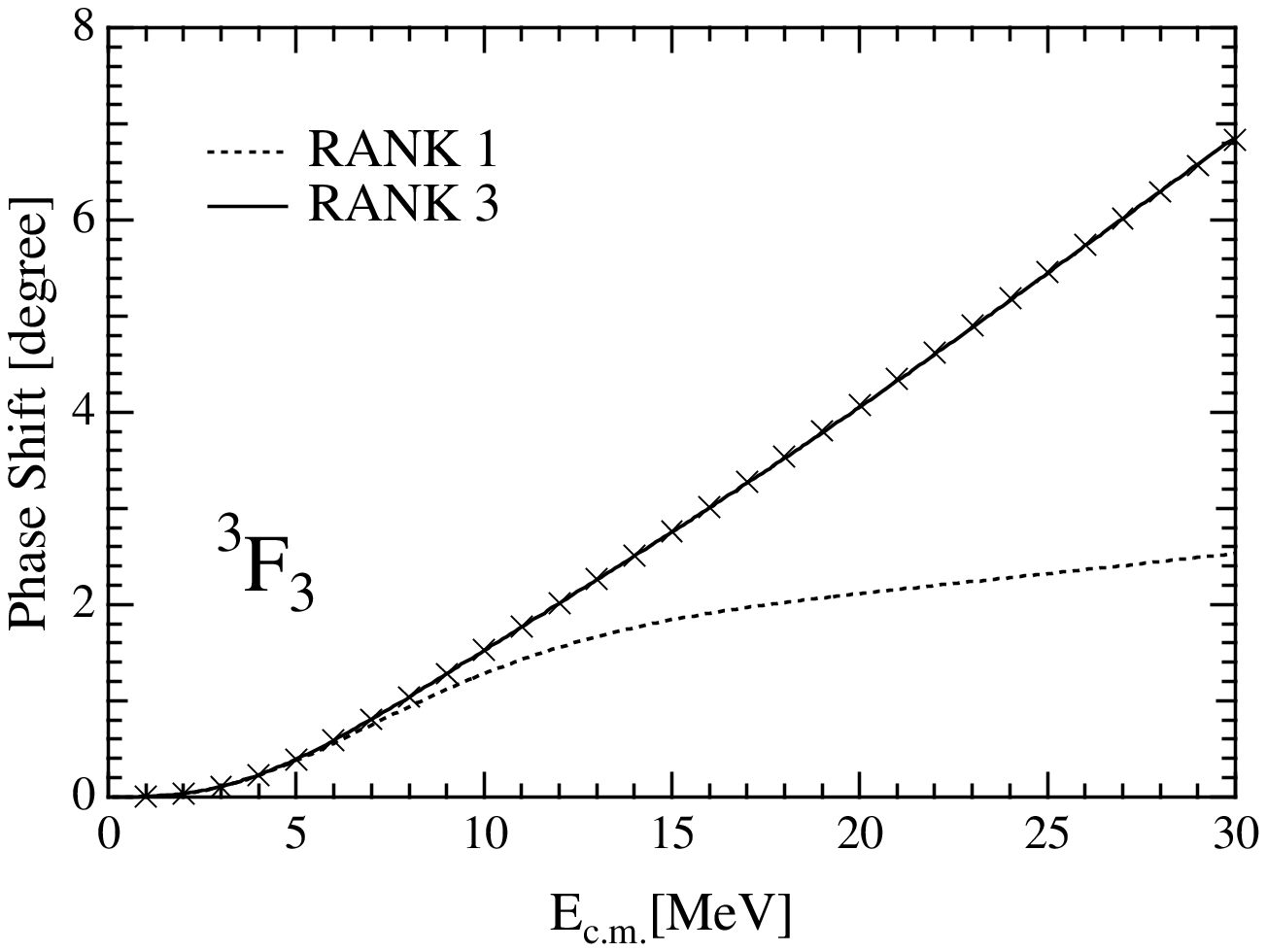}
\caption{Phase shifts for $^3$F$_2$ state without Coulomb effects. 
The crosses denote the RGM+OCM+LS result,
the dashed line the rank-1 result, and the solid line the rank-3 result, respectively.}
\end{figure}
%
%
\begin{table}
\caption{Parameters $a_{n,m}$ of the form factor of Eq. (4.4) 
for the $^3$F$_4$ potential. The first column exhibits the rank-1 parameters
with $^3$F$_4$ $\lambda_{11}=$-0.12982093$\times10^4$}
\begin{tabular}{lll}
\multicolumn{1}{c}{$a_{1,m}$}&\multicolumn{1}{c}{$a_{2,m}$}&\multicolumn{1}{c}{$a_{3,m}$}\\
\hline
-0.12296582                &-0.12705383$\times10^1$    &-0.13761116$\times10^1$\\
~0.31102049                &~0.26178431$\times10^1$    &~0.24457908$\times10^1$\\
-0.44801941                &-0.34259243$\times10^1$    &-0.29980301$\times10^1$\\
~0.39459099                &~0.25961039$\times10^1$    &~0.20157805$\times10^1$\\
-0.24377244                &-0.13430375$\times10^1$    &-0.87938384\\
~0.11259879                &~0.49776587                &~0.23706582\\
-0.40253696$\times10^{-1}$ &-0.13336950                &-0.22922059$\times10^{-1}$\\
~0.11322223$\times10^{-1}$ &~0.24641103$\times10^{-1}$ &-0.11752433$\times10^{-1}$\\
-0.25240891$\times10^{-2}$ &-0.25611382$\times10^{-2}$ &~0.68583670$\times10^{-2}$\\
~0.44711789$\times10^{-3}$ &-0.68617181$\times10^{-4}$ &-0.19848209$\times10^{-2}$\\
-0.62915174$\times10^{-4}$ &~0.81559745$\times10^{-4}$ &~0.38692325$\times10^{-3}$\\
~0.70141424$\times10^{-5}$ &-0.16636723$\times10^{-4}$ &-0.54555554$\times10^{-4}$\\
-0.61631030$\times10^{-6}$ &~0.20532442$\times10^{-5}$ &~0.56956135$\times10^{-5}$\\
~0.42302672$\times10^{-7}$ &-0.17453341$\times10^{-6}$ &-0.44240794$\times10^{-6}$\\
-0.22361094$\times10^{-8}$ &~0.10535983$\times10^{-7}$ &~0.25394314$\times10^{-7}$\\
~0.89005384$\times10^{-10}$&-0.45039190$\times10^{-9}$ &-0.10580995$\times10^{-8}$\\
-0.25735098$\times10^{-11}$&~0.13286312$\times10^{-10}$&~0.30978486$\times10^{-10}$\\
~0.50877050$\times10^{-13}$&-0.25608145$\times10^{-12}$&-0.60169351$\times10^{-12}$\\
-0.61321127$\times10^{-15}$&~0.28865188$\times10^{-14}$&~0.69356947$\times10^{-14}$\\
~0.33875686$\times10^{-17}$&-0.14345838$\times10^{-16}$&-0.35807821$\times10^{-16}$\\
\end{tabular}
\end{table}
%
%
\begin{table}
\caption{Coupling constants $\lambda_{ij}=\lambda_{ji}$ for the rank-3
$^3$F$_4$ potential.}
\begin{tabular}{c}
$\lambda_{11}=$-0.18792638$\times10^6$\\
$\lambda_{12}=$~0.45033170$\times10^5$\\
$\lambda_{13}=$-0.24918509$\times10^5$\\
$\lambda_{22}=$-0.11024146$\times10^5$\\
$\lambda_{23}=$~0.61549429$\times10^4$\\
$\lambda_{33}=$-0.34552195$\times10^4$\\
\end{tabular}
\end{table}
%
%
\begin{figure}[h]\centering
\psbox[size=0.5#1]{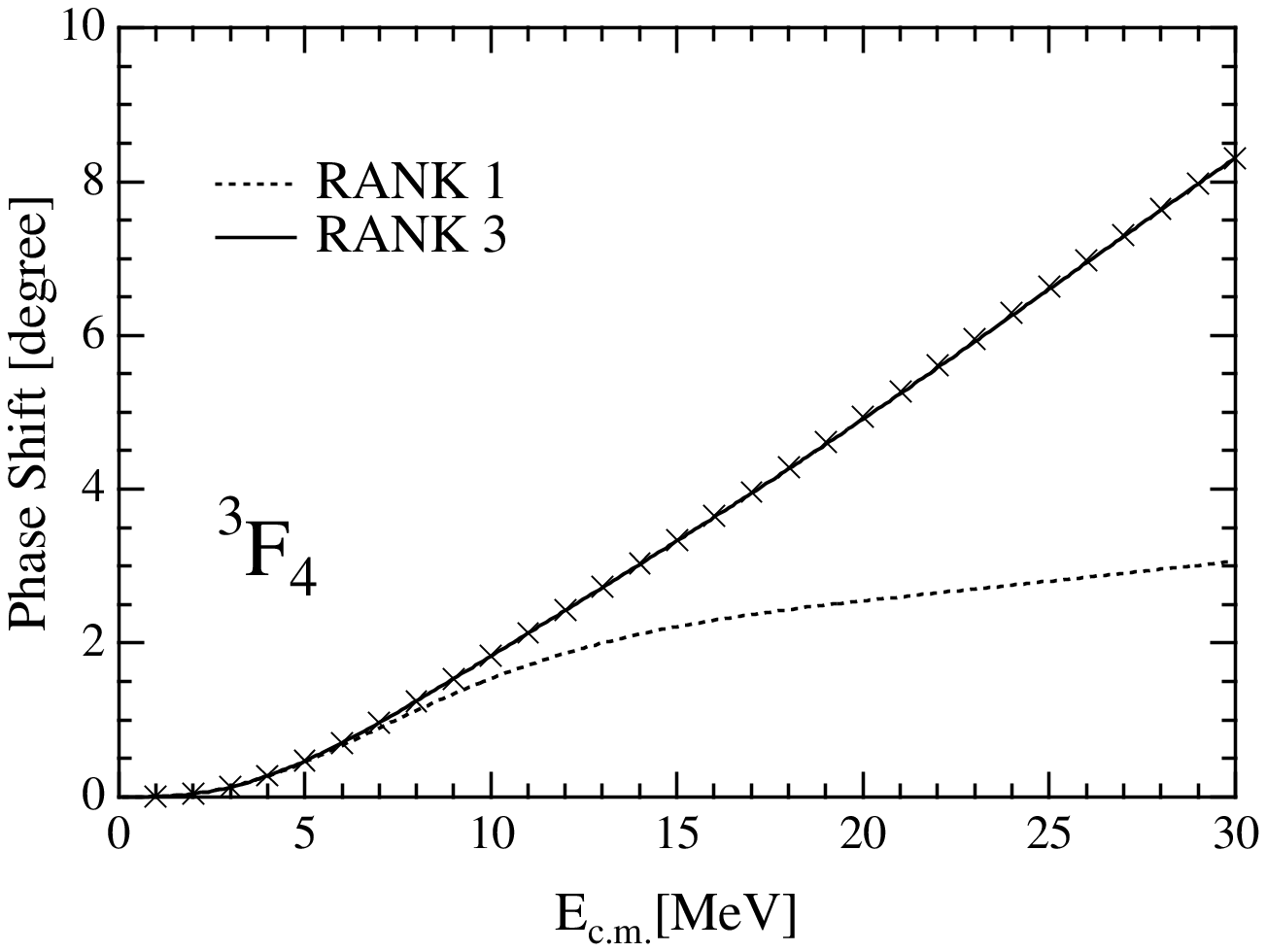}
\caption{Phase shifts for $^3$F$_4$ state without Coulomb effects. 
The crosses denote the RGM+OCM+LS result,
the dashed line the rank-1 result, and the solid line the rank-3  result, respectively.}
\end{figure}

In the higher energy region (above the break up threshold for 
$^3$He$\rightarrow$d+p), absorption effects stemming from 
inelasticity should be  considered. This requires the 
introduction of  an  imaginary component in the potential.
We modified our potential to reproduce 
the reflection parameters,  given by Yoshino {\it et al.}  \cite {yoshino}
at $E_{Lab}$=19.48\,MeV, using the anzatz
\widetext
\begin{equation}
	V^{\rm C(LS)}  =   \{ A C_0
	+A^* [C_1 (\mbox{\boldmath $L \cdot S$})
  	+C_2 (\mbox{\boldmath $L \cdot S$})^2
     	 +C_3 (\mbox{\boldmath $L \cdot S$})^3]\}V^{\rm MOCM},
\label{com}
\end{equation}
\narrowtext
where the parameters $C_0$, $C_1$, $C_2$, $C_3$ are the same as 
in the $V^{\rm OCM(LS)}$ potential.
And the complex factor $A$ is chosen as follows
\begin{equation}
	 A=1+i a\{1-e^{-bE_{th}^2}\}\theta(E_{th})
\label{complA}
\end{equation}
where $A^*$ is the complex conjugate of $A$,
and $a$=21.0, 
$b$=$2.50\times 10^{-5}$, respectively.
The break up threshold energy $E_{th}$ is given by $E_{th}=E-E_h+E_d$,
$E_h$ and $E_d$ are the $^3$He and the deuteron binding energies respectively,
while $\theta(x)$ is the step function.
Using this complex potential, we calculated the phase shifts   
and found that they change, at most, by a few degrees  in the region
of  19.48\,MeV. Therefore, we do not present these results 
in this paper.
%
%

\section{Discussion and Conclusions} 
We constructed the p-$^3$He effective potentials 
up to $L$=3 based on the RGM combined with the OCM technique which removes
the PFS. These models do not include either an LS or a tensor force,
which were not included in the nucleon-nucleon potential used
to construct the nucleon-trinucleon potential.
In the present work, we introduce the LS interaction phenomenologically 
at an intercluster level. This successfully  describes  the degeneracy in 
the spin--triplet 
($^3$P$_0$,$^3$P$_1$, $^3$P$_2$),
($^3$D$_1$, $^3$D$_2$,$^3$D$_3$),
and
($^3$F$_2$, $^3$F$_3$,$^3$F$_4$) channels.
The tensor force  could be similarly included. However, these forces 
could be safely omitted as their influence on the data
 was shown, in the recent work
of Yoshino {\it et al.}  \cite {yoshino}, to be very small.

In the higher energy region (above the break up threshold),
absorption effects can be included via Eq. (\ref{com}). Since, however,
our main concern is the construction of potentials
at low energies, little attention was given to
the construction of  complex potential (see forth Eq. (\ref{complA})).

The  theoretical (RGM)  phase shifts 
are well reproduced by our potentials. 
As far as the experimental phase shifts are concerned, 
it should be noted that the results obtained by various analyses are not in agreement to each other.
However, our potentials
fit the  most recent phase shifts \cite{yoshino} quite well.
Furthermore, we compared our results with those of Ref. \cite{Pisent},
which are based on a Yamaguchi type separable potential fitted 
to lower energy scattering data. The latter potentials reproduce  
well the differential cross sections at very low energies.

\begin{figure}[h]\centering
\psbox[size=0.6#1]{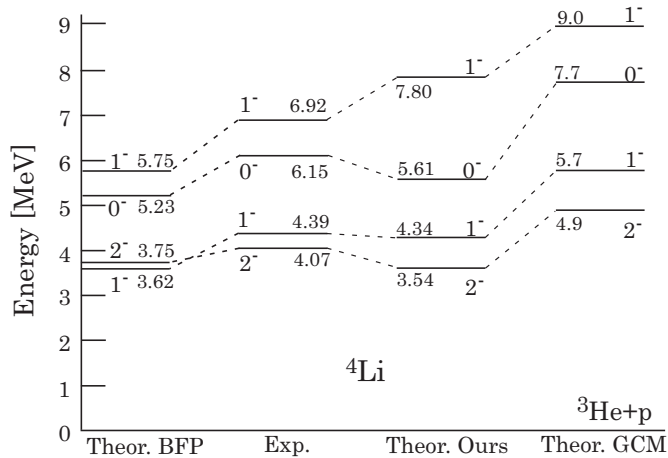}
\caption{$^4$Li resonance energy level above the p--$^3$He threshold.
The experimental data are  taken from Ref. [22], 
Theor. BFP  are the results of Ref. [10], Theor. GCM by Ref. [21], and Theor. Ours are the
results of  the present work.}
\label{spectrum}
\end{figure}

Special attention was paid to effects of non-central forces in each of the
negative parity states, $2^{-}$, $1^{-}$,  $0^{-}$.
We derived the energy spectrum by using the  
phase-shift data by Beltramin {\it et al.} (BFP) Ref. \cite{Pisent} 
employing the 
three-dimensional Spline function interpolation method.   
Our result is compared with the BFP spectrum and with the GCM 
calculation Ref. \cite{furutani}
as well as with the experimental data. 
The results are plotted in Fig. \ref{spectrum}.
The first and the second lowest levels of the BFP crossed each other.
The GCM calculation reproduces the ordering of four negative parity states
but the spectrum is totally shifted to a higher energy region,
while our results are in good agreement with the experimental
data in which the lowest level $2^{-}$ is due to the  $^3$P$_2$ 
state, the second lowest $1^{-}$ level to  $^3$P$_1$,
the $0^{-}$ level to $^3$P$_0$, and 
the excited $1^{-}$ level  to the $^1$P$_1$ state.

From the overall results, we may conclude  that our potentials
(RGM+OCM plus phenomenological LS force) reproduce the 
scattering and resonance state data of the p-$^3$He system
well. Concerning the OCM results, we note that they
are essentially on-shell equivalent to the RGM one.
However, we found that there are some differences, albeit small,
in the  phase shift  for the P and D waves.

Finally,  our EST separable  potentials can be used in few-cluster 
systems where p-$^3$He is involved.  One such reaction  is
$^3$He(d,p)$^4$He, which can be treated as a three-body system,
 p-n-$^3$He, within the Faddeev integral equations.
Moreover, they can be used for interpretation purposes
by  constructing local equivalent interactions which can 
provide us a further insight into the interaction between light clusters,
their characteristics concerning shape, range {\it etc.}, as well as 
their bound and  resonance structure. 
The understanding of the interaction between light clusters will pave the 
way for a better treatment of few-cluster systems and their
role in nuclear reactions  and primordial nucleosynthesis.  
%
\acknowledgments
The authors wish to thank H. Kamada, S. Nakamura, K. Samata, and T.Yamada for valuable discussions and technical supports.
This work was done using the computers in {\it FRCCS} of Science University of Tokyo,
{\it RIKEN}, National Astronomical Observatory of Japan, and National Institute for Fusion Science.

\appendix
\section{Resonating Group Method and the Orthogonality Condition Model}
In what follows we outline for convenience the RGM for the p-$^3$He system.
The total wave function for the p-$^3$He system is defined by
\begin{equation}
	\Psi={\cal A}[\phi_{\rm CL}\,\xi(\sigma,\tau)
        \,\chi(\mbox{\boldmath $R$}_N-\mbox{\boldmath $R$}_{\rm CL})] ,
\end{equation}
where ${\cal A}$ is the antisymmetrization operator, 
$\xi(\sigma,\tau)$ the isospin function,  
and $\chi(\mbox{\boldmath $R$}_N-\mbox{\boldmath $R$}_{\rm CL})$ the relative 
wave function between the clusters,  while 
$\phi_{\rm CL}$ is the internal wave function 
of the clusters which is defined by superimposing two Gaussian functions
\begin{eqnarray}\label{intwf}
	\phi_{\rm CL} & = & \exp[-\frac{1}{2}\alpha_1\sum_{i=1}^3
	(\mbox{\boldmath $r$}_i-\mbox{\boldmath $R$}_{\rm CL})^2]
\nonumber
\\
          & + & c\exp[-\frac{1}{2}\alpha_2\sum_{i=1}^{3}
	(\mbox{\boldmath $r$}_i-\mbox{\boldmath $R$}_{\rm CL})^2] .
\end{eqnarray}
Here $\mbox{\boldmath $R$}_{\rm CL}$ is the center of mass of the $^3$He cluster and
$\mbox{\boldmath $r$}_i$ are the coordinates of nucleons within the  $^3$He cluster.
The parameters $\alpha_1$, $\alpha_2$, and $c$ are taken from 
Ref. \cite{rgm}, namely, $\alpha_1=0.25$ fm$^{-2}$, $\alpha_2=0.71$ fm$^{-2}$, 
and $c=3.17$.
The Hamiltonian is given by
\begin{equation}
	H=H_0+\sum_{i>j}\,V_{ij},
\end{equation}
where $H_0$ is the total kinetic energy and  $V_{ij}$ are the interactions 
between the  $(i,j)$ nucleons. The relative function 
$\chi(\mbox{\boldmath $R$}_N-\mbox{\boldmath $R$}_{\rm CL})$ can be evaluated  variationally by considering
\begin{equation}
\label{henbun}
		\langle \delta \Psi | H - E' | \Psi \rangle = 0 ,
\end{equation}
where $E'$ is the total energy of  the whole system. In this model, 
the total energy is separated into the internal 
energy of the cluster $E_{int}$ and the relative energy between 
the colliding clusters $E$, {\em i.e.} $E'=E+E_{int}$.
Using Eq. (\ref{intwf}) in  Eq. (\ref{henbun}) one can obtain
the following  integro-differential equation,
\begin{equation}
\label{rgmeq}
	\left(\frac{\hbar^2}{2\mu}\nabla_{\mbox{\boldmath $r$}}^2-V_D(\mbox{\boldmath $r$})+E\right)
	\chi(\mbox{\boldmath $r$})
	=\int_{0}^{\infty} K(\mbox{\boldmath $r$},\mbox{\boldmath $r$}')
	\chi(\mbox{\boldmath $r$}')\, {\rm d}\mbox{\boldmath $r$}' ,
\end{equation}
where $V_D$ is the Direct local potential and $K({\bf r},{\bf r}')$ is 
the corresponding nonlocal one which consists of three terms
\begin{equation}
	K(\mbox{\boldmath $r$},\mbox{\boldmath $r$}')=
	K_T(\mbox{\boldmath $r$},\mbox{\boldmath $r$}')+
	K_V(\mbox{\boldmath $r$},\mbox{\boldmath $r$}')
	+E'{\cal N}(\mbox{\boldmath $r$},\mbox{\boldmath $r$}') ,
\end{equation}
where $K_T$  is  obtained from the kinetic energy term, $K_V$  from the 
potential term, and ${\cal N}$ is often referred to as the norm-integral kernel or the norm kernel.
In operator form  Eq. (\ref{rgmeq}) is  written as
\begin{equation}\label{rgmeq2}
	(h_0+V_D+K_T+K_V+E'{\cal N})\chi=E\chi
\end{equation}
where $E$ and $h_0$ are the relative energy and the kinetic operator 
between the  $^3$He cluster and the  proton, respectively.
The RGM (effective) potential can be identified to
\begin{equation}
\label{rgmpot}
	V^{\rm RGM} (E) \equiv V_D+K_T+K_V+E'{\cal N}=W+E{\cal N}
\end{equation}
with
\begin{equation}
	W=V_D+K_T+K_V+E_{int}{\cal N} .
\end{equation}
The normalization of the total wave function of the system is given by
\begin{equation}
	1 = \langle \Psi | \Psi \rangle =\langle \phi_a \phi_b \chi
    	| {\cal A} [\phi _a \phi_b \chi]\rangle
  	= \langle \chi | (1 - {\cal N}) |  \chi \rangle 
	\equiv \langle \psi | \psi \rangle ,
\end{equation}
where $\phi_a$ is the $^3$He cluster intrinsic wave function
while $\phi_b$ is a single nucleon wave function.
It is, however, known that  the norm of the relative function 
$\chi$ is not one and therefore one defines  the function $\psi$,
\begin{equation}
\label{renormwf}\label{renorm}
	\psi =\sqrt{1-{\cal N}}\chi\,,
\end{equation}
which  has the proper probability density interpretation as the physical wave function.

An additional problem with the RGM formalism is the 
existence of PFS. These states can be removed using the OCM technique
in which Eq. (\ref{rgmeq2}) is rewritten as follows,
\begin{equation}
	(h_0+V_D+K_T+K_V+E_{int}{\cal N})\chi = E(1-{\cal N}) \chi.
\end{equation}
Furthermore, using Eq. (\ref{renormwf}) one obtains
\widetext
\begin{equation}
	\frac{1}{\sqrt{1-{\cal N}}}(h_0+V_D+K_T+K_V+E_{int}
	{\cal N})\frac{1}{\sqrt{1-{\cal N}}}\psi = E \psi.
\end{equation}
Therefore,  the Pauli corrected intercluster potential is 
\begin{equation}
\label{ocmp1}
	V^{{\rm OCM}}=\frac{1}{\sqrt{1-{\cal N}}}\left (\,h_0+V_D+K_T
	+K_V+E_{int}{\cal N}\,\right)\frac{1}{\sqrt{1-{\cal N}}}-h_0\,.
\end{equation}
\narrowtext
A Pauli forbidden state is removed  when the corresponding eigen-value 
of $\sqrt{1-{\cal N}}$  vanishes.  It should be noted that in this
representation the energy dependence of the potential is eliminated.

The potential  $V^{{\rm OCM}}$ in 
 Eq. (\ref{ocmp1}) is given in  operator form. It can, however, be easily 
brought into a  more  suitable form  for numerical 
calculation.  For this, when the cluster wave function is given 
by one Gaussian term, the eigen-function of the norm kernel 
${\cal N}$ is found using the harmonic oscillator function 
\cite{aal}.  In the present case, however, where the $^3$He wave function is 
given by two Gaussian terms, the 
norm eigen-function is a  superposition 
of  harmonic oscillator functions. Now, the norm kernel 
${\cal N}$ is given by
\begin{equation}
\label{normk}
	{\cal N}=\sum_{ij=1}^{\infty}|U_i\rangle G_{ij} \langle U_{j}|
\end{equation}
where 
\begin{equation}
	G_{ij}\equiv \langle U_i|{\cal N}|U_j\rangle.
\end{equation}
The matrix elements of the 
norm kernel analytically while the matrix can be diagonalized numerically using the Jacobo method.
For this, Eq. (\ref{normk}) is written as
\begin{eqnarray}
	{\cal N} & = & \sum_{ijk=1}^{\infty}| U_i 
		\rangle a_{ik} \gamma_k a^{*}_{kj}
		\langle U_j |\nonumber
\\
             & = & \sum_{k=1}^{\infty}|\tilde{U}_k \rangle \gamma_k \langle
\tilde{U}_k |
\end{eqnarray}
where
\begin{equation}\label{hofex}
	|\tilde{U}_k \rangle \equiv \sum_{i=1}^{\infty} | U_i \rangle a_{ik}\,.
\end{equation}
Furthermore, the term $\gamma_k$ is the eigen-value of the norm kernel and
$\tilde{U}_k$ is the corresponding eigen-function.  
 Eq. (\ref{ocmp1}) then becomes
\widetext
\begin{equation}
\label{ocmp2}
	V^{{\rm OCM}}=\sum_{ij=1}^{\infty}| \tilde{U}_i \rangle
	[\frac{1}{\sqrt{1-\gamma_i}}(h_{0ij}+W_{ij})
	\frac{1}{\sqrt{1-\gamma_j}}-h_{0ij}]\langle \tilde{U}_j|
\end{equation}
\narrowtext
where 
\begin{eqnarray}
	h_{0ij} & \equiv & \langle \tilde{U}_i| h_0 | \tilde{U}_j \rangle\,,
\label{h0}
\\
	W_{ij}  & \equiv & \langle \tilde{U}_i| W | \tilde{U}_j \rangle\,.
\end{eqnarray}
%
%
\section{Momentum Representation}
In what follows we also present the above potential 
in momentum space suitable for the Alt-Grassberger-Sandhas (AGS)
equations \cite{AGS}. The momentum representation is introduced using 
the Fourier transforms,
\begin{eqnarray}
	{\cal F}\{f_\ell^{(0)}\} & = & 4\pi \int_0^\infty f_\ell^{(0)}
	\cdot j_\ell(kr)\cdot j_\ell(k'r)r^2 {\,\rm d}r\,,
\label{fouriertrans1}
\\
	{\cal F}\{f_\ell^{(n)}\} & = & 4\pi \int_0^\infty f_\ell^{(n)}
	\cdot j_\ell(kr)\cdot j_\ell(k'r') {\,\rm d}r{\,\rm d}r'\,.
\label{fouriertrans2}
\end{eqnarray}
The potential consist of the  following set of functions \cite{rgm}.
\widetext
\begin{eqnarray}
	f_\ell^{(0)}(a;r)& = & \exp(-a r^2)\label{gaussian1}\\
	f_\ell^{(1)}(a,a':r,r') & = & (-i)^\ell rr'\exp(-ar^2-a'r'^2) j_\ell(ibrr')
\label{gaussian2}
\\
	f_\ell^{(2)}(a,b:r,r') & = & (-i)^\ell
	rr'(r^2+r'^2)\exp(-ar^2-ar'^2)j_\ell(ibrr')
\label{gaussian3}
\\
	f_\ell^{(3)}(a,b:r,r') & = & (-i)^\ell
	rr'\exp(-ar^2-ar'^2)\{(ibrr')j_{L+1}(ibrr')-L\cdot j_\ell(ibrr')\}
\label{gaussian4}
\\
	f_\ell^{(4)}(a,a':r,r') & = & rr'\exp(-ar^2-a'r'^2)
\label{gaussian5}
\\
	f_\ell^{(5)}(a:r,r') & = & rr'(r^2+r'^2)\exp(-ar^2-ar'^2)
\label{gaussian6}
\\
	f_\ell^{(6)}(a,a',b,c,c':r,r') & = & (-i)^\ell rr'(cr^2+c'r'^2)\exp(-ar^2-a'r'^2)j_\ell(ibrr')
\label{gaussian7}
\\
	f_\ell^{(7)}(a,a',b:r,r') & = & -brr'\exp(-ar^2-a'r'^2)
	\{i^{L+1}rr'j_{L+1}(-ibrr')
	+i^\ell\frac{L}{b}j_\ell(-ibrr')\}
\label{gaussian8}
\end{eqnarray}
Using the definitions (\ref{fouriertrans1}) and (\ref{fouriertrans2})
we obtain  
\begin{eqnarray}
	{\cal F}\{f_\ell^{(0)}\} & = &
	(-1)^\ell(\frac{\pi}{a})^{3/2}\frac{2a}{kk'}\exp(-\frac{k^2+k'^2}
	{4a}){\cal J}_{L+1/2}(\frac{kk'}{2a})
\\
	{\cal F}\{f_\ell^{(1)}\} & = & 2\pi^2\frac{-1}
		{\sqrt{4aa'-b^2}}\frac{^1}{bkk'}
	\exp(-\frac{a'k^2+ak'^2}{4aa'-b^2})
	{\cal J}_{L+1/2}(-\frac{bkk'}{4aa'-b^2})
\\
	{\cal F}\{f_\ell^{(2)}\} & = & \frac{8abkk'}{(4a^2-b^2)^2}{\cal
	I}_{L+1}(a,a,b:k,k')
\nonumber
\\
               & + & [\frac{4a(2L+3)}{4a^2-b^2}-\frac{(4a^2+b^2)
		(k^2+k'^2)}{(4a^2-b^2)^2}]{\cal I}_\ell(a,a,b:k,k')
\\
	 {\cal F}\{f_\ell^{(3)}\} & = & -[L+\frac{(2L+3)b^2}{4a^2-b^2}
	-\frac{2ab^2(k^2+k'^2)} {4a^2-b^2)^2}]{\cal I}_\ell(a,a,b:k,k')
\nonumber
\\
               & - & \frac{(4a^2+b^2)bkk'}{(4a^2-b^2)^2}{\cal
	I}_{L+1}(a,a,b:k,k')
\\
	{\cal F}\{f_\ell^{(4)}\} & = & \frac{\pi^2}{4}(\frac{1}{aa'}
	)^{3/2}\exp(-\frac{a'k^2+ak'^2}{4aa'})
\\
	{\cal F}\{f_\ell^{(5)}\} & = &
	\frac{\pi^2}{4a^4}[3-\frac{(k^2+k'^2)}{4a}]
	\exp(-\frac{k^2+k'^2}{4a})
\\
	{\cal F}\{f_\ell^{(6)}\} & = & [c\{\frac{4a(L+3/2)}{4aa'-b^2}
	-\frac{4a'^2k^2+b^2k'^2}
	{(4aa'-b^2)^2}\}+c'\{\frac{4a(L+3/2)}{4aa'-b^2}
	-\frac{4a^2k'^2+b^2k^2}
	{(4aa'-b^2)^2}\}]{\cal I}_\ell(a,a',b:k,k')
\nonumber
\\
        & + & \frac{4bkk'(a'c+ac')}{(4aa'-b^2)^2}{\cal
	I}_{L+1}(a,a',b:k,k')
\\
	{\cal F}\{f_\ell^{(7)}\} & = &
	\{\frac{2b(L+3/2)}{4aa'-b^2}+\frac{L}{b}-\frac{2b(a'k^2+ak'^2)}
	{(4aa'-b^2)^2}\}{\cal I}_\ell(a,a',b:k,k')
	+  \frac{kk'(4aa'+b^2)}{(4aa'-b^2)^2}{\cal
	I}_{L+1}(a,a',b:k,k')\,.
\end{eqnarray}
\narrowtext
In the above
\begin{equation}
	{\cal I}_\ell(a,a',b:k,k')  \equiv  {\cal F}\{f_\ell^{(1)}\}
\end{equation}
while ${\cal J}_\ell$ is the Hyperbolic Spherical Bessel Function 
related to the Spherical Bessel Function $j_\ell$ by
\begin{equation}
	{\cal J}_{\ell+1/2}(x)=i^\ell \, x\, j_\ell(ix)\,.
\end{equation}
%


\begin{references}
\bibitem{Dstate}
	T. Uesaka {\it et al.}, Phys. Lett. {\bf B421}, 22 (1997);
        T. Uesaka {\it et al.}. Nucl. Phys. {\bf A684}, 606c (2001).
\bibitem{interfaddeev1} 
	S. Oryu, S. Gojuki, T. Hino, E. Uzu, and H. Kamada,
	Few-Body Systems Suppl. {\bf 10}, 371 (1999).
\bibitem{interfaddeev2} 
	S. Gojuki, H. Kamada, E. Uzu, and S. Oryu,
	Few-Body Systems Suppl. {\bf 12}, 501 (2000).
\bibitem{multifaddeev} 
	S. Gojuki, S. Nemoto, S. Oryu, E. Uzu, and H. Kamada,
	Nucl. Phys. {\bf A684} , 629c (2001).
\bibitem{Pot}
        B. S. Podmore and H. S. Sherif, in {\em Few-Body Problems 
	and Particle Physics}, ed. by R. J. Slobodrian {\em et al.} (Laval
        Univ. Press, Quebec City, Canada, 1995) p 517; H. S. Sherif and 
        B. S. Podmore,  in {\em Few Particle Problems in the Nuclear 
        Interaction}, Ed. I. Slaus {\em et al.}, (North-Holland publ, 1972)
        p. 691;  H. S. Sherif, Phys. Rev {\bf C19}, 1649 (1979).
\bibitem{Neu}
        V. G. Neudachin, A. A. Sakharuk, and Yu. F. Smirnov, 
        Sov. J. Part. Nucl. {\bf 23}, 210 (1992) (Fiz. Elem. 
        Chastits {\bf 23}, 479 (1992).
\bibitem{Oers}
        W. T. H. van Oers {\em et al.},  Phys. Rev.  C {\bf 65}, 390 (1982).
\bibitem{Sciav}
        R. Sciavilla, Phys. Rev. Lett. {\bf 65}, 835 (1990).
\bibitem{Fied}
        H. Fiedeldey, S. A. Sofianos, and G. Ellerkmann,
        Few-Body Systems {\bf 18}, 173 (1995).
\bibitem{Pisent}
        L. Beltramin, R. del Frate, and G. Pisent,
	Nucl. Phys. {\bf A442}, 266 (1985).
\bibitem{exp1} 
	T. A. Tombrello, C. M. Jones, G. C. Phillips, and J. L. Weil,
	Nucl. Phys. {\bf 39}, 541 (1962).
\bibitem{exp2} 
	T. A. Tombrello, Phys. Rev. {\bf 138}, B40 (1965).
\bibitem{exp3} 
	L. Drigo and G. Pisent, Nuovo Cimento {\bf 51}, 419 (1967).
\bibitem{exp4} 
	D. H. Mc Sherry and S. D. Baker, Phys. Rev. {\bf C1}, 888 (1970).
\bibitem{exp5} 
	R. Darves-Blanc, N. van Sen, J. Arvieux, and J. C. Gondrand,
	Nucl. Phys. {\bf A191}, 353 (1972)
\bibitem{exp6} 
	J. R. Morales, T. A. Cahill, and D. J. Shadoan, Phys. Rev.
	{\bf C11}, 1905 (1975).
\bibitem{exp7} 
	D. M\"{u}ller, R. Beckmann, and U. Holm, Nucl. Phys. 
	{\bf A311}, 1 (1978).
\bibitem{exp8} 
	H. Berg, W. Arnold, E. Huttel, H. H. Krause, J. Ulbricht,
	and G. Clausnitzer, Nucl. Phys. {\bf A334}, 21 (1980).
\bibitem{yoshino} 
	Y. Yoshino, V. Limkaisang, J. Nagata, H. Yoshino, and M.
	Matsuda, Prog. Theor. Phys. {\bf 103}, 107 (2000).
\bibitem{rgm} 
	I. Reichstein, D. R. Thompson, and Y. C. Tang, Phys. Rev.
	{\bf C3}, 2139 (1971).
\bibitem{furutani}
	H. Furutani, H. Horiuchi, and R. Tamagaki,
	Prog. Theor. Phys. {\bf 62}, 981 (1979)
	\bibitem{level}
	D. R. Tilley,H. R. Weller, and G. M. Hale, Nucl. Phys. {\bf A541}, 1 (1992).
\bibitem{cssm} 
        M. Teshigawara, K. Kat\={o}, and G. F. Filippov, 
		Prog. Theor. Phys, {\bf 92}, 79 (1994);
	M. Teshigawara, K. Kat\={o}, and G. F. Filippov private communication.
\bibitem{unified} 
	K. Wildermuth and Y. C. Tang, {\it A Unified Theory of the
	Nucleus} (Vieweg, Braunschweig, 1977).
\bibitem{ocm} 
	S. Saito, Prog. Theor. Phys. {\bf 40}, 893 (1968);
	S. Saito, Prog. Theor. Phys. {\bf 41}, 705 (1969).
\bibitem{aal} 
	S. Oryu, K. Samata, T. Suzuki, S. Nakamura, and H. Kamada,
	Few Body Systems {\bf 17}, 185 (1994).
\bibitem{lsp} 
	L. C. McIntyre and W. Haeberli, Nucl. Phys. {\bf A91}, 382 (1967).
\bibitem{EST}
	D. J. Ernst, C. M. Shakin, and R. M. Thaler,
	Phys. Rev. {\bf C8}, 46 (1973).
\bibitem{hackenbriich}
	P. Heiss and H. H. Hackenbroich, Nucl. Phys. {\bf A182}, 522 (1972)  
\bibitem{gse1} 
	S. Oryu, M. Araki, S. Satoh, Prog. Theor, Phys. Suppl.
	 {\bf 61}, 199 (1977).
\bibitem{gse2} 
	S. Oryu, Phys. Rev. {\bf C27}, 2500 (1983).
\bibitem{renorm} 
        V. G. Gorshkov, Soviet Phys. JETP {\bf 13}, 1037 (1961);
	M. L. Goldberger and K. M. Watson, ``Collision Theory''
	John Wiley and Sons, Inc. New York, London, Sydney 1964;
	E. O. Alt, W. Sandhas and H. Ziegelmann, Phys. Rev. {\bf C17},1981 (1978).
\bibitem{AGS}          
        P. Grassberger and W. Sandhas, Nucl. Phys. {\bf B2}, 181 (1967);
        E. O. Alt, P. Grassberger, and W. Sandhas, Nucl. Phys.  
        {\bf B2}, 167 (1967).
\end{references}
\end{document}